\documentclass{article}

\usepackage{xcolor}
\usepackage{authblk}

	\usepackage{graphicx}
	\usepackage{amsmath}
	\usepackage{amssymb}
	\usepackage{float}
	\usepackage{placeins}

	\definecolor{visprocessgray}{RGB}{127,127,127}

	\definecolor{RGcolor}{RGB}{49,121,198}

	\definecolor{newsleakunitcolor}{RGB}{144,144,144}
	
	\definecolor{frontbackendcolor}{RGB}{80,159,198}
	
	\definecolor{dataTypeColor}{RGB}{150,98,208}
	
	\definecolor{inputoutputcolor}{RGB}{22,166,3}
	
	\definecolor{apicolor}{RGB}{148,23,81}
	
	\definecolor{stepcolor}{RGB}{47,82,143}
	
	\definecolor{lisaRed}{RGB}{237,28,35}

	\definecolor{lisaBlue}{RGB}{62,70,200}

	\definecolor{lisaBlue2}{RGB}{1,162,232}

	\definecolor{lisaYellow}{RGB}{255,201,14}

	\definecolor{lisaGreen}{RGB}{33,177,76}

	\definecolor{colorbrewerorange}{RGB}{253,140,90}
	\definecolor{colorbreweryellow}{RGB}{255,192,1}
	\definecolor{colorbrewergreen}{RGB}{146,207,94}

	\definecolor{C1}{RGB}{231,69,51}
	\definecolor{C2}{RGB}{26,198,83}
	\definecolor{C3}{RGB}{56,110,165}
	\definecolor{C4}{RGB}{11,84,1}
	\definecolor{C5}{RGB}{253,148,7}
	\definecolor{C6}{RGB}{0,0,109}
	\definecolor{C7}{RGB}{230,28,121}
	\definecolor{C8}{RGB}{118,13,21}

	\definecolor{darkYellow}{RGB}{255,210,0}

	\definecolor{bc1}{HTML}{66C2A5}
	\definecolor{bc2}{HTML}{FC8D62}
	\definecolor{bc3}{HTML}{8DA0CB}

	\definecolor{transcriptblack}{RGB}{0,0,0}

	\definecolor{grayText}{RGB}{102,102,102}

	\definecolor{factorOne}{RGB}{255,47,146}
	\definecolor{factorTwo}{RGB}{3,176,240}

	\definecolor{darkblue}{HTML}{005CBB}
	\definecolor{darkgreen}{HTML}{307F25}
	\definecolorseries{gradient}{rgb}{last}{black!3}{black!50}
	\resetcolorseries[100]{gradient}

	\usepackage{color}
	\usepackage{booktabs}
	\usepackage{calc}
	\usepackage{colortbl}
	\usepackage{comment}
	\usepackage{ifthen}

	\newcommand{\mybarecce}[1]{%
		\textcolor{black}{\rule{0.25pt+1pt*\real{#1}}{1.5ex}}\hfill
	    \ifthenelse{\equal{#1}{0}}{\textcolor{black}{{\scriptsize #1}}}{\textcolor{black}{{\footnotesize #1}}}
	}

	\definecolor{cRanking}{HTML}{065700}

	\newcolumntype{a}{>{\columncolor{black!10}}p{6.4em}}
	\newcolumntype{b}{>{\columncolor{black!0}}p{6.4em}}

	\definecolor{orange}{RGB}{255,127,0}
	\newcommand{\orange}[1]{\textcolor{orange}{#1}}
	\definecolor{blue}{RGB}{0,128,255}

	\definecolor{lilac}{RGB}{158,188,218}

	\definecolor{myRed}{RGB}{177,36,24}

	\definecolor{myLightBlue}{RGB}{75,174,234}

	\definecolor{myLilac}{RGB}{191,162,209}

	\definecolor{myGreen}{RGB}{76,173,91}

	\newcommand{\replicatingExaminer}[1]{\emph{replicating examiner}\,}
	\newcommand{\originalExaminer}[1]{\emph{original examiner}\,}
	\newcommand{\originalExperiment}[1]{\emph{original experiment}\,}
	\newcommand{\replicatingExperiment}[1]{\emph{replicating experiment}\,}

	\definecolor{greenSC}{RGB}{0,176,80}

	\definecolor{lilac2}{RGB}{104,52,154}
	
	\definecolor{blue2}{RGB}{56,84,146}

	\newcommand{\sampleselector}[1]{\textsc{{\large S}ample{\Large S}elector}}
	\newcommand{\iv}[1]{independent variable}
	\newcommand{\gv}[1]{grouping variable}
	\newcommand{\cv}[1]{confounding variable}
	\newcommand{\con}[1]{constraint}
	\newcommand{\gss}[1]{goal sample size}
	\newcommand{\es}[1]{experiment sample}
	\newcommand{\evs}[1]{experiment variables}
	\newcommand{\dpdh}[1]{data-property-driven hypothesis}
	\newcommand{\erps}[1]{experiment-relevant properties}
	
	\definecolor{orangeAlena}{RGB}{237,125,49}
	\newcommand{\orangeAlenaColor}[1]{\textcolor{orangeAlena}{#1}}

	\usepackage{amsmath}
	\usepackage{amssymb}
	\usepackage{color}
	\usepackage[normalem]{ulem}
	\usepackage{booktabs}
	\usepackage{calc}
	\usepackage{overpic}
	\usepackage{colortbl}
	\usepackage{hyperref}
	\usepackage{gensymb}
	
	\usepackage[anythingbreaks,hyphenbreaks]{breakurl}
	\usepackage{makecell}

	\extrafloats{100}

	\usepackage{subfig}

\usepackage[subfigure]{tocloft}

	\usepackage{enumitem}
	\definecolor{yellow_min_cirlcle}{RGB}{255,192,0}
	
	\definecolor{blue_max_cirlcle}{RGB}{0,176,240}
	
	\definecolor{red_large_cirlcle}{RGB}{255,0,0}
	
	\usepackage{tocloft}
	\usepackage{geometry}
	\newcommand{\listequationsname}{List of Equations}
	\newlistof{myequations}{equ}{\listequationsname}
	\newcommand{\myequations}[1]{%
	\addcontentsline{equ}{myequations}{\protect\numberline{\theequation}#1}\par}
	\usepackage{float}
	\usepackage{aliascnt}
	\newaliascnt{eqfloat}{equation}
	\newfloat{eqfloat}{h}{eqflts}
	\floatname{eqfloat}{Equation}

	\newcommand*{\ORGeqfloat}{}
	\let\ORGeqfloat\eqfloat
	\def\eqfloat{%
	  \let\ORIGINALcaption\caption
	  \def\caption{%
	    \addtocounter{equation}{-1}%
	    \ORIGINALcaption
	  }%
	  \ORGeqfloat
	}

	\definecolor{orange}{RGB}{235,125,0}
	\definecolor{green}{RGB}{146,208,80}
	\definecolor{darkblue}{RGB}{34,75,140}
	\definecolor{pink}{RGB}{252, 214, 255}
	\newcommand{\orangeAlena}[1]{\textcolor{orange}{#1}}
	\newcommand{\greenAlena}[1]{\textcolor{green}{#1}}
	\newcommand{\darkblueAlena}[1]{\textcolor{darkblue}{#1}}
	\newcommand{\pinkAlena}[1]{\textcolor{pink}{#1}}

	\definecolor{GTblue}{RGB}{38, 7, 255}
	\newcommand{\GTblueColor}[1]{\textcolor{GTblue}{#1}}

	\definecolor{HLdifferencesgreen}{RGB}{127, 172, 85}
	\newcommand{\HLdifferencesgreenColor}[1]{\textcolor{HLdifferencesgreen}{#1}}

	\definecolor{prblue}{RGB}{52, 115, 159}
	\newcommand{\PRBluecolor}[1]{\textcolor{prblue}{#1}}
	\definecolor{pryellow}{RGB}{252, 132, 56}
	\newcommand{\PRYellowcolor}[1]{\textcolor{pryellow}{#1}}
	\definecolor{prgreen}{RGB}{44, 146, 43}
	\newcommand{\PRGreencolor}[1]{\textcolor{prgreen}{#1}}
	\definecolor{prred}{RGB}{216, 57, 62}
	\newcommand{\PRRedcolor}[1]{\textcolor{prred}{#1}}

	\definecolor{prorange}{RGB}{255, 143, 43}
	\newcommand{\PROrangecolor}[1]{\textcolor{pryellow}{#1}}

	\definecolor{randomGT}{RGB}{164, 205, 234}
	\newcommand{\randomGTcolor}[1]{\textcolor{randomGT}{#1}}
	\definecolor{humanLikeDetected}{RGB}{138, 206, 127}
	\newcommand{\humanLikeDetectedcolor}[1]{\textcolor{humanLikeDetected}{#1}}

	\newcommand{\datasetSparseAGone}{\includegraphics[scale=0.55]{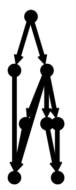}}
	\newcommand{\datasetSparseAGtwo}{\includegraphics[scale=0.55]{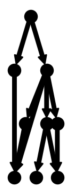}}
	\newcommand{\datasetSparseAGdiff}{\includegraphics[scale=0.55]{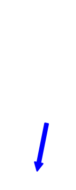}}
	\newcommand{\datasetSparseBGone}{\includegraphics[scale=0.55]{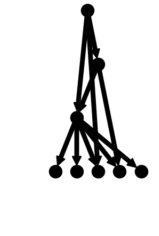}}
	\newcommand{\datasetSparseBGtwo}{\includegraphics[scale=0.55]{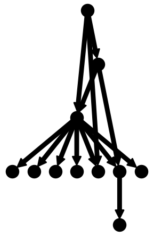}}
	\newcommand{\datasetSparseBGdiff}{\includegraphics[scale=0.55]{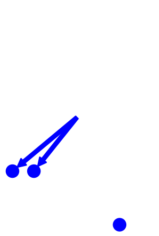}}
	\newcommand{\datasetSparseCGone}{\includegraphics[scale=0.55]{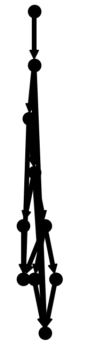}}
	\newcommand{\datasetSparseCGtwo}{\includegraphics[scale=0.55]{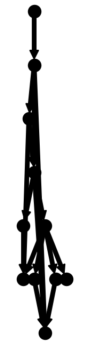}}
	\newcommand{\datasetSparseCGdiff}{\includegraphics[scale=0.55]{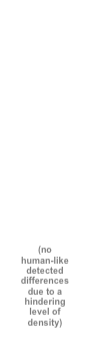}}
	\newcommand{\datasetSparseDGone}{\includegraphics[scale=0.55]{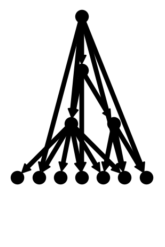}}
	\newcommand{\datasetSparseDGtwo}{\includegraphics[scale=0.55]{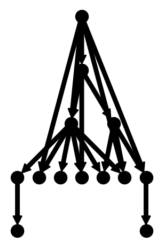}}
	\newcommand{\datasetSparseDGdiff}{\includegraphics[scale=0.55]{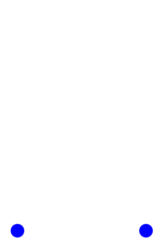}}

\begin{document}

	\title{Learning Human Detected Differences in Directed Acyclic Graphs}

	\author[1]{Kathrin Guckes (n\'{e}e Ballweg)}{Kathrin.Ballweg@gris.tu-darmstadt.de}
	\author[1]{Alena Beyer}{alena.beyer@gmail.com}
	\author[2]{Prof. Margit Pohl}{margit.pohl@tuwien.ac.at}
	\author[3]{Prof. Tatiana von Landesberger}{landesberger@cs.uni-koeln.de}

	\affil[1]{Graphical Interactive Systems Group, Technical University Darmstadt}

	\affil[2]{Informatics, TU Wien}


	\affil[3]{Visualisierung und Visual Analytics, University of Cologne}

	\maketitle

	\begin{abstract}
	Prior research has shown that human perception of similarity differs from mathematical measures in visual comparison tasks, including those involving directed acyclic graphs. This divergence can lead to missed differences and skepticism about algorithmic results. To address this, we aim to learn the structural differences humans detect in graphs visually. We want to visualize these human-detected differences alongside actual changes, enhancing credibility and aiding users in spotting overlooked differences. Our approach aligns with recent research in machine learning capturing human behavior. We provide a data augmentation algorithm, a dataset, and a machine learning model to support this task. This work fills a gap in learning differences in directed acyclic graphs and contributes to better comparative visualizations.
	\end{abstract}

\section{Introduction} 
\label{sec:learning_perceived_structural_differences_in_directed_acyclic_graphs}

From previous work (cf. i.a. \cite{10.1145/2858036.2858155,Fuchs:2014,Klippel:2009,di2020evaluating,bernstein2005similar,doi:10.1207/s15327906mbr27014,Peterson2017}), including ours \cite{DiagramsDifferencePerception,10.1145/3335082.3335083,10.1007/978-3-319-73915-1_20,JGAA-467}, we know that the human similarity notion diverges from the mathematical similarity. However, visual interactive systems supporting the task of visual comparison commonly use graph theoretical, i.e., mathematical definitions of similarity or commonalities and differences (cf. e.g., \cite{10.1145/1201775.882291,Rufiange2013,Liu2017a,bremm2011interactive,10.1145/2254556.2254654,10.1109/TVCG.2007.70521,10.5555/1992917.1992947,Andrews2009a}). This divergence can negatively impact the insights humans gained from comparative visualizations -- in our case directed acyclic graphs. The humans' insights can be biased since the human visual system and the following cognitive processes are not perfect and humans, for instance, overlook some differences \cite{Franconeri2014}. Insights can even be nullified since humans start the question the algorithmic result as they have a different notion of what the result should be -- e.g., what the differences are. Human biases due to imperfections of the human visual system and the following cognitive processes are rather an issue for the purely visual comparative visualizations (cf. i.a. \cite{Dimara2018,Wall2018,Padilla2018,Valdez2018}). According to von Landesberger \cite{TatianaHabil}, these are comparative visualizations which solely show the data items to be compared -- like \cite{10.1109/TVCG.2004.39} or \cite{dlr37368}. The credibility of algorithmic results is especially an issue for algorithmic or algorithmically enhanced visual comparison solutions. While in the former, following the definition of von Landesberger \cite{TatianaHabil}, algorithms compute results in a backend, not apparent to the user, and these results may be visualized, in the latter the algorithms take over a supportive role for the comparative visualization of the data items to be compared. They i.a. compute the common and/or distinct parts of graphs to ease the human to come to a final notion of their similarity like the work of Archambault et al. \cite{archambault2009structural} or Munzner et al. \cite{10.1145/1201775.882291}.

\begin{figure}[tb]
  \centering
    \includegraphics[width=.9\textwidth]{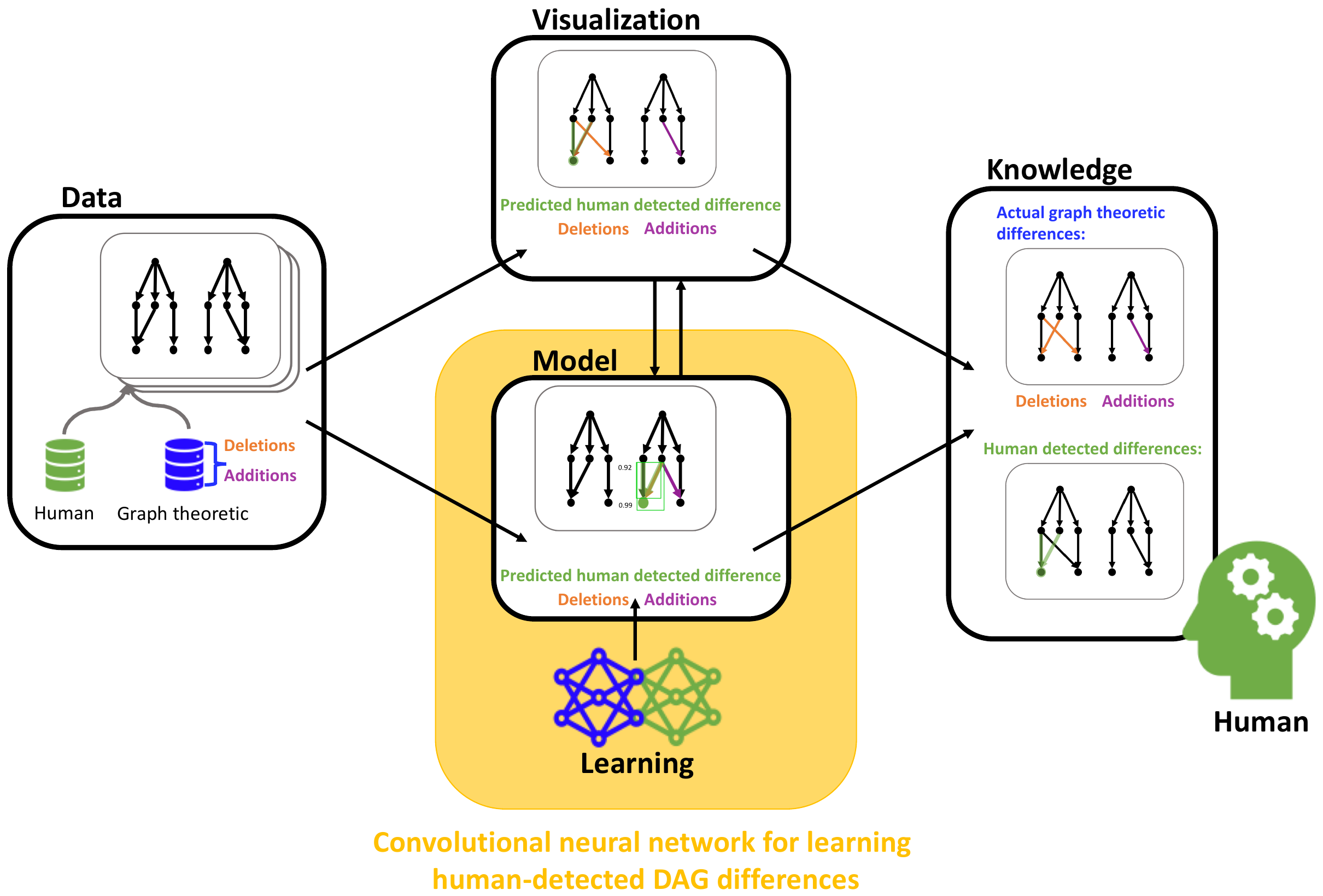}
  \caption[Visual analytics pipeline (cf. \cite{Keim2008}) to illustrate the benefit of our convolutional neural network, which learned to predict human detected differences in directed acyclic graphs]{Visual analytics pipeline, adapted from Keim et al. \cite{Keim2008}, to illustrate the benefit of our convolutional neural network, which is able to predict human detected differences in directed acyclic graphs, in a visual interactive system supporting comparisons.}
  \label{fig:pictures_DACH_DifferenceLearning_Support_of_our_ML_Model}
\end{figure}

To mitigate these potential issues for pairwise juxtaposed comparative visualizations of directed acyclic graphs, we learned the structural differences humans detect. By human detected structural differences we mean which graph elements -- nodes and/or edges -- change from one directed acyclic graph to the other according to humans visually comparing the two directed acyclic graphs. Basically, the human detected structural differences follow the same definition as the graph theoretically determined structural differences (cf. Section~\ref{sub:definitions_and_research_context} -- \texttt{Definitions}, \texttt{Structural Differences}). However, they are not calculated by a mathematical measure, they are the result of the human notion of what is common and distinctive in the directed acyclic graphs. Potential changes are additions and/or deletions. Our goal for learning the human detected structural differences is to finally be able to visualize the human detected difference in combination with the actual change on a graph element level -- i.e., which graph elements were actually added and/or deleted. With respect to our core idea, the actual graph theoretical measure is interchangeable. We decided for the actual graph changes as these are the core elements for any other graph theoretical measure. Our resulting machine learning model can be used in algorithmically enhanced visual comparison systems as Figure~\ref{fig:pictures_DACH_DifferenceLearning_Support_of_our_ML_Model} illustrates. We are firmly convinced that this approach can mitigate the afore explained issues as:
\begin{enumerate}
	\item due to the visualization of the actual changes on a graph element level the human user would recognize which change she presumably would have overlooked or over- resp. underestimated,
	\item there is the combination of the human notion of differences and a mathematical one -- this can mitigate the credibility issues.
\end{enumerate}
We find substantiation of our claim in the recent research of the visualization domain in which researcher successfully captured human brushing and selection behavior and the human notion of correlation resp. similarity with machine learning (cf. i.a. \cite{Ma2018,wohler2019learning,Demiralp2014,2020_intent,ML4VISsurvey}). For instance, Wöhler \cite{wohler2019learning} and Ma et al. \cite{Ma2018} were able to capture perceived correlation in scatterplots resp. the human similarity notion of scatterplots. This research has goals comparable to ours. Its underlying principle is to train a machine learning model on data captured while the user completes a certain task. Consequently, the machine learning model learns the patterns of human behavior and is able to predict the human behavior for new, previously unseen data. The empirical investigation of the benefit of such an algorithmically enhanced visual comparison system, however, is subject to future work as the primary focus of our work was to provide the trained machine learning model together with a suitable training and test dataset. Both of which, to the best of our knowledge, were not existing before.

To the best of our knowledge there is no work yet aiming to learn which differences humans detect in pairs of directed acyclic graphs visualized as node-link diagrams. With our work on this topic we contribute:
\begin{enumerate}
	\item \textbf{A data augmentation algorithm -- the DFS-algorithm}\\
	Our data augmentation algorithm provides an approach to reach the dataset sizes machine learning requires. Further, it shows how the knowledge from empirical studies can be incorporated into application-driven solutions which support the human user with her tasks. Admittedly, here we have a certain indirection. Our DFS-algorithm has a direct impact on the learning phase of our neural network. There, the network learns features and patterns characterizing the differences humans detect. Finally, the trained network is supporting the human user with her tasks -- i.a. by allowing the human to see which differences she presumably spotted and what changes actually happened.
	\item \textbf{A training and test dataset consisting of both tree-like and sparse directed acyclic graphs}\\
	As already proven by benchmark datasets from other domains (cf. e.g., \cite{Lin2014,10.5555/998687.1007045}), they remarkable foster the replicability and comparability of research and its results. Further they substantially ease the conduction of experiments thus the data needed is already there.
	\item \textbf{A machine learning approach for learning human detected structural differences in directed acyclic graphs and a trained model for tree-like and sparse directed acyclic graphs}\\
	Used in a visual comparison system, it will lead to a task convenience for humans. Change detection, as we know from the current body of work (cf. i.a. \cite{Franconeri2014}), is generally a task which humans are slow to complete, so, by allowing the human to see which differences she presumably spotted and what changes actually happened there will be less cognitive load necessary to spot the differences. Finally, there will be more cognitive capacities available for the higher-level tasks, for instance, drawing insights based on the graph changes happened. Also, it may have positive effects on human biases lie overlooking changes.
\end{enumerate}

%
%
%

\subsection{Definitions and Research Context} 
\label{sub:definitions_and_research_context}
In this Section we introduce definitions of core terms which we will use in the remainder of this Section. Furthermore, we provide an overview over the context in which our research is embedded. Due to the fact that our work is located in both the research domain of visualization and machine learning our research context is quite broad. From the field of visualization it is related to visual comparison, the task of change detection, graph differences, graph drawing, and dynamic graphs. From the machine learning domain, neural networks will be in the focus of our work as we aim to learn human detected structural differences. We focus on neural networks as they already achieved outstanding results in learning visual features. The learning of the human detected differences and the learning of visual features are closely related since visual features and their changes characterize a change or hinder a change's detection. In this Section, we provide an overview over the field of machine learning itself to clarify basic terms and how neural networks fit into the field of machine learning. As this paper is located in the visualization research domain, it may be the case that an overview over the field of machine learning is necessary for the reader to understand the remainder of our work.

\subsubsection{Definitions} 
\label{ssub:definitions}
\textbf{Graph theory} is a research domain concerned with graphs. Graphs are structural data which represents relational data. Relational data occurs in many areas, such as molecules in chemistry, brain research \cite{Ma2018a, Zalesky2010} or process models \cite{Gall2015, Girschick2006, Kriglstein2013}. Often, graphs are denoted as networks as well -- e.g., brain graphs that model the structural connectivity of the brain are commonly denoted as brain networks \cite{Grattarola2019, Robinson2010, Zalesky2010}. Since we work with directed acyclic graphs, we define the following terms:

\textbf{Graph, Attribute, Label.} A graph is a tuple $G=(V,E)$ where $V$ is the set of nodes and $E\subseteq(V,V)$ is the set of edges \cite{Tittmann2003}. Nodes represent entities. Edges represent the entities' relations. Edges $e=(s,d)$ connect two nodes $s,d$ with $(s,d)\in V$. Both nodes and edges can have additional information that further describes their properties \cite{Hewapathirana2019}, e.g., the amount of capital a company holds. Such additional information is called an \textbf{attribute}.
A node attribute can be a label $L$. This label is a unique identifier of the node. The label usually is an Integer ranging from $1$ to the number of nodes \cite{Hegeman2018}. Labels are pivotal for graph visualization as many graph layout algorithms use labels for assigning unique positions to nodes.

\textbf{Directed.} Edges $e$ are directed if $E$ contains ordered tuples $e=(s,d)$ and $e$ is coming from $s$ and going to $d$. A graph is called directed if $E$ only contains directed edges \cite{Tittmann2003}.

\textbf{Path, Walk, Cycle, Acyclic.} A path of $G=(V,E)$ is a finite, ordered sequence with
\begin{eqfloat}
	\begin{equation}
    	v_1, (v_1, v_2), v_2, (v_2, v_3), v_3, ..., v_{k-1}, (v_{k-1}, v_k), v_k
	\end{equation}
	\caption{Path equation}
	\label{eq:path}
\end{eqfloat}
\myequations{Path equation}
and $v \in V$ \cite{Tittmann2003}. If all vertices occur once in a path, the path is called a walk. If at least one vertex occurs twice in the path, the path is called cycle. A graph is called acyclic if there exist no cycle.

\textbf{Directed Acyclic Graph.} So, a directed acyclic graph is a directed graph with no directed cycles.

\textbf{Connected.} Any two nodes $v_1, v_2$ of a graph are denoted as connected if at least one path between $v_1, ..., v_2$ exist \cite{Tversky2004}.

\textbf{Graph Size.} Graphs can vary in their size. Graph size is known to influence humans working with a graph visualization \cite{10.1007/978-3-642-36763-2_42, YOGHOURDJIAN2018264}. Usually, the graph size is given by the number of nodes $|V|$ in a graph \cite{doi:10.1111/j.1467-8659.2011.01898.x}.

\textbf{Subgraph.} A subset $V_{sub}, E_{sub}$  with $V_{sub} \subset V$ and $E_{sub} \subset E$ of $G=(V,E)$ with $V_{sub} \subset V$ and $E_{sub} \subset E$ is denoted as a subgraph of $G$. We. employ connected directed acyclic graphs of small graph size. As differences in directed acyclic graphs consists of parts of the directed acyclic graphs, differences are subgraphs.

\textbf{Structural Differences.} Node and connectivity changes, i.e., changes in the nodes relations, are denoted as structural changes as they change the directed acyclic graph's structure \cite{Hewapathirana2019}. The structure of a directed acyclic graph, or topology \cite{Zhou2009}, describes how many nodes -- entities -- exist in the directed acyclic graph, at which positions and how their relations are \cite{10.1145/1201775.882291}. Structural changes commonly are insertions and/or deletions of nodes and/or edges \cite{Treude2007,Giacomo2015}. As structural changes lead to differences in directed acyclic graphs, we call the structural changes in this work also differences.

\subsubsection{Graph Drawings} 
\label{ssub:graph_drawings}
When a graph is visualized, it is transformed into a so called graph drawing \cite{Giacomo2015}. A graph drawing $D(G)$ is defined as a coplanar 2D visual representation which assigns a position $(x_i, y_i)$ to each node $v_i \in V$ with $i = 0, ..., n; n = |V|$ \cite{purchase2002metrics}. Common types of graph drawings are (adjacency) matrices or node-link diagrams (cf. e.g., \cite{Ghoniem2005,Chen2019}). Matrices show the nodes in rows and columns. A colored cell is the visual representation of a connection, an edge, between a pair of nodes \cite{doi:10.1111/cgf.12791}. A node-link diagram usually visualizes nodes as circles and edges as lines \cite{meirelles2013design,lima2014book}. In case of a directed graph being visualized as a node-link diagram, the edges usually are lines with arrow heads \cite{meirelles2013design,lima2014book}. However, there is also a substantial amount of related work which investigates alternate designs -- especially for edges (cf. i.a. \cite{flowMaps,10.1145/1518701.1519054,5742390,holten:hal-00696823,10.1007/978-3-642-31223-6_34,10.1007/978-3-642-36763-2_40}).

There are further types of graph drawings -- e.g., adjacency lists \cite{6812198,doi:10.1111/cgf.12791}. In an adjacency list edges are just indirectly shown. They show edges as a list of connected node tuples \cite{Koutra2017}.

\paragraph{Our Design Choice.} 
\label{par:our_design_choice}
We decide to follow the standard approach of visualizing directed node-link diagrams: nodes as circles, edges as lines with arrow heads. We are aware that the standard approach can have disadvantages like visual clutter around the node due to to arrow heads of the incident edges (c.f. e.g., \cite{10.1145/1518701.1519054}). However, other design alternatives are also not free of the visual clutter problem as our work in \cite{guckes2} shows. So, we belief that humans being used to the standard design reducing the risk of misreading our graphs outweighs the potential disadvantage.

\emph{\textbf{Layout Algorithms.}} Layout algorithms do the position assignment for each node $v_i$ throughout the transformation of the abstract graph into a graph drawing. There are various existing layout algorithms. This i.a. results from the necessity of different layout algorithms for different graph types (hierarchies, small world networks, etc.) or different optimization goals (graph aesthetics (edge crossings, edge length/angle, etc.), separating the graph's clusters, etc.). Generally, layout algorithms aim to visualize the graph so that it is well readable based on the various optimization goals \cite{TatianaHabil}. There is also the goal to preserve the mental map. This is especially interesting for changing dynamic graphs \cite{doi:10.1111/cgf.12791,Diehl2002}. The term ``mental map'' denotes the mental representation of the graph humans form by looking at and working with graph drawings \cite{munznerDesingPrinciples}. The preservation of the mental map usually means the preservation of the node positions \cite{10.5555/1378337.1378354}. For comparing changing, dynamic graphs or detecting changes in them a preserved mental map is pivotal as in case already existing nodes would change their position this would suggest more changes which are actually there \cite{misue1995layout,purchase2007important,Purchase08,10.5555/1378337.1378354}. This, in turn, would create additional unnecessary cognitive load of the user to figure this out or in the worst case would even bias the user insights in case she would not realize that in actuality just the nodes' positions have changed.

{\scshape Our Layout Algorithm Choice.} The choice and suitability of a layout algorithm depends on the purpose or the task to be supported \cite{archambault2009structural}. Our work is located in the areas of visual pairwise comparison, change detection, and changing directed acyclic graphs. Consequently, we choose a Sugiyama-like hierarchical layout \cite{sugiyama1981methods} which preserves the mental map. We choose a Sugiyama-like layout since Burch et al. showed that this hierarchical layout type outperforms others \cite{6065011,6596142}. The mental map preservation is necessary for us due to the rationals discussed above. Furthermore, the results of machine learning-based difference detection on images considerably improve, if the images are registered \cite{1395984,Ji2019,archambault2009structural}. A layout algorithm preserving the mental map achieves the registration of the images. Thus, our decision of mental map preservation is also necessary and relevant due to the benefits for our future machine learning approach.

\subsubsection{Visual Comparison and the Change Detection Task} 
\label{ssub:visual_comparison_the_change_detection_task_and_dynamic_graphs}
\paragraph{Psychological Perspective.} 
\label{par:psychological_perspective}
\begin{figure}
  \centering
    \includegraphics[width=.3\textwidth]{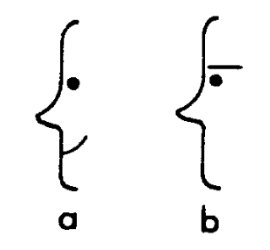}
  \caption[Example for the set-theoretical matching function of Tversky's pairwise similarity judgments model]{A set of two faces. Determining the set of distinctive features of $a$ compared to $b$ based on the notion of a set-theoretical matching function of Tversky's pairwise similarity model this results in the smiling mouth. The Figure is an excerpt of Tversky's Figure in \cite{tversky1977features}.}
  \label{fig:pictures_DACH_DifferenceLearning_Tversky_Example}
\end{figure}

Here, we also employ Tverky's model as the theoretical concept of how the human similarity notion for visual comparison works. Even if we treat the actual process as a black box, in the behaviorist style, model theoretical knowledge is important for i.a.
\begin{itemize}
	\item the definition of the visual features which are used for the machine learning,
	\item the modeling of the detected differences for the creation of a training and test dataset (cf. Section~\ref{sub:dataset_creation_algorithm_for_enriching_dag_data_with_human_like_detected_graph_differences}),
	\item and the learning of the change detection process patterns to predict what changes humans presumably detect as different for new, previously unseen directed acyclic graph pairs (cf. Section~\ref{sub:a_convolutional_neural_network_for_learning_human_detected_differences_in_dags}).
\end{itemize}

Tversky describes the pairwise similarity judgment as a feature matching process on the basis of ``the set-theoretical notion of a matching function'' \cite{tversky1977features}. In his model he assumes \cite{tversky1977features}:
\begin{enumerate}
	\item It exists an ordinal matching function $s(a,b)$ for all distinct objects in $\delta$ -- the set of objects to be compared. This matching function provides an ordering of the objects regarding their similarity; e.g., $s(a,b)>s(c,d)$ expresses that $a$ is more similar to $b$ than $c$ is to $d$, but it is unknown by how much. Humans usually do this ordering with verbal expressions \cite{Umbach_Similaroty_Classification}.
	\item It is possible to predict an interval scale $S$ that preserves the ordering of the ordinal scale $s$ and expresses the similarity as the contrast of the measures of common and distinctive features for all objects in $\delta$ (a.k.a contrast model). There is also the possibility to express the similarity as a weighted contrast. Tversky states that the results improve with an additional similarity judgment made by the participants used as the weighting factor.
\end{enumerate}
In the feature matching process humans compare pairwise objects and match their features whether they are common or distinct. The features of each object form a feature set. Assuming that there are two objects in $\delta = {a, b}$ then they are expressed by their features resp. feature sets as follows: $\delta = {A, B}$. Figure~\ref{fig:pictures_DACH_DifferenceLearning_Tversky_Example} shows an example of two faces as the objects $a$ and $b$ in $\delta$. Determining the distinct features of $A$, $A - B$, results in the smiling mouth as all other features are identical. Usually, humans have a notion of similarity of the pairwise objects if they share common features and have less distinct features. Set operations will play an important role for our data augmentation algorithm described in Section~\ref{sub:dataset_creation_algorithm_for_enriching_dag_data_with_human_like_detected_graph_differences}.
\FloatBarrier

Tversky further states that the features can be low- or high-level ones such as orientation of patterns formed by lines or points or color (low-level features), or components such as eyes or the shape or complexity of something (high-level features) \cite{tversky1977features}. Tversky determines that in advance to the task -- here change detection -- humans extract from the data a limited list of features based on which they perform the task \cite{tversky1977features}. Performing the task means calculating the the set of common and distinctive features to detect the changes in the directed acyclic graph pairs as a sub-task of visual comparison. Further, Tversky \cite{tversky1977features} showed that the weight of the used features depend on the task. For similarity judgements, common features are perceived stronger than distinctive features whereas for comparisons with respect to differences resp. for change detection distinctive features receive the higher weights.

Tversky's assumptions and hypotheses are supported by other well-known literature on similarity -- this also applies to more recent literature \cite{brunswig1910vergleichen,Umbach_Similaroty_Classification}. Furthermore, his model is able to explain the human similarity notion for a variety of visual comparisons \cite{tversky1977features,TverksyPoster}.

\emph{\textbf{Our Feature Choice.}} From our work on the influence factors for visual comparisons with respect to differences \cite{10.1145/3335082.3335083,DiagramsDifferencePerception}, we can see that humans detect the differences in a directed acyclic graph on the graph elements' level -- i.e., based on the nodes and edges of the two directed acyclic graphs. So, following Tverksy's theory, the nodes and edges of our directed acyclic graphs are the features based on which humans determine the set of common and distinctive features. We can further learn that the detection of a certain change is affected by the influence factors of visual comparisons with respect to differences (cf. \cite{10.1145/3335082.3335083,DiagramsDifferencePerception}). While, for instance, a shape change or a newly introduced edge crossing fosters the supports the detection of change, a high density hinders the change's detection.


\paragraph{Visualization Domain's Perspective.} 
\label{par:visualization_domain_s_perspective}

\begin{figure}[tb]
  \centering
    \includegraphics[width=.9\textwidth]{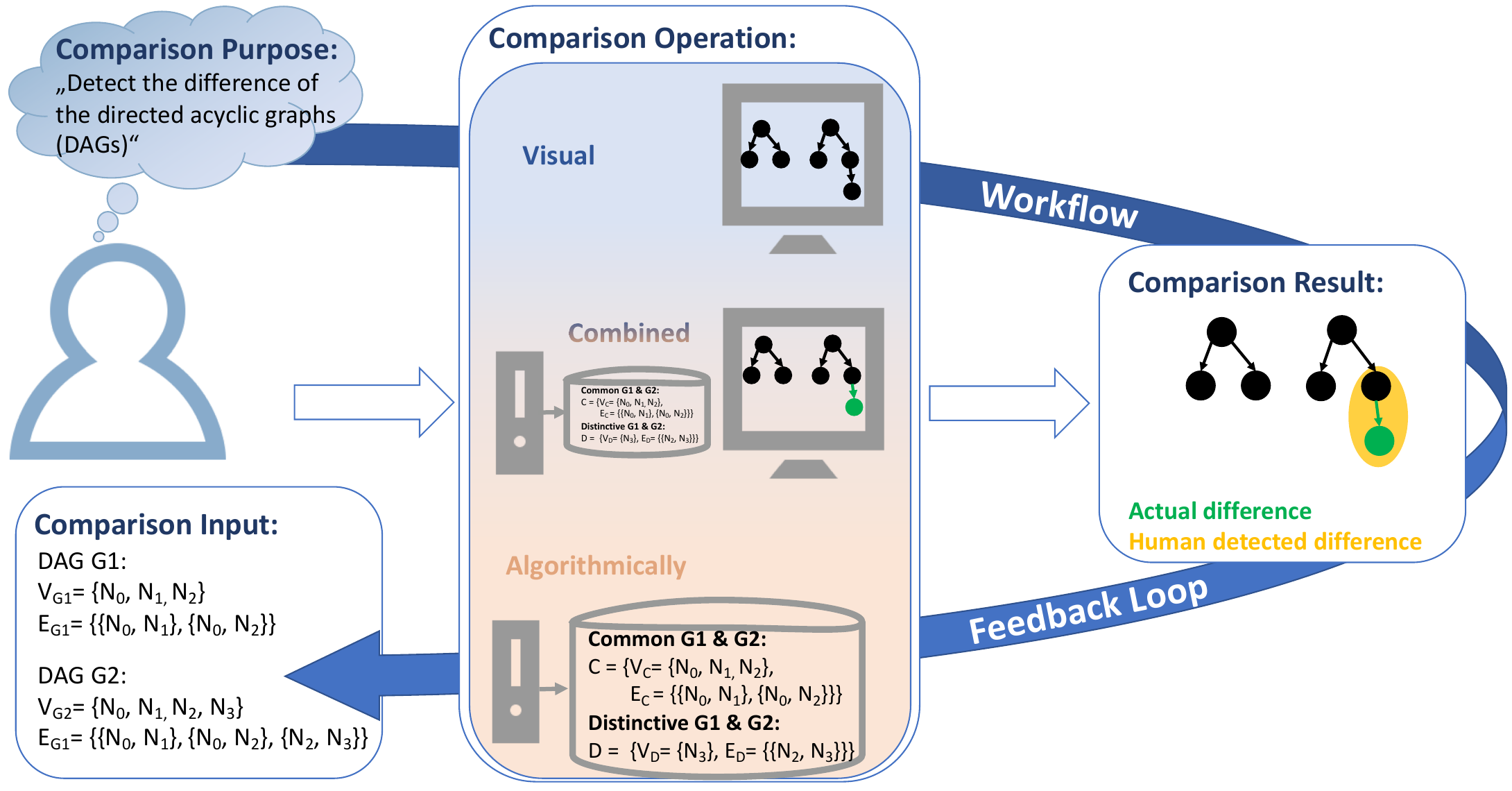}
  \caption[The visual comparison workflow according to von Landesberger \cite{TatianaHabil}]{The visual comparison workflow according to von Landesberger \cite{TatianaHabil} (Figure based on original Figure from \cite{TatianaHabil}).}
  \label{fig:pictures_DACH_VisComp_ChangeDetection}
\end{figure}

\emph{\textbf{Visual Comparison.}} Following von Landesberger \cite{TatianaHabil}, who extends the considerations and the taxonomy of Gleicher et al. \cite{10.1177/1473871611416549}, visual comparisons is a workflow as sketched in Figure~\ref{fig:pictures_DACH_VisComp_ChangeDetection}. The \texttt{comparison purpose} is equivalent to the concrete comparison task -- in our case, the visual comparison with the aim to detect changes. Moreover, von Landesberger \cite{TatianaHabil} defines the \texttt{comparison input}. It is the data to be compared. The actual comparison is conducted by the \texttt{comparison operation} which leads to a final \texttt{comparison result} which usually is visualized as well. According to von Landesberger there are three types of \texttt{comparison operations} -- visual, algorithmically, or a combined version. Purely visual \texttt{comparison operations} solely show the data items to be compared \cite{TatianaHabil} -- like \cite{10.1109/TVCG.2004.39} or \cite{dlr37368}. Here, it can be that the user can express her \texttt{comparison result} via interaction or she has to remember her \texttt{result} for later recall. Algorithmic \texttt{comparison operations} compute results in a backend, not apparent to the user, and these results may be visualized as the \texttt{comparison result} \cite{TatianaHabil}. Combined approaches either use visualization to, for instance, support the configuration of the respective algorithmic \texttt{comparison operation}, or they use algorithms to ease otherwise complex tasks \cite{TatianaHabil,Srinivasan2018a}. The work of Archambault et al. \cite{archambault2009structural} and Munzner et al. \cite{10.1145/1201775.882291} are a prime example for such algorithmically enhanced visual comparison solutions. They i.a. compute common and/or distinct parts of the graphs to ease the human to come to a final notion of the graphs' similarity. Especially for the visual comparison of graphs and related tasks such as change detection cognitive load is a pivotal aspect. Algorithms, like for the work of Archambault \cite{archambault2009structural} or Munzner et al. \cite{10.1145/1201775.882291}, reduce the cognitive load by reducing, e.g., the number of data items to compare or the complexity \cite{TatianaHabil,Srinivasan2018a}. Huang et al. showed that visual analysis becomes slower and more error-prone if the complexity of the task increases \cite{AlenaPaper,vahanCognitiveLoad}. Algorithms which are used in existing algorithmically enhanced visual comparison systems are i.a.:
\begin{itemize}
	\item similarity measures (e.g., \cite{10.1145/1201775.882291,bremm2011interactive})
	\item difference maps (e.g., \cite{Rufiange2013,archambault2009structural})
	\item pattern matching (e.g., \cite{2020_intent})
	\item difference highlighting (e.g., \cite{10.1145/1201775.882291})
	\item structural brushing (e.g., \cite{10.1145/1201775.882291})
	\item graph matching (e.g., \cite{10.1145/1201775.882291,10.1145/2254556.2254654})
\end{itemize}
To the best of our knowledge, there is currently no approach like ours for graphs, specifically for directed acyclic graphs, which aims to capture the human notion of differences for the change detection task with machine learning. The workflow can provide the user with the option to steer the process by giving feedback (cf. Figure~\ref{fig:pictures_DACH_VisComp_ChangeDetection} -- \texttt{Feedback Loop}).

{\scshape Our Comparison Input.} The comparison input for our research goal of learning human detected structural differences are pairs of directed acyclic graphs.
Algorithmically, we implement the correspondence between the directed acyclic graph pairs via unique node labels. The fixed mental map of our chosen layout algorithm, cf. Section~\ref{sub:definitions_and_research_context} -- \texttt{Graph Drawings}, \texttt{Our Layout Algorithm Choice}, is able to visually transmit the correspondences even though we do not show the nodes' unique labels. Research on dynamic graphs and/or the mental map and its stability have already shown this potential of a fixed mental map (cf. i.a. \cite{10.1007/978-3-642-36763-2_42,bach2014graphdiaries}). For our convolutional neural network we use the visualizations of the directed acyclic graph pairs as the comparison input (cf. Section~\ref{sub:change_detection_as_a_machine_learning_problem} and~\ref{sub:a_convolutional_neural_network_for_learning_human_detected_differences_in_dags} for details).

{\scshape Our Comparison Operation.} A comparison operation comes to the final comparison result by processing the data to be compared depending on the task \cite{TatianaHabil}. Possible types are visual, algorithmically or combined. Learning human-detected structural differences processes pairs of visualized directed acyclic graphs with a learning algorithm. So, out comparison operation is an algorithmic one. As Figure~\ref{fig:pictures_DACH_DifferenceLearning_Support_of_our_ML_Model} illustrates, the outcome of our comparison operation can be used in an algorithmically enhanced visual comparison system


\emph{\textbf{Change Detection Task.}} Also from the perspective of the visualization domain the detection of changes is sub-task of visual comparison \cite{TatianaHabil,doi:10.1111/j.1467-8659.2011.01898.x,10.1145/1168149.1168168,Akoglu2015,Zhang2017}. To complete the task of change detections humans have to compare a set of data items -- in our case a pair of directed acyclic graphs -- and detect the changes \cite{Zhang2017}. The detection of structural changes needs the humans to spot added and/or deleted nodes and/or edges.
\begin{figure}[tb]
  \centering
    \includegraphics[width=.9\textwidth]{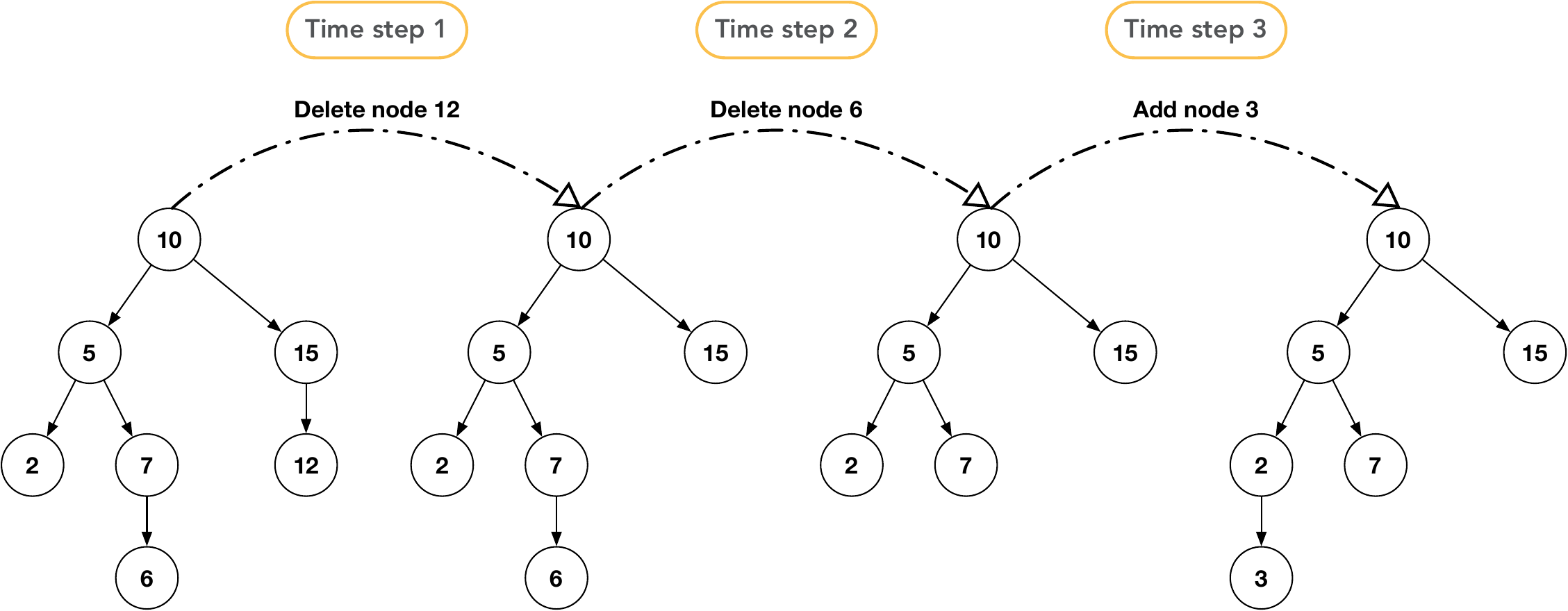}
  \caption[Dynamic graph graph with structural changes from \cite{Giacomo2015}]{Dynamic graph graph with structural changes -- additions and deletions of nodes -- from \cite{Giacomo2015} (Figure based on original Figure from \cite{Giacomo2015}).}
  \label{fig:pictures_DACH_DynamicGraph_VisComp}
\end{figure}

The change detection task often also occurs in the context of dynamic graph analysis or drawing (cf. \cite{Akoglu2015,10.1007/978-3-642-36763-2_42}). While visual comparison often deals with pairwise comparison, dynamic graph analysis resp. drawing deals with a series of graphs \cite{TatianaHabil}. Pairwise visual comparison, in our case of directed acyclic graphs, can be transferred into the domain of dynamic graph analysis resp. drawing. Let $G_1$ (base graph) and $G_2$ (alternative) be a pair of directed acyclic graphs differing by $N$ node and/or edge changes. Given that per time step $t$, starting with $t_0$, $N$ changes happen then it is $(G_{1}, G_{2}) = (G_{t_0}, G_{t_{0+N}})$. For the dynamic graph, shown in Figure~\ref{fig:pictures_DACH_DynamicGraph_VisComp}, the changes are node deletions (node $12$ and $6$) and an addition (node $3$). If the first graph in Figure~\ref{fig:pictures_DACH_DynamicGraph_VisComp} is considered as the base graph $G_1$ then all other graphs at \texttt{time step 1}, \texttt{2}, and \texttt{3} can be the alternative $G_2$ with either $1$, $2$, or $3$ changes. Hence, research in the domain of dynamic graph analysis resp. dynamic graph drawing is applicable tp pairwise visual comparison. Since our topic is grounded in pairwise visual comparison of directed acyclic graphs, we mainly consider related work on visual comparison. Still, some subtopics of our work are more extensively explored in related domains, e.g., graph drawing. Change detection tasks in combination with graph drawing have been explored in recent years both in machine learning and in perception (cf. i.a. \cite{10.1145/3335082.3335083,DiagramsDifferencePerception,archambault2009structural,CayeDaudt2018,Ji2019,10.1007/978-3-030-11012-3_10}).

%

\subsubsection{Overview of the Area of Machine Learning} 
\label{ssub:overview_of_the_area_of_machine_learning}
Neural networks (neural networks) are a class of machine learning (machine learning) algorithms. machine learning algorithms learn and solve a certain task based on the evaluation of a large amount of data. Bishop et al. provide a comprehensive introduction to machine learning in their book ``Pattern Recognition and Machine Learning'' \cite{bishop2006pattern}. machine learning algorithms are widely employed in the most diverse areas. Example areas are computer vision or graph learning. machine learning for graphs covers i.a. link prediction \cite{Zhou2009}, anomaly detection \cite{Akoglu2015}, or classification \cite{Niepert2016}.

\paragraph{Machine Learning in General} 
\label{par:_acrshort_ml_in_general}
A machine learning algorithm -- or model -- is inseparable of its input data and its loss function. The algorithm learns from the data via multiple iterations aiming to reduce the loss function's result in each iteration. This process is also called optimization. The optimization objective determines the loss function of choice. Example objectives are i.a. regression or classification. While the objective for a regression problem is the optimization of continuous values \cite{bishop2006pattern} classification optimizes discrete values -- i.e., the class labels -- for, e.g., node classification \cite{Bhagat2011}. In machine learning algorithms can learn supervised or unsupervised. Supervised machine learning algorithms receive receive feedback on their performance after each iteration. This feedback is also called target. Further, the target is defined in advance, e.g., via user input. The input data $x$ and the target vector $t$ are fed into the algorithm. Based on this input, the machine learning algorithm is a function $y(x)$ which puts out a vector similar to $t$ \cite{bishop2006pattern}. The final algorithm together with its parameter configurations results from a training (learning) and a testing (inference) phase. For the training phase, the training dataset is used as the input $x$. The afore explained loss function $L_i(y(x), t)$ is the quantification of the quantification of the model's accuracy in solving the task it was designed for in the current iteration $i$. Once the training phase is completed, the machine learning model is tested with a separate test dataset. The model's ability to generalize -- i.e., prediction accuracy on new, previously unseen data -- is likely to differ from the results for the training dataset. Maximizing the diversity of the data used for training increases the machine learning algorithm's ability for generalization \cite{Agarwal2018}. Unsupervised algorithms follow, in general, the same procedure. But they learn without the feedback $t$.

\emph{\textbf{Classification.}} Classification algorithms are versatile. The current body of work divides the algorithms  into traditional and non-traditional algorithms depending on the structure of the data (cf. \cite{Bhagat2011,10.5555/303568.303704}).In general, difference learning is expressible as a classification problem. The changes -- differences -- are classified as additions or deletions.\\
Traditional algorithms for node classification model classifiers based on iterations over graph features while non-traditional algorithms are are based on random walks \cite{Bhagat2011}. In the image classification domain traditional algorithms depend on manually created features \cite{10.5555/303568.303704}. Traditional image classification algorithms i.a. include support vector machines, decision trees, and K-nearest neighbor algorithms \cite{bishop2006pattern}. For decision trees the required manually created features are the rule set and for support vector machines it is the kernel function. As traditional learning algorithms require a rule set resp. a-priori defined knowledge -- the features -- they are also called knowledge-based algorithms.\\
In related work, the term ``non-traditional machine learning algorithms'' is often used for neural network algorithms. Opposed to traditional algorithms, non-traditional ones avoid manual feature creation and they work data-driven -- i.e., they learn the features directly from the input dataset. While for traditional algorithms it is the challenge to find the right features, non-traditional algorithms face the challenge of hyperparameter optimization \cite{10.5555/303568.303704}. These are parameters which cannot be learned directly from the input data and are usually fixed before the actual learning phase starts. Example for such hyperparameters are the learning rate or the topology of a neural network.

\emph{\textbf{Neural Networks.}} A neural network consists of multiple layers of neurons. Neurons are expressed as parameters -- weights $w$ and biases $b$. During training, these parameters are optimized for the items $D$ of the training dataset. The weights and biases of each certain neuron are grouped together to a vector and thereby build a layer of a neural network. The layers can be combined in various ways. This is then called the network architecture. The information flow through the neural network goes in two directions: 1) feed-forward, 2) feed-backward. The forward propagation calculates the neural network's loss function for the given training data. The backward-pass pipes the loss function's results back into the network to update the neuron's weights ($w$) and the biases ($b$). neural networks might be a possible way to approach our learning problem since they are not dependent on features defined in advance which is challenging in the context of learning differences -- especially human detected ones -- as these are not extensively studied (cf. \cite{Grattarola2019,DiagramsDifferencePerception,10.1145/3335082.3335083}).

{\scshape Deep Neural Networks.} A deep neural network has multiple layers. A two-layer neural network is modeled by the following equation
\begin{eqfloat}
	\begin{equation}
    	y_k(x, w) = \sigma(\sum\limits_{j=1}^{D} w_j*h(\sum\limits_{i=1}^{N} (w_{ji}x_i + b_j) + b_k
	\end{equation}
	\caption{Equation modeling a two-layer neural network.}
		\label{eq:twoLayerNN}
\end{eqfloat}
\myequations{Equation modeling a two-layer neural network}
with $k=1, ..., K$ and $K$ being the number of outputs and $\sigma(.)$ being a non-linear activation function. The parameter $h$ denotes the output of a previous layer, called hidden function, and $N$ denotes the number of hidden layers.

{\scshape Epochs, Batch Size, and Iterations.} The pivotal parameters for training a neural network are the epochs and the batch size. The number of epochs denote the number of times a neural network processes the entire training dataset. The batch size denotes the number of training samples -- subsets of the training dataset -- the neural network works through before updating its model parameters. The network's complexity and the hardware performance influence the batch size. The number of iterations commonly denote the training time. The number of iterations is defined by the following equation

\begin{eqfloat}
	\begin{equation}
    	iterations = \frac{training size}{batch size} * epochs
	\end{equation}
	\caption{Equation determining the number of iterations for the training of a neural network.}
		\label{eq:Iterations}
\end{eqfloat}
\myequations{Equation -- number of iterations for the training of a neural network}

\emph{\textbf{Backpropagation Algorithm, Gradient Descent Optimization and Learning Rate.}}
The training of a neural network consists of a forward and a backward pass (backpropagation) for each data item. During the forward pass the input data is processed by each layer. The result of the last network layer is the result of the forward pass. Its quality is measured by a loss function. The training goal is to minimize that loss function.  A common optimization method is gradient descent optimization. It optimizes based on the subtraction of the loss function's gradient from the current loss result (step size). The step size is controlled by the learning rate. The optimizer specifies how much the weights must change at the network's last layer. This information is used by the previous layer to calculate its necessary change so that those can update their weights. This is done backwards through the whole network. The algorithm is called backpropagation.

\emph{\textbf{Convolutional Neural Networks.}} Dumoulin et al. \cite{Dumoulin2016} provide a comprehensive introduction to neural networks. neural networks are a type of neural networks \cite{10.5555/303568.303704}. neural networks gained attention since they achieved outstanding results on images, e.g., detecting objects in images \cite{7485869}. Today, they are used on images and graphs as well as for various tasks -- i.a., difference detection which is the task of our focus \cite{7485869,Wu2019,Zhang:2019aa}. neural networks have multiple layers performing convolutional operations on the extracted features -- i.e, the deeper the layer the more complex the features. So, the first, or shallow, layers learn low- and mid-level features whereas the deeper layers learn the high-level, i.e., more complex, features \cite{7780459,Zeiler2014}. Low-lever features are, for instance lines or points. Examples for high-level features is the silhouette of a face or the shape of a visualization like a graph (cf. \cite{10.1007/978-3-319-73915-1_20,JGAA-467}). Features aggregate structural information -- low- or high-level -- about parts of the respective data item. Tversky et al.'s \cite{tversky1977features} definition of features from a psychological model-theoretic point of view on the human similarity notion and the notion of commonalities and differences correspond to that. This means the mathematical model of neural networks, or more precisely neural networks, and the psychological model are compatible which, in turn, makes neural networks a suitable approach for modeling the human notion of differences in directed acyclic graphs.\\
\textbf{\emph{Our Approach Decision.}} Consequently, we anchored our search for a suitable learning approach in the field of convolutional neural networks. We present details on our approach decision in Section~\ref{sub:change_detection_as_a_machine_learning_problem} and~\ref{sub:a_convolutional_neural_network_for_learning_human_detected_differences_in_dags}.

\emph{\textbf{Visual Features of Difference Detection in Graphs.}} Since graphs may differ by one or more nodes and/or edges, graph differences are structures containing single or multiple nodes and/or edges or multiple connected nodes and edges. We call these connected nodes and edges connected components in the remainder of our work. As we work with node-link diagrams and chose to follow the standard way of visualizing a node-link diagram \cite{lima2014book,meirelles2013design} -- nodes as circles and edges as lines with arrow heads -- the shapes providing the visual features are circles, rect-, and triangles. Today, neural networks are usually trained large datasets containing these shapes. Examples for such large datasets are ImageNet \cite{Deng2009} with $1000.000$ or the COCO dataset \cite{Lin2014} wit $118.000$ training images.

\subsubsection{Related Approaches for Capturing Human Behavior, Notions, or Perception} 
\label{ssub:related_approaches}
In recent years, various approaches have been proposed which capture human behavior, notions, or perception to understand how humans work with visualizations and to improve the analyst's support by the visualization system \cite{Demiralp2014,2020_intent,Ma2018,Saket2018,wohler2019learning}. These approaches consider both data and visualization characteristics and the characteristics of the human behavior, notion, or perception.

Capturing human behavior, notions, or perception describes the process of mapping human process to algorithms. Modeling human factors or processes as an algorithm strives for the goal to support the user with her task in an enhanced manner. Various research domains pursue this goal. The algorithm of Bourbakis et al. \cite{Bourbakis2002}, for instance, captures human perception to depict image differences. Bourbakis et al. \cite{Bourbakis2002} also employ the similarity model of Tversky \cite{tversky1977features} -- i.e., they employ the same model as we do. In general, there are two different paradigms for the capturing of a human process \cite{Saket2018}:
\begin{itemize}
	\item knowledge-driven modeling
	\item data-driven modeling
\end{itemize}
The knowledge-driven approach injects knowledge on the human cognition, reasoning, or perception in the algorithm with a set of rules. The approach of Bourbakis et al. \cite{Bourbakis2002} is an example for knowledge-driven modeling. Data-driven modeling, also known as machine learning, trains based on data gathered from the human. The data collection can happen via user studies, experiments, or tracking of user interactions with a visual analytics system. The result of the data-driven modeling is a trained model which can be saved to disk and thus be used again later on. The possibility to reuse these models later is also beneficial for scientific replicability \cite{2020_intent}. Currently, the most modeling approaches in the visualization domain follow the data-driven trend as many other sciences \cite{2020_intent}. The advantage of a data-driven approach is that no manual feature engineering is needed anymore as opposed to for the creation of the rule set of knowledge-based approaches \cite{Saket2018}. Gadhave et al. \cite{2020_intent} capture human brushing and selection behavior for visualized scatterplots based on event data tracked by their visual analytics system. The authors are convinced that the prediction of human brushing and selection behavior is the prerequisite for capturing high-level reasoning processes -- in their case: the user intent. The authors believe that this prerequisite is crucial thus such higher-level processes are in general complex and not yet completely understood. Scatternet, the approach of Ma et al. \cite{Ma2018}, learned the human notion of scatterplot similarity based on data from humans. A scatterplot triple -- anchor, similar scatterplot, and dissimilar scatterplot -- with the respective labels ((dis)similar)is the input of the neural network. The neural network learns based in the visualized scatterplots -- i.e., the images. Consequently, Scatternet \cite{Ma2018} depends on the scatterplots' visual design; e.g., the sizes of the scatterplot dots \cite{wohler2019learning}. Wöhler et al. \cite{wohler2019learning} train a neural network to learn the human notion of correlation in scatterplots. Their neural network's input are pairs of images labeled by humans. Wöhler et al. \cite{wohler2019learning} employ both image-based and vector-based\footnote{These are networks which work with the data which is part of the learning objective in a not visualized manner.} networks which enables them to achieve a certain independence of the scatterplots' visual design. Ma et al. \cite{Ma2018} and Wöhler et al. \cite{wohler2019learning} both learn subjective metric relative to the entire data items. This is different to our learning objective. We are interested in learning what are the structural changes which humans detect when they compare two visualized directed acyclic graphs.\\
As we will also use a neural network, see Section~\ref{sub:change_detection_as_a_machine_learning_problem} and~\ref{sub:a_convolutional_neural_network_for_learning_human_detected_differences_in_dags} for details, we will also follow the data-driven modeling paradigm. We will use a neural network as its approach of feature extraction is in line with Tversky's model of how humans use features to solve the task of change detection (cf. Section \texttt{Visual Comparison and the Change Detection Task} -- \texttt{Psychological Perspective}). We will use a knowledge-driven approach for data augmentation to achieve the training dataset size which is needed to train a generalizable and robust model (cf. Section~\ref{sub:dataset_creation_algorithm_for_enriching_dag_data_with_human_like_detected_graph_differences}).

\subsection{A Data Augmentation Algorithm for Enriching Directed Acyclic Graph Data with Human-Like Detected Structural Differences} 
\label{sub:dataset_creation_algorithm_for_enriching_dag_data_with_human_like_detected_graph_differences}
To the best of our knowledge, there is currently no annotated dataset for learning human detected differences of directed acyclic graph pairs. In general machine learning algorithms need a large amount of training data. Current networks for pairwise image change detection use $1900$ image pairs \cite{Ji2019}. Perception capturing networks use dataset sizes of $10.000$ \cite{10.1007/978-3-030-35802-0_38} to $\geq50.000$ \cite{Ma2018} data items. To learn human-detected differences, we need a huge amount of directed acyclic graph pairs annotated with the differences humans detected. The annotated data of our difference comparison studies \cite{DiagramsDifferencePerception,10.1145/3335082.3335083} are far too few. From this studies we had $915$ annotated directed acyclic graph pairs ($62\text{ participants } * 15\text{ directed acyclic graph pairs} = 915$ (We omitted one directed acyclic graph pair because of the tracked coordinates being erroneous.)). A dataset with the size of $915$ os too small for the training of a robust and general neural network. Options for increasing the dataset size are:
\begin{enumerate}
	\item increase the number of participants
	\item increase the number of data items per participant
\end{enumerate}

Option 1 is a challenging endeavor since participant acquisition is always hard \cite{10.1007/978-3-319-66435-4_5}. Also crowd-sourcing, which is the standard way of increasing the number of participants, is problematic. Due to the specificity of the change detection task and its required subtasks -- object identification and visual search -- it is not a typical crowd-sourcing task since crowd-sourcing tasks are typically micro-tasks which change detection definitely is not \cite{10.1007/978-3-319-66435-4_5}. The number of data items per participant, however, is constrained by the cognitive capabilities of humans \cite{Pohl2019}. Cognitive load models allow for an estimation of these cognitive limits of humans \cite{AlenaPaper}. Huang et al. \cite{AlenaPaper} found that humans' cognitive load is influenced by various factors. Examples are task complexity, visual complexity, and data complexity. So, given that the question for differences is considerably more specific \cite{tversky1977features} -- the participants have to specifically look for what sets the directed acyclic graphs apart -- a considerably increased number of directed acyclic graph pairs would overwhelm the participants. So, in conclusion, also option 2 is not feasible.

Due to the impracticalities of option 1 and 2, a solution beyond increasing the number of participants or the number of data items per participants is needed. Our solution of choice is a knowledge-based augmentation of directed acyclic graph pairs to construct a large set of directed acyclic graph pairs annotated with human-like detected differences. By human-like we mean difference annotations which are similar to those which an actual human would do. The similarity is achievable by using a knowledge-based approach. In our case, the knowledge base are our studies on difference-coined comparisons (cf. \cite{10.1007/978-3-319-73915-1_20,JGAA-467}). This allows a feasible combination of the necessity of a large training dataset and the annotations of the directed acyclic graph pairs with the human-like detected difference. As a knowledge source, we use our difference-coined comparison studies \cite{DiagramsDifferencePerception,10.1145/3335082.3335083}. Since our algorithm is based on the influence factors of comparison with respect to difference, we call our algorithm the \textbf{D}ifference \textbf{F}actors \textbf{S}imulation (DFS) algorithm. 

%

\subsubsection{DFS-Algorithm -- Overview} 
\label{ssub:dfs_algorithm_overview}
\begin{figure}[tb]
  \centering
    \includegraphics[width=.9\textwidth]{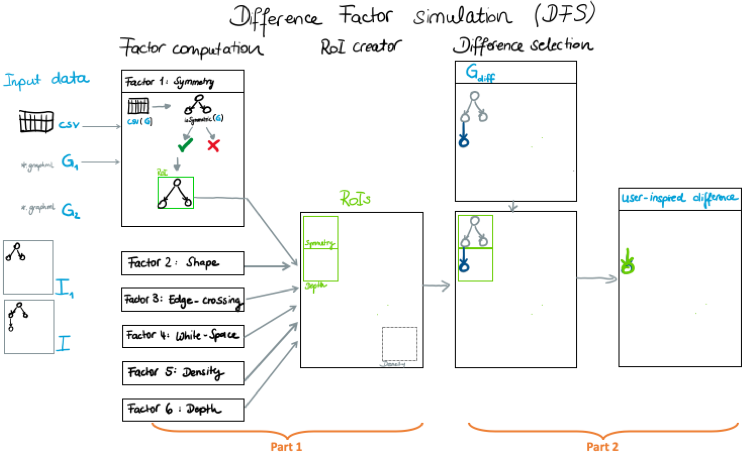}
  \caption[\textbf{D}ifference \textbf{F}actors \textbf{S}imulation (DFS) algorithm -- schematic representation]{\textbf{D}ifference \textbf{F}actors \textbf{S}imulation (DFS) algorithm -- schematic representation. (Figure based on original Figure from \cite{AlenaMA})}
  \label{fig:pictures_DACH_DifferenceLearning_AlgoSchema}
\end{figure}

In our studies on comparison with respect to differences, we found that eight factors influence the human similarity notion \cite{10.1007/978-3-319-73915-1_20,JGAA-467}. As our participants mentioned these factors with varying frequency, we only consider the dominant ones (cf. Section~\ref{ssub:influence_factors_for_detecting_changes}). Consequently, we do not consider the factors ``groups'' and ``arrowheads''.

For difference-coined comparisons, the influence factors describe regions of attention. We can anticipate this from participant statements where they specifically talk about paying attention to a specific region of the directed acyclic graphs because of (not) finding one of the influence factors there. An example is the statement of participant P3 and symmetry: ``If I see [...] symmetry [...] then I use it to orient myself and to check whether it is the same [in the other graph].'' Participants further describe regions resp. influence factors as supportive or hindering for spotting a change. Hindering influence factors are, for instance, density and edge-crossing. These factors have a known tendency to make visualizations less readable \cite{Bertini2005,YOGHOURDJIAN2018264,Zhang2017}. Supportive are changes in depth or shape.

Based on the evaluation of the participant statements of our difference-coined comparison studies and the consultation of related work -- i.a. \cite{6327278,Franconeri2014} -- we can say that the change detection process of humans is a two part one. First, humans identify regions of interest (RoI) by focusing on them. RoIs are regions of a directed acyclic graph visualized as a node-link diagram with specific features -- i.e., regions where the influence factors for difference-coined comparisons have occurred. Also other domains like perception and machine learning use the term RoI. We follow the definition of the machine learning domain where a RoI is a region of special interest and of fixed spatial extent. Second, they compare these RoIs in both directed acyclic graphs and check for the existence of differences within these RoIs. The RoIs of both directed acyclic graphs of the pair are related and differences are only considered if they are located inside the respective RoI.\\
So, our DFS algorithm also has two parts (cf. Figure~\ref{fig:pictures_DACH_DifferenceLearning_AlgoSchema}):
\begin{enumerate}
	\item \textbf{\orangeAlenaColor{Part 1}: RoI detection}\\
	In the DFS algorithm, we first compute for all dominant influence factors of difference-coined comparisons the RoIs (cf. Figure~\ref{fig:pictures_DACH_DifferenceLearning_AlgoSchema} -- \texttt{Factor computation}, Section~\ref{ssub:influence_factors_for_detecting_changes}).
	\item \textbf{\orangeAlenaColor{Part 2}: Comparison of the related RoIs in both directed acyclic graphs of the directed acyclic graph pair to identify the differences within the respective RoI}\\
	Second, we add all differences which we locate inside one of the calculated RoIs to the annotations set (cf. Figure~\ref{fig:pictures_DACH_DifferenceLearning_AlgoSchema} -- \texttt{Difference selection}). We encode the differences which we identified in the RoIs as annotations of the nodes and/or edges: $[0:\text{ no difference}, 1:\text{ added/difference}]$.
\end{enumerate}

\paragraph{Supportive vs. Hindering Factors.} 
\label{par:supportive_vs_hindering_factors}
We differentiate RoIs detected by supportive and hindering factors via tagging the respective RoIs as \texttt{supportive} or \texttt{hindering}. The set of difference annotations, $G_{diff}$, consists of sets of differences inside \texttt{supportive} and \texttt{hindering} RoIs (cf. Figure~\ref{fig:pictures_DACH_DifferenceLearning_AlgoSchema} -- \texttt{Difference selection}, $G_{diff}$):
\begin{itemize}
	\item \textbf{supportive RoIs:}
	\begin{itemize}
		\item $V_{supportive}$ for nodes, also called vertices, from supportive RoIs
		\item $E_{supportive}$ for edges from supportive RoIs
	\end{itemize}
	\item \textbf{hindering RoIs:}
	\begin{itemize}
		\item $V_{hindering}$ for nodes, also called vertices, from hindering RoIs
		\item $E_{hindering}$ for edges from hindering RoIs
	\end{itemize}
\end{itemize}
By inside we mean all nodes and edges which are inside the RoI box with their position.  Graph differences -- added/deleted nodes and/or edges -- can be both in the hindering and in the supportive set since it is possible that one and the same differences falls in a RoI box detected based on both a hindering and a supportive factor. So, for example, a difference in depth which also falls into a high density area will be both in the hindering and the supportive set of $G_{diff}$. It is \texttt{supportive} due to the factor depth and hindering due to the high density (cf. Section \texttt{Depth}, \texttt{Density}).

\paragraph{RoI Box Size.} 
\label{par:roi_box_size}
The individual factor determines the size of the RoI box. Factors like shape relate to the entire directed acyclic graph, so the box size is set to the entire area of the visualized directed acyclic graph. Again others like white space accumulate regions -- here: white pixels -- i.e., the RoI box size is of the size of the finally accumulated region. The RoI box results from the bounding box calculation for the respective area. A bounding box is defined as the minimum/maximum extent of an area -- $bbox = [x_{min}, x_{max}, y_{min}, y_{max}]$. If the factor is calculated based on the visualized directed acyclic graph's structure, the bounding box is defined by the minimum/maximum of all node positions. In case the bounding box is calculated pixel-based, it is defined by the minimum/maximum of all pixel positions of the respective area. It uses the Python library networkx (nx)\footnote{https://networkx.github.io/} for the calculations on the graph structure.

The factor edge crossing is a special case. It results in a fixation point around which the visualized directed acyclic graph is inspected. The fixation point is the intersection of the edges. A suitable model for the area around this fixation point is the foveal part of the human useful field of view since this is the part of the useful field of view where shape and sharp contours are processed \cite{Heun2012,Mackworth1976}. The foveal diameter is $5,2^{\circ}-6^{\circ}$  \cite{Heun2012,peripheralVision}. By following the definition of Norman H. Mackworth \cite{Mackworth1976} and setting maximum image angle $\omega$ to $6^{\circ}$, the formula for the maximum image diagonal becomes a legitimate approximation of the diagonal of foveal useful field of view bounding box (cf. Equation~\ref{eq:usefulFieldOfView}):

\begin{eqfloat}[h]
	\begin{equation}
		\begin{aligned}
    	d = 2 * f * tan(\frac{\omega}{2})\\
		\\
		\omega=\text{ visual angle} = 6^{\circ}\\
		f=\text{ monitor distance} = 700mm\\\text{ as visual search experiments usually work with this distance \cite{Pflugshaupt2009}}\\
		d=\text{ image diagonal}\\
		\\
		\Rightarrow d = 2 * f * tan(\frac{\omega}{2})\\
		\textcolor{white}{\Rightarrow} d = 2 * 700mm * tan(\frac{6}{2})\\
		\textcolor{white}{\Rightarrow} d = 2 * 700mm * tan(3)\\
		\textcolor{white}{\Rightarrow} d = \mathbf{73mm}\\
		\end{aligned}
	\end{equation}
	\caption{DFS algorithm -- approximation of the foveal useful field of view bounding box diagonal. By following the definition of Norman H. Mackworth \cite{Mackworth1976} and setting maximum image angle $\omega$ to $6^{\circ}$, the formula for the maximum image diagonal becomes a legitimate approximation of the diagonal of foveal useful field of view bounding box.}
		\label{eq:usefulFieldOfView}
\end{eqfloat}
\myequations{DFS algorithm -- approximation of the foveal useful field of view bounding box diagonal}

Since the human field of view is circular, it's bounding box is a square. This leads to the following formula and result for the box's width and height (cf. Equation~\ref{eq:usefulFieldOfViewBoundingBoxWidthHeight}):

\begin{eqfloat}[h]
	\begin{equation}
		\begin{aligned}
    	x_{min} = focusPoint(x) - \frac{\frac{d}{2}}{\sqrt{2}}\\
		x_{max} = focusPoint(x) + \frac{\frac{d}{2}}{\sqrt{2}}\\
		y_{min} = focusPoint(y) - \frac{\frac{d}{2}}{\sqrt{2}}\\
		y_{max} = focusPoint(x) + \frac{\frac{d}{2}}{\sqrt{2}}\\
		\\
		\text{focusPoint can be any point on the canvas, for this calculation we assume:} focusPoint(0,0)\\
		d = 73mm \text{(cf. Equation~\eqref{eq:usefulFieldOfView})}\\
		\\
		\Rightarrow x_{min} = focusPoint(0) - \frac{\frac{73mm}{2}}{\sqrt{2}}\approx -25.81mm\\
		\textcolor{white}{\Rightarrow} x_{max} = focusPoint(0) + \frac{\frac{73mm}{2}}{\sqrt{2}} \approx 25.81mm\\
		\textcolor{white}{\Rightarrow} y_{min} = focusPoint(0) - \frac{\frac{73mm}{2}}{\sqrt{2}} \approx -25.81mm\\
		\textcolor{white}{\Rightarrow} y_{max} = focusPoint(0) + \frac{\frac{73mm}{2}}{\sqrt{2}}\approx 25.81mm\\
		\\
		\Rightarrow boxHeight = |x_{min}| + x_{max} \approx \mathbf{51.62mm}\\
		\Rightarrow boxHeight = |y_{min}| + y_{max} \approx \mathbf{51.62mm}\\
		\end{aligned}
	\end{equation}
	\caption{DFS algorithm -- calculation of the foveal field of view's bounding box' width and height.}
		\label{eq:usefulFieldOfViewBoundingBoxWidthHeight}
\end{eqfloat}
\myequations{DFS algorithm -- calculation of the foveal field of view's bounding box' width and height}

Algorithmically, each RoI is a tuple -- $(isSupportive, box)$. The boolean parameter $isSupportive$ indicates whether the RoI results from supportive (true) or a hindering (false) influence factor. The bounding box $box$ is given by the coordinates $x_{min}, x_{max}, y_{min}, y_{max}$. 

\paragraph{Construction of the Set of Human-Like-Detected Differences.} 
\label{par:construction_of_the_set_of_human_detected_differences}
As RoI boxes model the features humans use for their comparison and as we know from Tversky's model, see Section~\ref{ssub:visual_comparison_the_change_detection_task_and_dynamic_graphs} -- \texttt{Psychological Perspective}, that these features are combined by set operations, we also employ set operations to construct our set of human-detected differences $G_{diffHumanLikeDetected}$ (cf. Equation~\eqref{eq:humanDetectedDifferences}). It is necessary to subtract the set of hindering node resp. edges differences from the set of supportive node resp. edge differences since, as discussed in Section \texttt{Supportive vs. Hindering Factors}, one and the same difference can be inside both a \texttt{supportive} and \texttt{hindering} RoI. Due to the fact that the hindering factors make it difficult to detect the change, it is legitimate to assume that a change in a hindering RoI is not spotted, therefore the subtraction of the supportive and hindering set is also legitimate.

\begin{eqfloat}
	\begin{equation}
		\begin{aligned}
    	E_{diffHumanLikeDetected} = E_{supportive} \setminus E_{hindering}\\
		V_{diffHumanLikeDetected} = V_{supportive} \setminus V_{hindering}\\
		\Rightarrow G_{diffHumanLikeDetected} = E_{diffHumanLikeDetected} \cup V_{diffHumanLikeDetected}\\
		\\
		G_{diffHumanLikeDetected} = \text{ set of human-detected graph differences}\\
		E_{diffHumanLikeDetected} = \text{ set of human-detected edge differences}\\
		V_{diffHumanLikeDetected} = \text{ set of human-detected node differences}\\
		E/V_{supportive} = \text{ set of edge/node differences from supportive RoIs}\\
		E/V_{hindering} = \text{ set of edge/node differences from hindering RoIs}
		\end{aligned}
	\end{equation}
	\caption{DFS algorithm -- set operations to construct the set of human-detected differences $G_{diffHumanLikeDetected}$. We employ set operations since we know from Tversky's model that features used for visual comparison are combined in sets and are thus modeled via set operations.}
		\label{eq:humanDetectedDifferences}
\end{eqfloat}
\myequations{DFS algorithm -- set operations to construct the set of human-detected differences}

Algorithmically, we check for each box whether it is supportive or hindering and then which differences it has. Herewith, we build the supportive and hindering sets of node ($N_{supportive}, N_{hindering}$) and edge ($E_{supportive}, E_{hindering}$) differences. If no RoI box exists around and actually existing graph difference, the difference is ignored by the algorithm since there is no RoI box indicating that the human would have spotted that difference. It is only natural for humans to overlook something \cite{Franconeri2014,Pflugshaupt2009}. The DFS algorithm only respects difference which are partially or entirely inside a RoI box. We allow for differences being partially inside a RoI box to consider the characteristics of human perception. Human perception is guided by principles known from Gestalt research domain -- i.a. the law of proximity and continuity \cite{4658147}. Following these two Gestalt laws, we assume that follow the difference which is partially inside the the box to spot the existing difference as a whole. Additionally, our model of the field of view bounding box is relatively restrictive. Humans also spot changes in the paracentral part of the useful field of view \cite{peripheralVision} which substantiated us allowing for differences which are partially inside a RoI box.



\subsubsection{Influence Factors for Detecting Changes} 
\label{ssub:influence_factors_for_detecting_changes}
%
%
The dominant influence factors for difference-coined comparisons can be computed
\begin{enumerate}
	\item based on the graph data using the graph data ($G=(V;E)$),
	\item based on the visualized node-link diagram -- i.e., the image using pixel data,
	\item or as a combination of both.
\end{enumerate}
In general all difference factors are computable both in the graph and in the image space. The space of choice is a question of computational efficiency and the availability of information in the graph space.

The graph generation tool GraphCreator (cf. \cite{10.1007/978-3-319-73915-1_20} for details) which we used again here, calculates a set of graph properties in advance. Some of which are useful for the calculation of some of the difference factors so this information is given to the DFS algorithm as an input parameter (cf. Figure~\ref{fig:pictures_DACH_DifferenceLearning_AlgoSchema} -- \texttt{csv}).







\paragraph{Visual Symmetry.} 
\label{par:visual_symmetry}

\textbf{Factor Explanation.} The most dominant influence factor is symmetry. Our participants explained that visual symmetry is supportive for spotting changes. Furthermore, participants assessed the symmetry of each directed acyclic graph separately and then used this result to spot the differences. For details please refer to \cite{10.1007/978-3-319-73915-1_20,JGAA-467}.

\textbf{Algorithmic Calculation.} For symmetry, we use the pre-calculated properties of GraphCreator. Since GraphCreator uses the nodes positions which are stored as additional information within the graph structure, visual symmetry is calculated in the graph space. After checking for symmetry, we calculate the bounding box. In case of the influence factor \texttt{symmetry}, this is a bounding box encompassing the entire directed acyclic graph. Since participants of our difference-coined comparison studies explained that they first determine symmetry for the base graph and then for the alternative, we calculate the symmetry RoI for each directed acyclic graph separately. Consequently, each directed acyclic graph pair has two symmetry RoI boxes. 


\paragraph{Shape.} 
\label{par:shape}
\begin{figure}[tb]
  \centering
    \includegraphics[width=.45\textwidth]{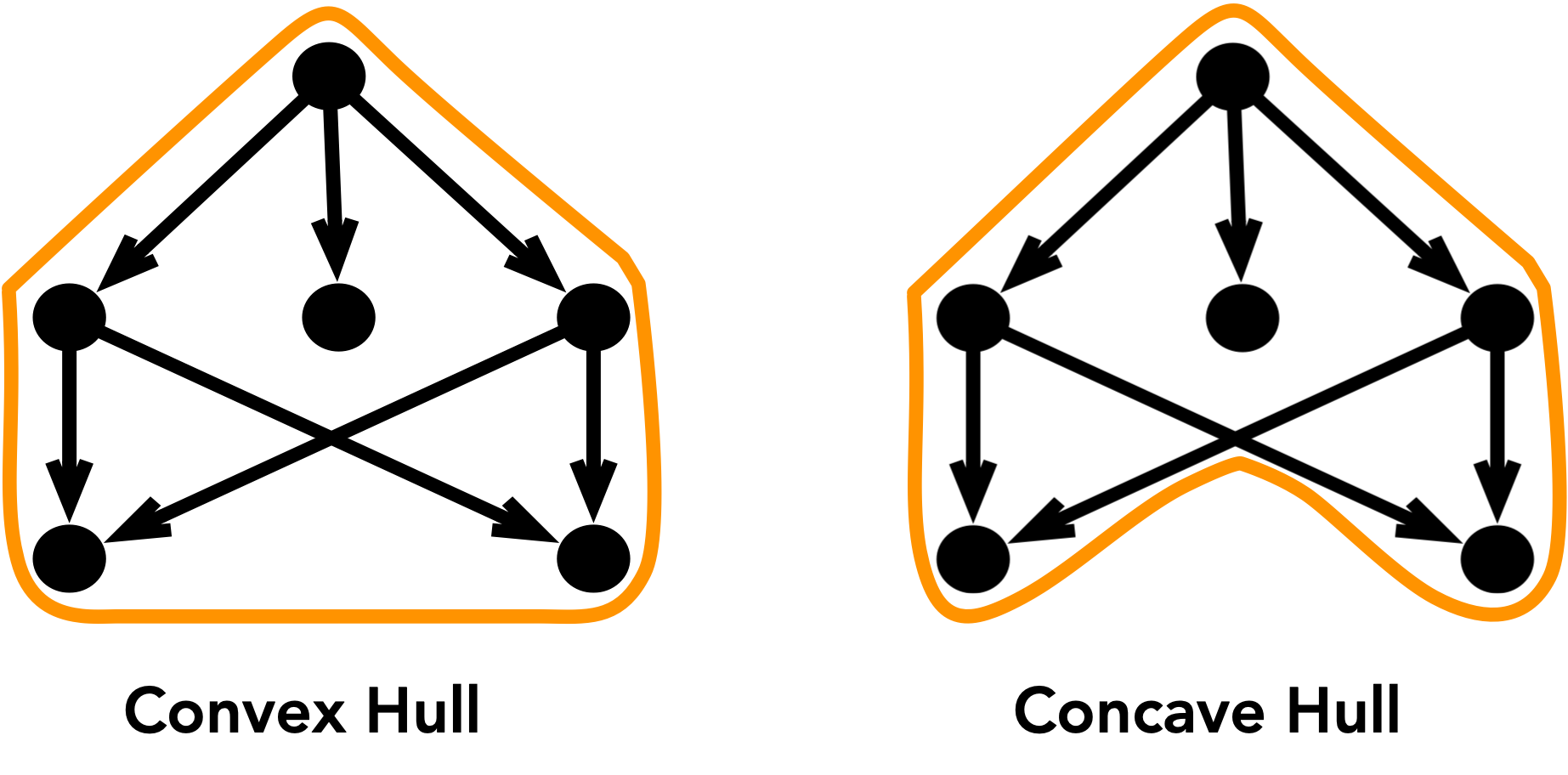}
  \caption[Convex and concave hull -- illustrated on an example directed acyclic graph]{Convex and concave hull -- illustrated on an example directed acyclic graph.}
  \label{fig:pictures_DACH_DifferenceLearning_Convex_vs_ConcaveHull}
\end{figure}

\textbf{Factor Explanation.} The participants considered the factor \texttt{shape} as very supportive. From the transcriptions it was clear that many participants searched for apparent shapes to aid their comparison and spotting the differences. So, participants repeatedly mentioned geometric shapes such as quadrilaterals, rhombi, or triangles and even used more vague descriptions such as fountain-pen nib, peak of a mountain, or smiley face to describe certain parts of a directed acyclic graph. The participants checked for the shape of both directed acyclic graphs and then compared the shapes' difference. For details please refer to \cite{10.1007/978-3-319-73915-1_20,JGAA-467}.

\textbf{Algorithmic Calculation.} The shape calculations resp. the shape difference calculations are based on the calculations of the graphs' outer hull since our participants denoted the directed acyclic graphs' shape also as the shape of the directed acyclic graphs'  silhouette and the silhouette is commonly denoted as the hull (cf. i.a. \cite{Laurentini1994,4658147,Zhang2017}). While there are varying opinions on which hull type -- convex or concave -- to use, we found throughout the development of our shape change enhancing layout for visual comparison with a user study that for our small directed acyclic graphs that there is a recognizable tendency for the concave hull. The concave hull, as opposed to the convex hull, considers also inner graph areas for the hull path. Figure~\ref{fig:pictures_DACH_DifferenceLearning_Convex_vs_ConcaveHull} shows an example. $11$ out of $20$ participants drew a concave hull around our small directed acyclic graphs when we asked them to draw the directed acyclic graphs' shape. Six participants drew convex hulls and three participants were outliers. They, for instance, applied affine transformations to the visualized directed acyclic graphs to achieve more complex formations like something which resembles the infinity symbol ($\infty$).\\
Given a base graph $G_1$ and an alternative $G_2$, we first compute the concave hulls for both directed acyclic graphs: $hull_1$, $hull_2$. 
Second, we compare the hulls and compute their area difference. 
Both concave hulls can differ in more than one area thus the difference area can consist of more than one difference area and, consequently the bounding boxes of all difference areas are defined as supportive RoIs.

%


\paragraph{Edge Crossing.} 
\label{par:edge_crossing}

\textbf{Factor Explanation.} Edge crossing denotes the fact that visualized edges may intersect each other. Our participants interpreted the factor edge crossing both supportive and hindering for the detection of differences in the directed acyclic graph pairs. The majority of our participants considered edge crossings as supportive ($17$ statements for supportive and $8$ for hindering). For details please refer to \cite{10.1007/978-3-319-73915-1_20,JGAA-467}.

\textbf{Algorithmic Calculation.} For the calculation of the edge crossings of our visualized directed acyclic graphs, we resort to the results of GraphCreator. GraphCreator calculates all edge crossing positions of the visualized directed acyclic graphs. Around each crossing position we position a field of view box as explained in Section \texttt{RoI Box Size}. This field of view box defines the RoI. Due to the fact that approximately $\frac{2}{3}$ of our participants found edge crossings supportive for the detection of changes and approximately $\frac{1}{3}$ found it hindering, we introduced a random parameter that decides whether the respective edge crossing RoI is considered as supportive or hindering. The probabilities of the random parameter result from the statement frequencies -- $70\%$ supportive, $30\%$ hindering. The edge crossing RoIs are calculated for each directed acyclic graph individually. 



\paragraph{Depth.} 
\label{par:depth}

\textbf{Factor Explanation.} The depth, or number of layers, was mainly a reason for why participants spotted a change. In general, participants considered a new layer to be clearly visible. For details please refer to \cite{10.1007/978-3-319-73915-1_20,JGAA-467}.

\textbf{Algorithmic Calculation.} To calculate the depth of the directed acyclic graphs resp. their depth difference we first get the depths and the nodes per layer for the base graph $G_1$ and for the alternative $G_2$ from the GraphCreator. We can determine the nodes per layer from the nodes' y coordinate as nodes on the same layer share the same y coordinate. Then we calculate the depth difference -- $depth_2-depth_1$ -- and the difference graph $G_{diff}$ which includes all nodes and their edges with a depth greater than $depth_1$. The resulting RoI is a bounding box of the difference graph $G_{diff}$. 

%

\paragraph{Density.} 
\label{par:density}

\textbf{Factor Explanation.} Commonly, density in the context of graphs is defined as the ratio of the number of edges and the number of nodes \cite{YOGHOURDJIAN2018264}. As we work with visualized directed acyclic graphs, density gets reinterpreted to image regions with a high rate of colored pixels \cite{Zhang2017,5613437} -- \includegraphics[height=0.4cm]{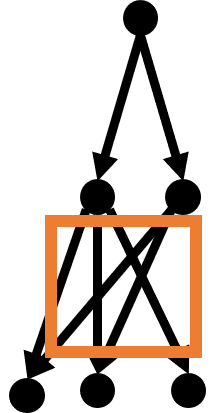}. Our participants felt that dense regions hindered the identification of differences. Less dense areas, however, facilitated spotting differences in the directed acyclic graphs.

\textbf{Algorithmic Calculation.} For the density calculations we move a field of view box, cf. Section \texttt{RoI Box Size}, over the directed acyclic graphs' node-link visualization. We approximated the pixel density threshold used for the decision on density based on research on scatterplot density by Bertini et al. \cite{Bertini2005}. We had to use this approximation, since, to the best of our knowledge, it is currently not known in related work what the precise human notion of pixel density with respect to graph density is. It is fair to draw inspiration from scatterplot research thus the nodes of graphs and scatterplots are comparable to each other. The graph edges make the visualizations pixel-wise even more dense. Bertini et al. \cite{Bertini2005} define the following density classes: areas with $\leq10\%$ of colored pixels are considered as very sparse, areas with $\leq20\%$ of colored pixels are considered as sparse up to areas with $\leq80\%$ (dense) resp. $\leq90\%$ of colored pixels which are considered as very dense. As we also want to take the statement ``Less dense areas helped to spot a difference'' into account for our model, we are looking for an upper bound of the threshold of less dense areas and a lower bound of the threshold for dense areas. Following the research of Bertini er al. \cite{Bertini2005}, we use their sparse area threshold as an upper bound of the threshold for less dense areas and the one for dense areas as the lower bound for dense areas. However, in contrast to Bertini er al. \cite{Bertini2005}, we divide those thresholds by two. The reason is that we traverse top-rooted directed acyclic graphs which have nearly a triangular shape with a rectangular sliding window. Therefore, a density of $20\%$ or $80\%$ as in Bertini would require that $20\%$ or $80\%$ of the pixel space in the rectangle is filled with gray elements. This is never achieved when we are near the outer hull of the directed acyclic graph due to the triangular shape. Since the triangular area covers about half of the rectangular area, we define our density as half of $20\%$ and $80\%$ resulting in density thresholds of $10\%$ and $40\%$. With these thresholds we calculate the density by transforming the image into a binary image and calculating the histogram sum per field of view box position. The histogram sum is highest if the field of view box only contains white pixels since white has the highest RGB value -- $255$. 

\paragraph{White Space.} 
\label{par:white_space}

\begin{figure}[tb]
    \centering
    \includegraphics[width=0.4\linewidth]{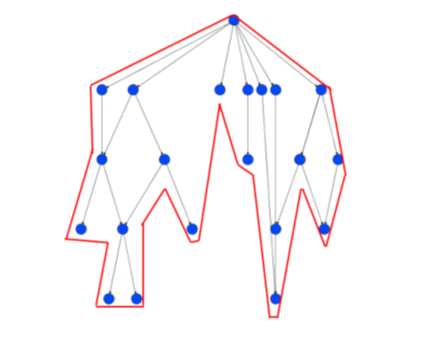}
        \includegraphics[width=0.3\linewidth]{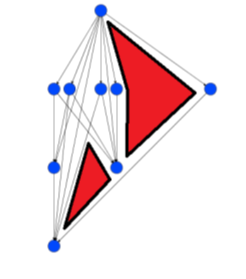}
    \caption[Shape in the graph visualization domain]{Shape in the graph visualization domain -- outer shape (left -- red line), inner shape (right -- red areas). (Figure based on original Figures from \cite{MarcMA})}
    \label{fig:shapetypes}
\end{figure}

\textbf{Factor Explanation.} White space areas result from laying out the directed acyclic graphs and visualizing them as node-link diagrams (cf. Figure~\ref{fig:shapetypes}). These areas are surrounded by graph elements but to not contain graph elements. According to our participants changes within those white space areas -- i.e., graph elements getting added in those areas or graph elements getting deleted and thus leading to white space areas -- helped to spot differences of the two directed acyclic graphs.

\textbf{Algorithmic Calculation.} As we could find with our user study data presented in \cite{guckes1} that participants switch from a convex hull to a concave hull and herewith consider white space areas near to the outer hull of the directed acyclic graph, we calculated the ratio of the difference of the concave hull pixels to the convex hull pixels (also called white space pixels) to convex hull pixels for those directed acyclic graphs where the participants switched from a convex hull to a concave hull. We found that if the difference is greater than $30\%$ of the convex hull participants switch the hull type. Thus, we initially calculate this ratio and check whether it is greater than $30\%$. For the hull calculations we use the algorithms which we already used for the work presented in \cite{guckes1}. We traverse the remaining areas resp. directed acyclic graphs where we do not find any hull difference similar to the density influence factor calculation with a sliding window approach and calculate the histogram of the sliding window. In case the histogram sum is greater than the white space threshold -- i.e., more than $30\%$ of the pixels have to be white -- we define the sliding window content is ``white space''. Here and in contrast to density, we accumulate the consecutive white space areas found with our sliding window approach. We calculate the white space only for areas inside the concave hull since white space areas are surrounded by graph elements but do not contain graph elements. These white space areas define the inner shape of a directed acyclic graph (cf. \cite{guckes1}). Since we accumulate the white space areas, the sliding window is not the field of view box. It is a convolutional filter kernel \cite{Huang2014}. Typical kernel sizes are odd numbers \cite{Jamro2001}. We set the kernel size to $k = 3$ to avoid traversing every pixel ($\widehat{=}k = 1$). This would result in processing every pixel. Furthermore, a traversal with $k=1$ would be very computationally expensive. A kernel size of $>3$ would result in a pixelated white space mask. So, $k = 3$ is a compromise between computational feasibility and accuracy. 

\subsubsection{Evaluation of the DFS-Algorithm} 
\label{ssub:evaluation_of_the_dfs_algorithm}

We have the following hypotheses regarding the human-like detected differences created by our DFS algorithm:
\begin{itemize}
	\item \textbf{Hypothesis 1:}\\
	We expect the total number of human-like detected differences to be smaller than the total number of actually existing graph differences (GT differences) since the DFS algorithm only selects differences which are (partially) inside a RoI box. Consequently, the human-like detected differences are a subset of the GT differences. As our DFS algorithm does not introduce actually non-existing differences, we expect the total number of human-like detected differences to be smaller than the GT differences' total number.
	\item \textbf{Hypothesis 2:}\\
	Our DFS algorithm filters out more edge than node differences. As the factors density and edge-crossing which are considered as hindering are edge related, we assume that edge differences are more often filtered out due to them being in a hindering RoI box.
	\item \textbf{Hypothesis 3:}\\
	With increasing density, the total number if human-like detected differences decreases. The reason is that density is a hindering factor for difference detection and thus also our DFS algorithm considers that.
\end{itemize}

For the evaluation of the DFS algorithm and our hypotheses, we
\begin{enumerate}
	\item compared the human-like detected differences with the GT differences and
	\item evaluated whether we can confirm our hypotheses.
\end{enumerate}
We used a total number of $10.000$ GT and human-like detected differences. Machine learners commonly employ this dataset size for machine learning models which capture human behavior \cite{Ma2018,wohler2019learning} or factors of the graph drawing domain \cite{10.1007/978-3-030-35802-0_38}.

\begin{figure}[tb]
  \centering
    \includegraphics[width=.9\textwidth]{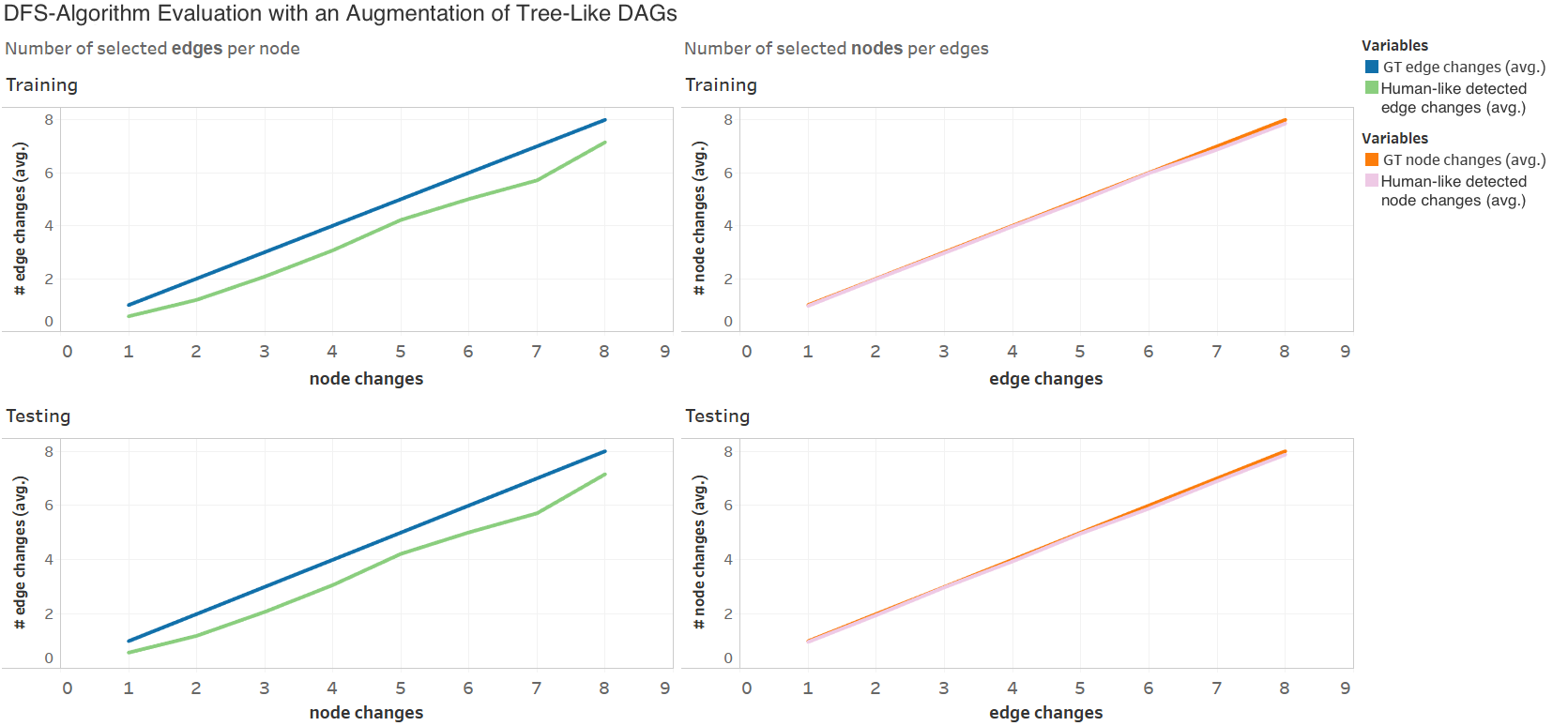}
  \caption[DFS algorithm -- algorithm evaluation with tree-like directed acyclic graphs]{DFS algorithm -- algorithm evaluation with tree-like directed acyclic graphs. The left column shows that \greenAlena{only edges are filtered out} for our set of human-like detected differences. While the \greenAlena{human-like detected edge differences} on average contain one edge less than the \darkblueAlena{GT edge differences}, the node differences of the \pinkAlena{\textbf{human-like detected}} and the \orangeAlena{GT} differences show no difference. We drew the lines slightly offset so that both are still visible. Herewith, we are able to confirm hypothesis 1 and 2. (Figure based on original Figure from \cite{AlenaMA})}
  \label{fig:pictures_DACH_DifferenceLearning_TreeLikeData}
\end{figure}

\begin{figure}[tb]
  \centering
    \includegraphics[width=.9\textwidth]{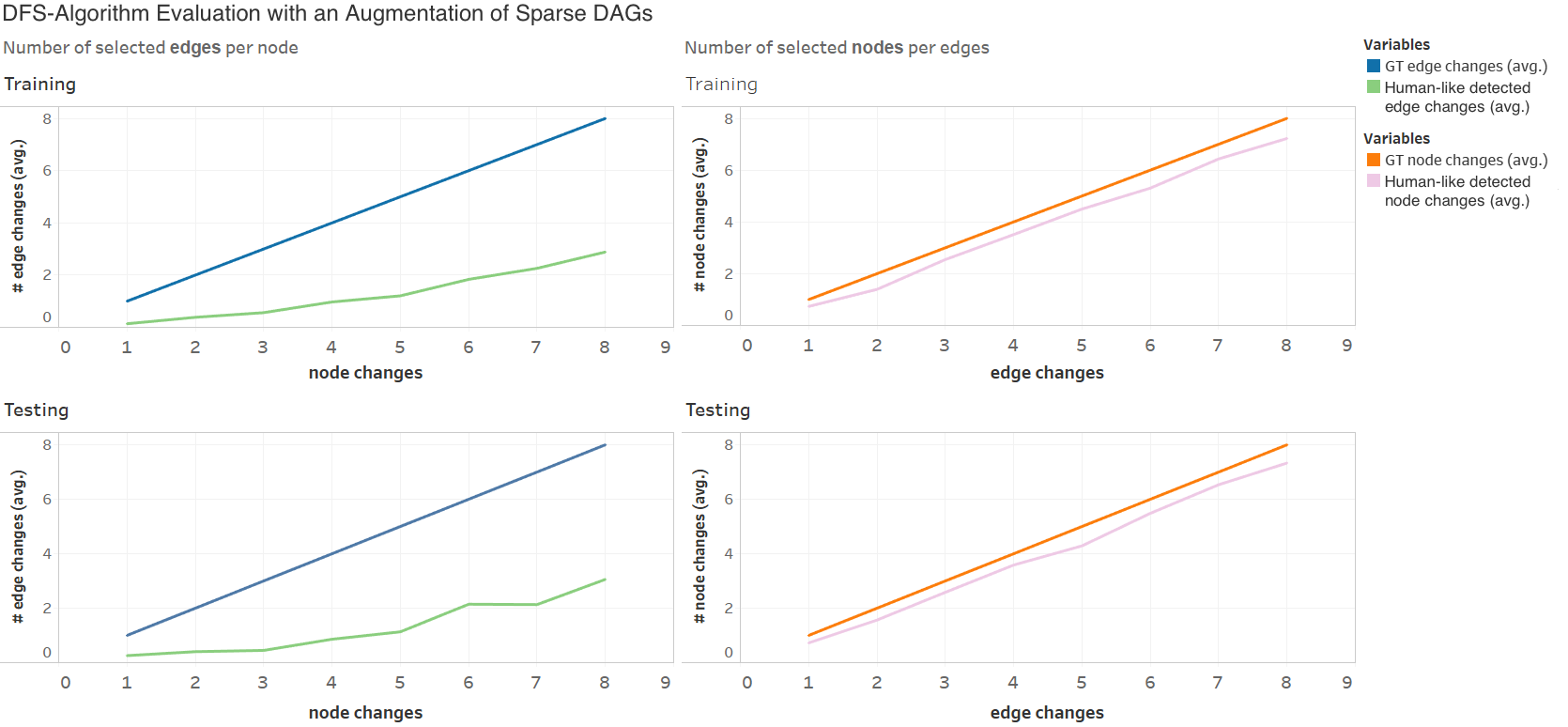}
  \caption[DFS algorithm -- algorithm evaluation with sparse directed acyclic graphs]{DFS algorithm -- algorithm evaluation with sparse directed acyclic graphs. While for \greenAlena{human-like detected} difference of tree-like directed acyclic graphs the average number of edge difference has a constant offset of one compared to the \darkblueAlena{GT} differences, the average number of \greenAlena{human-like detected} edge difference of sparse directed acyclic graphs decreases the more the more node changes appear (cf. Figure~\ref{fig:pictures_DACH_DifferenceLearning_SparseData}). For the sparse directed acyclic graphs, our DFS algorithm also detects less node changes for the set of \pinkAlena{\textbf{human-like detected}} node differenced. Compared to the \orangeAlena{GT} node differences, they have a constant offset of $0.5$. These findings clearly show that an increased density leads to an increase in hindering RoIs and thus the number of spotted node and edge differences decrease. In conclusion, we also can confirm hypothesis 3. (Figure based on original Figure from \cite{AlenaMA})}
  \label{fig:pictures_DACH_DifferenceLearning_SparseData}
\end{figure}

The Figures~\ref{fig:pictures_DACH_DifferenceLearning_TreeLikeData} and~\ref{fig:pictures_DACH_DifferenceLearning_SparseData} visualize the evaluation result. The left column shows in both Figures the average number of edge differences per node differences. The right column of both Figures shows the vice versa scenario -- the average number of node differences per edge differences. As we follow the findings of S. Franconeri \cite{Franconeri2014} and thus only consider one to eight changes, only the value range of $1-8$ is covered for both the x- and the y axis (cf. Section~\ref{sub:a_dataset_for_learning_human_detected_differences_in_dags_based_on_graph_data_enriched_with_human_inspired_detected_graph_difference}). For details on the directed acyclic graphs' sizes please also refer to Section~\ref{sub:a_dataset_for_learning_human_detected_differences_in_dags_based_on_graph_data_enriched_with_human_inspired_detected_graph_difference}.

Figure~\ref{fig:pictures_DACH_DifferenceLearning_TreeLikeData} shows that \greenAlena{only edges are filtered out }-- both for the training and the test dataset. While the \greenAlena{human-like detected edge differences} on average contain one edge less than the \darkblueAlena{GT edge differences}, the node differences of the \pinkAlena{\textbf{human-like detected}} and the \orangeAlena{GT} differences show no difference. We drew the lines slightly offset so that both are still visible. Herewith, we are able to confirm hypothesis 1 and 2.

While for \greenAlena{human-like detected} difference of tree-like directed acyclic graphs the average number of edge difference has a constant offset of one compared to the \darkblueAlena{GT} differences, the average number of \greenAlena{human-like detected} edge difference of sparse directed acyclic graphs decreases the more the more node changes appear (cf. Figure~\ref{fig:pictures_DACH_DifferenceLearning_SparseData}). For the sparse directed acyclic graphs, our DFS algorithm also detects less node changes for the set of \pinkAlena{\textbf{human-like detected}} node differenced. Compared to the \orangeAlena{GT} node differences, they have a constant offset of $0.5$. These findings clearly show that an increased density leads to an increase in hindering RoIs and thus the number of spotted node and edge differences decrease. In conclusion, we also can confirm hypothesis 3.
\FloatBarrier

\subsection{A Dataset for Learning Human Detected Structural  Differences in Directed Acyclic Graphs} 
\label{sub:a_dataset_for_learning_human_detected_differences_in_dags_based_on_graph_data_enriched_with_human_inspired_detected_graph_difference}

\begin{figure}[tb]
  \centering
    \includegraphics[width=.5\textwidth]{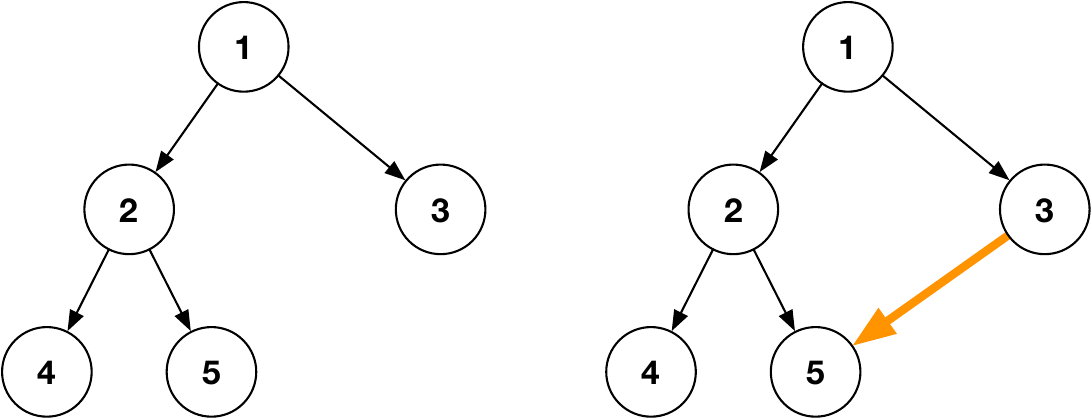}
  \caption[Example visual pairwise comparison of two graphs]{Example visual pairwise comparison of two graphs -- an edge (orange) is added to the right graph.}
  \label{fig:pictures_DACH_Introduction_AddedEdge}
\end{figure}

The creation of a suitable dataset is a challenge for empirical research and machine learning alike. Cunningham et al. \cite{wallraven2011experimental} explain that the challenge resides in the selection of appropriate data items or, as they call them, stimuli. Appropriate is a data item, following Cunningham et al. \cite{wallraven2011experimental}, if it is relevant to the research question and if it is most possibly controlled to avoid confounding effects in the experiment. Still the actual selection remains challenging since the possible dimensions characterizing data items are huge. When we consider the example of the added edge of Figure~\ref{fig:pictures_DACH_Introduction_AddedEdge} and how this change could be interpreted, the potential range of dimensions becomes roughly sketched:  Graph theoretically, the change in the right graph in Figure~\ref{fig:pictures_DACH_Introduction_AddedEdge} (orange edge) can be interpreted as an added edge. So, all black parts are common for both the left and the right graph and just the orange edge is different. The change of the orange edge can, however, also be interpreted as node 5 now having two parent nodes. This means, under this interpretation, not only the orange edge is different but also node 5. Another legitimate graph theoretical interpretation is that the graph originally being a tree became a directed acyclic graph. So, basically the entire graph is different. And these are only examples of the possible graph theoretical interpretations. The visual and human interpretations are not considered in this specific example of the Introduction. Consequently, it is impossible to cover all possible dimensions \cite{wallraven2011experimental}. In addition to that, there is also the experiment specificity and generalizability tradeoff \cite{wallraven2011experimental}. A dataset has to be specific enough to enable the examiner to answer her research question. It also has to be generalizable -- i.e., it has to be a subset of the entire population of possibly suitable data items of the experiment \cite{brunswig1910vergleichen} -- in order to ensure the generalizability of the experimental results.

The challenge of stimulus selection and the specificity-generalizability-tradeoff also applies for training and test datasets in the machine learning domain \cite{wallraven2011experimental}. The trained model is supposed to reliably solve a certain task for all previously seen data (training data), however, as well for previously unseen data (test data or live data when the model is in use). The former, solving a specific task for specific data, machine learners call robustness \cite{Keskar2016}, while they also call the latter generalizability. To achieve robustness, machine learners commonly apply data augmentation \cite{Keskar2016}. It enlarges the database and thus the total number data items presented to the machine learning model.

\subsubsection{Graph Size, Density, and Number of Differences} 
\label{ssub:graph_size_density_and_number_of_differences}
Relevant graph properties influencing the human task completion on visualized graphs are the graphs' size, their density \cite{YOGHOURDJIAN2018264}. The graph size is defined based on the number of nodes and density based on the number of edges in the graph. Due to the task at hand -- change detection or also called visual comparison with respect to differences -- the number of changes is pivotal as well \cite{AlenaPaper,Landesberger:2017,JGAA-467,DiagramsDifferencePerception,10.1145/3335082.3335083,10.1007/978-3-319-73915-1_20}. Consequently, we created our dataset for learning human detected structural differences in directed acyclic graphs considering these properties. Thus, in the following we elaborate on them in more detail.

\paragraph{Graph Size.} 
\label{par:graph_size}
Research on cognitive load found that humans have limited resources for completing visual comparison tasks such as difference detection \cite{Franconeri2014}. The detection of differences is a slow and cognitive demanding task \cite{Franconeri2014} which raises the necessity to constrain the graph size. To the best of our knowledge, there is currently no research discussing the optimal graph size for experiments on visual comparisons with respect to differences. However, there is research which is able to provide guidance for deciding on the graph size. Yoghourdjian et al. \cite{YOGHOURDJIAN2018264} give an encompassing overview of graph sizes and densities used in experiments and human-centered user studies over the last years. Huang et al. \cite{AlenaPaper} found in their experiments on cognitive load that already small graphs, $25$ nodes, load to such a high cognitive load that humans' error rate and task completion time remarkably increase. Given that insight and given the complexity of difference detection \cite{Franconeri2014}, it is advisable for us to use graphs which are smaller than the small graphs of Huang et al. \cite{AlenaPaper}. Alper et al. \cite{Alper2013} substantiate that with their findings that node-link diagrams are especially affected by graph size. Hlawatsch et al. \cite{6812198} evaluate static visualizations of dynamic weighted graphs and employ also graphs smaller than those of Huang et al. \cite{AlenaPaper} ($8$ and $20$ nodes) to avoid interaction or animation. Blythe \cite{Blythe1995}, Purchase \cite{purchaseaesthetics}, and Kieffer et al. \cite{Kieffer2016} also use graphs smaller than those of Huang et al. to avoid overwhelming their participants.

In conclusion, the choice of the graph size is a decision where there is no clear answer to it yet. Since we also employ static node-link diagram visualizations, we follow the argumentation of Hlawatsch et al. \cite{6812198} at set the maximum graph size to $20$. Since our research question is located in the psychological domain as well, our goal is also to avoid cognitive overload which is why we decided to also follow the argumentation of Blythe and ours of our commonalities-coined comparison studies and thus set the lower bound of our graph size to $6$.

\paragraph{Density.} 
\label{par:density}
Density is a graph property important to consider since it generally affects the readability of graphs \cite{Purchase:1997a} and the human chances to spot a difference \cite{DiagramsDifferencePerception,10.1145/3335082.3335083}. With their state of the art review of experiments and user studies on visualized graphs, Yoghourdjian et al. \cite{YOGHOURDJIAN2018264} found that usually graphs with a density of $\leq20\%$ are used. There is more than one option to measure density. The suitability of the respective density formula depends i.a. on the purpose  and the graph type \cite{10.1145/1168149.1168167}. Usually, the network efficiency is used as a measure for density \cite{Zhang2017}. The network efficiency is defined as the ratio of the existing edges $|E|$ of the graph and all possible edges of the graph ($|V|*(|V|-1)$): $\frac{|E|}{|V|*(|V|-1)}$ \cite{Zhang2017}. Since the denominator is the set of all possible edges which would include i.a. edges leading to circles, this metric is not applicable for our directed acyclic graphs. This leads to us using the linear density metric -- $d = \frac{|E|}{|V|}$. Linear density is also a frequently used density metric \cite{10.1145/1168149.1168167}. Yoghourdjian et al. \cite{YOGHOURDJIAN2018264} also use density to build their density classes: $d_{tree-like}=[0-1]$, $d_{sparse} = (1-2])$, $d_{dense} = (2-4]$

The choice of the density range might be even a tougher decision than the graph size since even lesser studies justify their graph density or in general research optimal density ranges for specific tasks or other contexts. Some papers (cf. e.g., \cite{Alper2013,Moscovich2009}) argue that the participants' error rate and task completion time remarkably increase when the density is increased a bit. These papers, however, use graphs with a size of $1000$ nodes which makes their results at best only remotely comparable to the graphs we are aiming for. Consequently, we also consulted the same body of work for our density decision which we consulted for the decision on the graph size. The majority uses similar densities -- $d\approx[1-2]$. Huang et al. \cite{AlenaPaper} explain that they chose a density between $\approx1-2$ to manage the complexity of the participants' task. The other authors did not justify their choice, however, we could calculate their graphs' density by ourselves. In Addition to that, Yoghourdjian et al. \cite{YOGHOURDJIAN2018264} state that sparse graphs are well-suited for ``structural identification tasks''. As our task at hand is learning human detected structural differences in the context of difference-coined visual comparisons, we decided to definitely use sparse graphs. For the definition of ``sparse'', we follow the one of. Yoghourdjian et al. \cite{YOGHOURDJIAN2018264} -- $d_{sparse} = (1-2])$. We keep the sparse density class as our density upper bound since the vast majority of experiments and user studies Yoghourdjian et al. \cite{YOGHOURDJIAN2018264} present in their state of the art review use a density of $\leq2$. We add the tree-like class to avoid oversimplification of the problem for our machine learning algorithm. For this class, we also follow the definition of Yoghourdjian et al. \cite{YOGHOURDJIAN2018264} -- $d_{tree-like}=[0-1]$.


\paragraph{Number of Differences.} 
\label{par:number_of_differences}
From the current body of work on cognitive limits, cf. e.g., \cite{Franconeri2014}, we know that the cognitive resources needed for tasks like difference detections are limited. Thus, it is pivotal for us to keep the number of changes in our directed acyclic graph pairs in limits which would not overwhelm participants and would avoid biased results. We follow the results of Burkell et al. \cite{Burkell1997}. The authors found that the upper bound of object selection are five to eight locations simultaneously. Since object locations correspond to change locations in a visualized directed acyclic graph, we set our range of changes to $[1-8]$. Our upper bound follows Burkell et al. \cite{Burkell1997}. the lower bound is to achieve altered directed acyclic graphs which also respect the density classes. These limits already find application in the visualization domain (cf. e.g., \cite{10.1007/978-3-642-36763-2_42}).

\paragraph{Final directed acyclic graph Properties Setting.} 
\label{par:final_dag_properties_setting}
We summarize our final setting in Table~\ref{tab:graphsize}. For details on the rationals of the upper and lower bounds please refer to Section~\ref{sub:a_dataset_for_learning_human_detected_differences_in_dags_based_on_graph_data_enriched_with_human_inspired_detected_graph_difference}.

\begin{table}[H]
    \centering
    \begin{tabular}{|c|c|c|}
        \hline
         Graph Size & Linear Density & Number of Differences \\
         \hline
        $6-20$ nodes& $0 - 2$ & $1-8$ changes\\
        \hline
    \end{tabular}
    \caption[Properties of the directed acyclic graphs of our dataset to learn human detected structural differences]{Properties of the directed acyclic graphs of our dataset to learn human detected structural differences of directed acyclic graphs. For details on the rationals of the upper and lower bounds please refer to Section~\ref{sub:a_dataset_for_learning_human_detected_differences_in_dags_based_on_graph_data_enriched_with_human_inspired_detected_graph_difference}. (Table taken from \cite{AlenaMA})}
    \label{tab:graphsize}
\end{table}


\subsubsection{Dataset Creation} 
\label{ssub:dataset_creation}
\begin{figure}[tb]
  \centering
    \includegraphics[width=.7\textwidth]{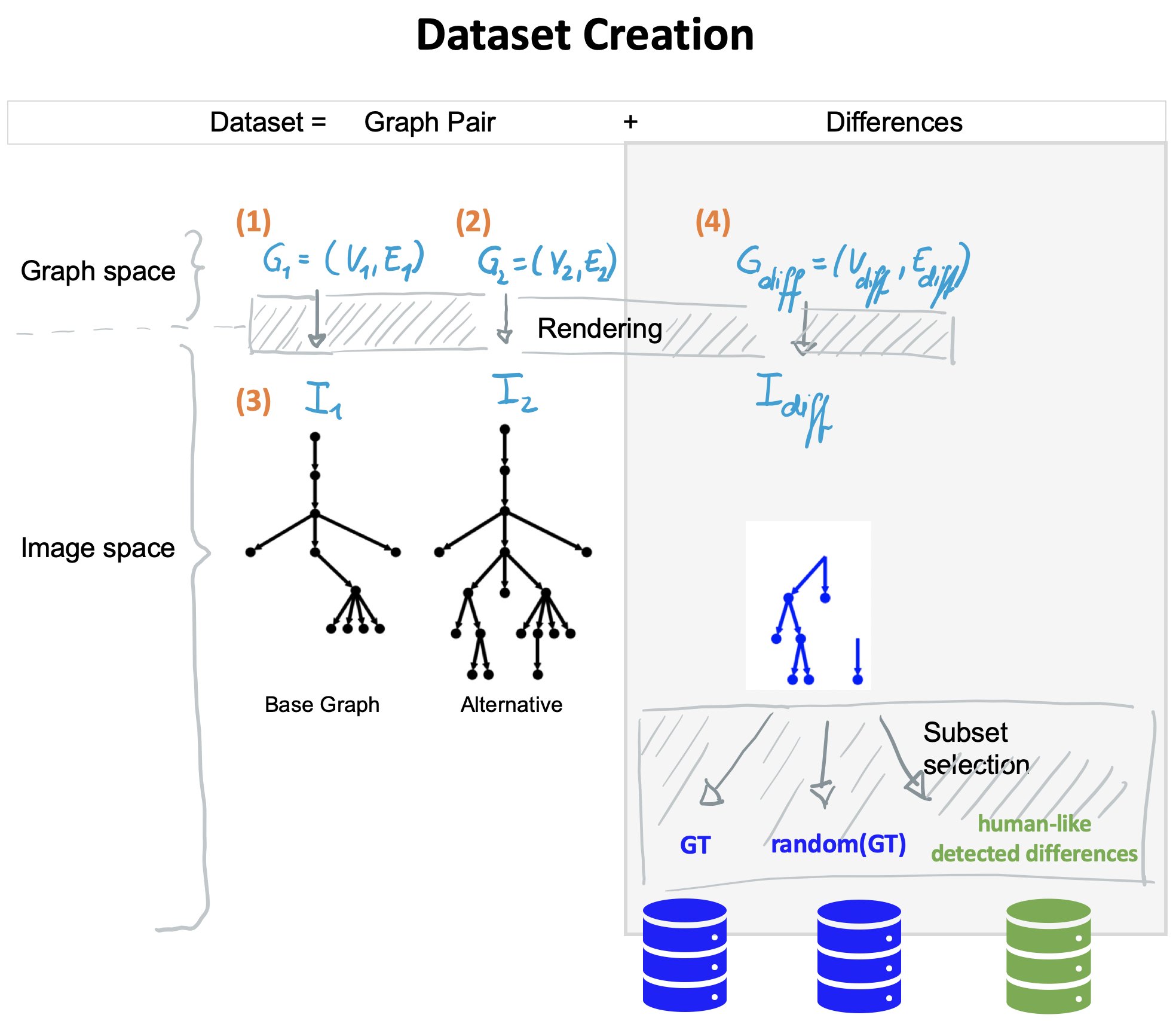}
  \caption{Dataset creation process for our training and test dataset to learn human detected structural differences. (Figure based on original Figure from \cite{AlenaMA})}
  \label{fig:pictures_DACH_DifferenceLearning_DatasetCreation}
\end{figure}

Our dataset creation process has four steps (cf. Figure~\ref{fig:pictures_DACH_DifferenceLearning_DatasetCreation}):
\begin{enumerate}
	\item[\orange{(1)}] Creation of the base graphs ($G_{1}$)
	\item[\orange{(2)}] Creation of the alternative graph ($G_{2}$)
	\item[\orange{(3)}] Image generation of the node-link visualizations of $G_{1}$ and $G_{2}$ ($I_{1}$, $I_{2}$)
	\item[\orange{(4)}] Difference generation
	\begin{enumerate}
		\item \HLdifferencesgreenColor{human-like detected differences}
		\item ground truth (GT) differences -- \GTblueColor{GT} and \GTblueColor{random(GT)}
	\end{enumerate}
\end{enumerate}
\FloatBarrier

\paragraph{Base Graph Creation \orange{(1)}. } 
\label{par:base_graph_creation_orange_1}
First, we create all possible directed acyclic graphs with a size of six to $12$ nodes. Second, we had to filter resp. sample the created base graphs to i.a. attribute the base graphs to the density classes and to ensure that they show the influence factors of difference-coined visual comparisons.

To construct the tree-like density class, we filter all created base graphs for a density of $d \leq 1$. For the class of sparse directed acyclic graphs we do the same just for a density of $1 < d \leq 2$. As we work with all possibilities, the directed acyclic graphs with a density of $1$ contain a high redundancy. To reduce the redundancy, i.e., maximize the data diversity\footnote{needed to ensure the robustness of the machine learning model to be learned \cite{Keskar2016}}, and to ensure that the selected tree-like base graphs show the influence factors of the human similarity notion for difference-coined comparisons we conduct a selective filtering based on the influence factors (cf. \cite{Chen2019} for procedural details). For the selective filtering, we consider the remaining influence factors as follows:
\begin{itemize}
	\item \textbf{Symmetry}\\
	We only filter symmetric directed acyclic graphs since symmetry is easier to break than to create \cite{10.1007/978-3-319-73915-1_20}.
	\item \textbf{Shape}\\
	The shape has no influence on the selection of the base graphs. It takes effect only when there is a change in shape between the base graph and the alternative graph.
	\item \textbf{Edge Crossing}\\
	Due to us focusing on additions as graph changes, cf. Section \texttt{Alternative Graph Creation} for details, existing edge crossings of a base graph can never be resolved due to no edge deletions. Consequently, we only select base graphs with no edge crossings and produce the variation necessary for machine learning -- edge crossings and no edge crossings -- via the creation of the alternative graphs.
	\item \textbf{Depth}\\
	We cut off the upper and lower $25\%$ quartile to filter for the most common depth and to allow for depth changes when the alternative graphs are created.
	\item \textbf{Whitespace}\\
	Whitespace has no influence on the selection of the base graphs. It takes effect only when there is a change in it between the base graph and the alternative graph.
\end{itemize}
As the number of possible edges depends on the density to be achieved \cite{diestel2017graphentheorie} and the number of directed acyclic graphs increases exponentially with the number of edges there are considerably more sparse base graphs than tree-like base graphs. Consequently, we employ a random sampling for those graphs \cite{Chen2019}.


\paragraph{Alternative Graph Creation \orange{(2)}.} 
\label{par:alternative_graph_creation_orange_2}
From the base graphs, we generate directed acyclic graph pairs by adding up to eight graph elements -- i.e., nodes and/or edges -- to the base graph to create an alternative graph. Here, we use again GraphCreator (cf. \cite{10.1007/978-3-319-73915-1_20} for details). We focus on additions since we identified the influence factors also for additions due to Tversky's insight that the direction of comparison is not invariant \cite{Tversky2004}. This implies that one cannot assume that the influence factors of visual comparisons of data items which differ due to additions also hold for visual comparisons of data items which differ due to deletions. This is also in line with our findings from our preliminary study on the human similarity notion of star-shaped graphs \cite{Landesberger:2017}. There we found that participants similarity perception varies remarkably depending on whether there are additions or deletions in the graph pairs. We apply to both the base and the alternative graph a Sugiyama-like hierarchical layout which preserves the mental map via first laying out a union graph of both and second removing the additions when laying out the base graph. Consequently, an exact node matching between the base ($G_{1}$) and the alternative ($G_{2}$) graph is ensured.

Finally, we randomly sample the just created directed acyclic graph pairs to achieve a dataset size of $10.000$. Machine learners commonly employ this dataset size for machine learning models which capture human behavior \cite{Ma2018,wohler2019learning} or factors of the graph drawing domain \cite{10.1007/978-3-030-35802-0_38}. Since the base graphs have $6-12$ nodes for both the tree-like and the sparse directed acyclic graphs and $5-10$ (tree-like) resp. $8-17$ (sparse) edges, all pairs' directed acyclic graphs have $6-12$ nodes and $5-10$ (tree-like) resp. $8-17$ (sparse) edges.

\paragraph{Image Generation \orange{(3)}.} 
\label{par:image_generation}
For the generation of the node-link visualizations we again use the networkx\footnote{https://networkx.github.io/} library which allows for an easy conversion of graph-structured data to visualizations or, generally speaking, images. The library fetches the node positions from the graph-structured data where the final layout result is stored. We store\footnote{Implementation can be found in the \texttt{convert.py} Python script.} the images with a fixed size of $800x800$px and fix node/edge color (black), node size ($300$), edge width ($2$), and difference node/edge color (blue) since visual design is known to influence the human processing a graph visualization (cf. e.g., \cite{5742390}). We evaluated also smaller images sizes as their advantage would have been a reduction of the training time. However, the resulting lower image resolution lead to pixelated segmentation masks. Consequently, we refrained from lower image sizes.

\paragraph{Difference Generation \orange{(4)}.} 
\label{par:difference_generation}

\begin{figure}[tb]
    \centering
    \scalebox{.5}{\includegraphics{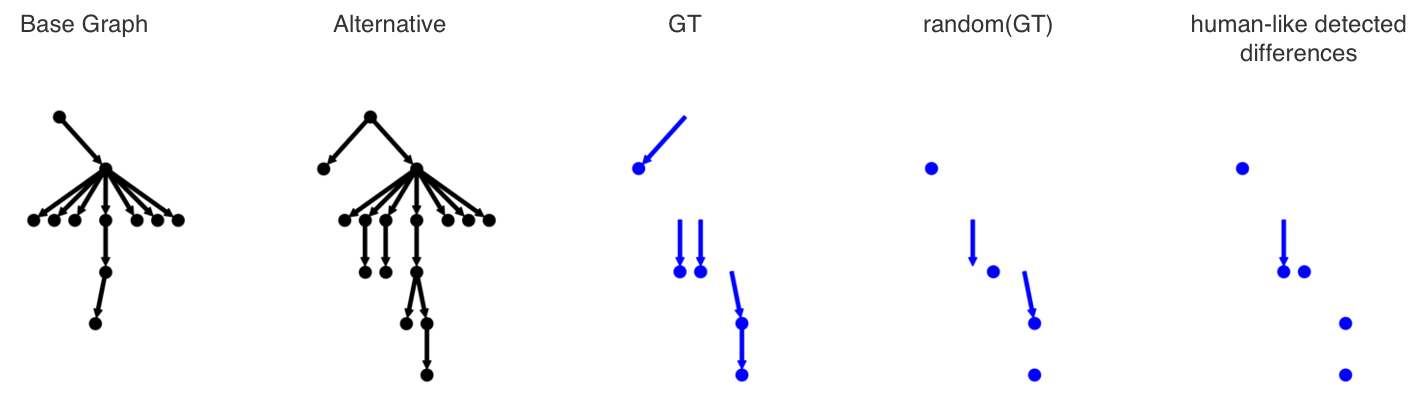}}
    \caption[Exemplar directed acyclic graph pair showing the variations between the GT, random(GT), and human-like detected differences]{Exemplar directed acyclic graph pair showing the variations between the GT, random(GT), and human-like detected differences. (Figure based on original Figure from \cite{AlenaMA})}
    \label{fig:trainingSample}
\end{figure}

\begin{figure}[tb]
    \centering
    \includegraphics[scale=0.2]{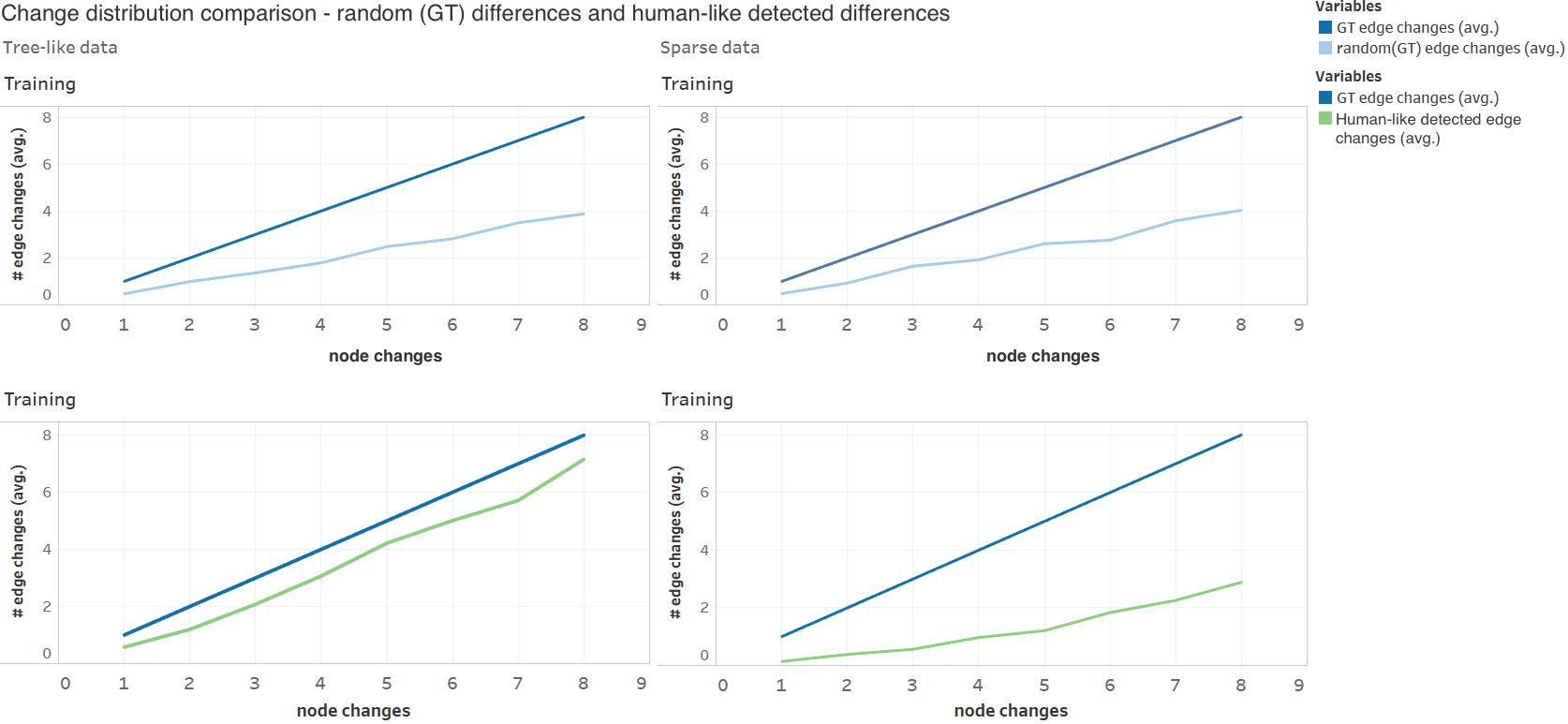}
    \caption[Change distributions of node/edge changes average for GT, random(GT) and human-like detected differences]{Change distributions of node/edge changes average for GT, random(GT) and human-like detected differences. (Figure based on original Figure from \cite{AlenaMA})}
    \label{fig:dataDistribution}
\end{figure}

We encode the differences which we identified in the RoIs as annotations of the nodes and/or edges: $[0:\text{ no difference}, 1:\text{ added/difference}]$.

\textbf{Ground Truth Differences (\GTblueColor{GT}).} We calculate the ground truth differences using the difference map approach of Archambault et al. \cite{Archambault:2011} since it is an approach specifically designed for structural graph differences. We store the result in $G_{diff}$. The Python script responsible for these calculations is the \texttt{GraphDifferences.py} script. The image-based differences are calculated by the \\ \texttt{ImageDifferences.py} script which stores its in $I_{diff}$.

\textbf{Random Ground Truth Differences (\GTblueColor{random(GT)}).} During the domain-specific fine-tuning phase (cf. Section~\ref{sub:a_convolutional_neural_network_for_learning_human_detected_differences_in_dags} -- \texttt{Domain-Specific Fine-Tuning and Final Architecture} for details), we were operating under the idea that a two-stage fine-tuning -- 1) fine-tuning on GT dataset, 2) fine-tuning on human-like detected differences dataset -- is beneficial as features characterizing nodes and edges can be learned from the larger GT differences dataset and then be used throughout the adaption to the differences humans would detect. As especially real human detected difference data may be limited the fine-tuning on the GT difference dataset would enlarge the size of the training data. However, we had to realize that human detected difference data resp. the human-like detected difference dataset constructed with our DFS-algorithm differs in one key property which hindered our architecture to adapt from the GT difference to the human-like detected differences: due to the influence factors with hindering effect, the differences resulting from humans are more disconnected (cf. Section~\ref{sub:dataset_creation_algorithm_for_enriching_dag_data_with_human_like_detected_graph_differences} for details on the rationals). We also model this with our DFS-algorithm (cf. Section~\ref{sub:dataset_creation_algorithm_for_enriching_dag_data_with_human_like_detected_graph_differences} for details). Therefore, we decided to randomly sample the GT differences to get more disconnected changes\footnote{The creation of random(GT) is implemented with the function \texttt{compute\_user\_differences} in \texttt{GraphUserDifference}.}. The random sampling only takes a subset of nodes and edges of the GT differences in $G_{diff}$. The number of nodes to be randomly selected is defined by the parameter $n_{random}$. $n_{random}$ is uniformly distributed over $[1,8]$ to maximize the diversity of changes in the node set. The same holds for the parameter $e_{random}$ which determines the number of edges which shall be randomly sampled. Since the selection is random the differences of random(GT) can be single edges and/or nodes or connected changes -- e.g., a node and an incident edge (cf. Figure~\ref{fig:trainingSample}). By sampling over $[1, number(GT differences)]$ we ensure that at least one changes is taken and that there are connected changes.\\
Figure~\ref{fig:trainingSample} shows based on an exemplar data item that the random(GT) difference and the human-like detected differences resemble each other more than the GT differences and the human-like detected differences. To check the random(GT) dataset, we also compared the change distribution of the \humanLikeDetectedcolor{human-like detected differences dataset} and the \randomGTcolor{\textbf{random(GT) dataset}}. Figure~\ref{fig:dataDistribution} visualizes the result. We can see that there is a certain difference in the change distribution of the \humanLikeDetectedcolor{human-like detected differences dataset} and the \randomGTcolor{\textbf{random(GT) dataset}}. This is due to the sampling's validity check that $n_{random}$ and $e_{random}$ shall not be greater than $n$ and $e$ -- the number of actual node and edge changes. This would try to sample more changes than actually exist. As this is not possible, the validity check replaces the $random$ parameter with the number of actual node resp. edge changes. This affects the uniform distribution over $[1,8]$ and leads to the difference in the change distributions. Since other distributions like the triangular distribution were also affected by this indispensable validity check and the most diverging slopes -- the slope of the tree-like \randomGTcolor{\textbf{random(GT) dataset}} and the one of the tree-like \humanLikeDetectedcolor{human-like detected differences dataset} -- remain in a comparable range of $[0,1)$ as opposed to an approximately constant slope of $1$ for the tree-like \humanLikeDetectedcolor{human-like detected differences dataset}, we decided to keep the uniform distribution for the creation of the random(GT) dataset.

\textbf{\HLdifferencesgreenColor{Human-Like Detected Differences}.}
For the human-like detected differences only a subset of the GT differences is further processed. This subset contains all nodes and edges selected by our DFS-algorithm because they are located inside a supportive RoI (cf. Section \texttt{DFS-Algorithm -- Overview} for details). The functions responsible for filtering only those nodes and edges which are located inside a supportive RoI are \texttt{clamp\_nodes} and \texttt{clamp\_edges} in \texttt{GraphUserDifferences}.

We created our training and test dataset based on the afore explained process. Table~\ref{tbl:datasetSparse} shows an excerpt of the training dataset's sparse directed acyclic graphs.

  \vspace{-0.6cm}
\begin{table}
    \centering
	  \vspace{-0.6cm}
    \begin{tabular}{|c|ccc|}
       \hline
        Dataset Size & Base Graph ($G_1$) & Alternative Graph ($G_2$) & $G_{diffHumanLikeDetected}$  \\
        \hline
        & \begin{minipage}{0.2\textwidth} \vspace{2mm} \centering \datasetSparseAGone \vspace{2mm} \end{minipage} &  \begin{minipage}{0.2\textwidth} \vspace{2mm} \datasetSparseAGtwo \vspace{2mm} \end{minipage} & \begin{minipage}{0.2\textwidth} \vspace{2mm} \datasetSparseAGdiff \vspace{2mm} \end{minipage} \\
        \cline{2-4}
        10000 &  \begin{minipage}{0.2\textwidth} \vspace{2mm} \centering \datasetSparseBGone \vspace{2mm} \end{minipage} &  \begin{minipage}{0.2\textwidth} \vspace{2mm} \centering \datasetSparseBGtwo \vspace{2mm} \end{minipage} &  \begin{minipage}{0.2\textwidth} \vspace{2mm} \centering \datasetSparseBGdiff \vspace{2mm} \end{minipage} \\
        \cline{2-4}
         & \begin{minipage}{0.2\textwidth} \vspace{2mm} \centering \datasetSparseCGone \vspace{2mm} \end{minipage} & \begin{minipage}{0.2\textwidth} \vspace{2mm} \centering \datasetSparseCGtwo \vspace{2mm} \end{minipage} & \begin{minipage}{0.2\textwidth} \vspace{2mm} \centering \datasetSparseCGdiff \vspace{2mm} \end{minipage} \\
         \cline{2-4}
         & \begin{minipage}{0.2\textwidth} \vspace{2mm} \centering \datasetSparseDGone \vspace{2mm} \end{minipage} & \begin{minipage}{0.2\textwidth} \vspace{2mm} \centering \datasetSparseDGtwo \vspace{2mm} \end{minipage} & \begin{minipage}{0.2\textwidth} \vspace{2mm} \centering \datasetSparseDGdiff \vspace{2mm} \end{minipage} \\
        \hline
    \end{tabular}
    \caption[Excerpt of the sparse training dataset for learning human detected structural differences]{Training dataset for learning human detected structural differences -- four exemplary sparse directed acyclic graph pairs including user-inspired differences. (Table based on original Table from \cite{AlenaMA})}
    \label{tbl:datasetSparse}
\end{table}
\FloatBarrier


\subsection{Change Detection as a Machine Learning Problem} 
\label{sub:change_detection_as_a_machine_learning_problem}
Since the human detected differences do not match entirely with the actual graph differences -- ground truth differences (GT) -- traditional methods like segmentation are not sufficient \cite{HUSSAIN201391}. Traditional methods identify actually appearing differences either in the graph space or in the image space. Graph space methods usually employ graph matching \cite{10.1145/1201775.882291,Girschick2006,archambault2009structural}. Traditional image space methods rely on change detection via a series of preprocessing steps such as segmentation \cite{HUSSAIN201391,1395984,Ji2019}. The result is a change mask. Radke et al. \cite{1395984} give an encompassing overview in their state of the art review. The traditional graph- and image-based methods have in common that they aim to identify all appearing differences within the two graphs or images \cite{1395984}. However, our aim is to be able to predict which differences humans detect when they compare two visualized directed acyclic graphs. So, we do not want to detect all present differences but just the ones human spot.As the human detected differences are a subset of the actually appearing structural differences, our research goal corresponds to subset detection \cite{1395984}. Subset detection requires semantic classification of image regions and thus goes beyond the traditional approaches \cite{1395984}. Commonly, machine learning approaches successfully solve the semantic classification problem \cite{Ji2019}. So, our research goal also requires a learning based approach.

Machine learning approaches depend on the objective function and on the data used for learning since the objective and the data type determine which learning approach and loss function is suitable \cite{bishop2006pattern}. We covered all data relevant aspects in Section~\ref{sub:dataset_creation_algorithm_for_enriching_dag_data_with_human_like_detected_graph_differences} and ~\ref{sub:a_dataset_for_learning_human_detected_differences_in_dags_based_on_graph_data_enriched_with_human_inspired_detected_graph_difference}. In this Section, we discuss the objective function of change detection and give an overview over potential learning approaches.

\subsubsection{Objective Function} 
\label{ssub:objective_function}
Given a directed acyclic graph pair, our objective is to detect nodes and edges which humans detect as different between the base graph ($G_{1}$) and the alternative ($G_{2}$) of the directed acyclic graph pair. Differences are represented as annotations of the nodes and edges. $0$ means ``no difference'' while $1$ means ``change/addition''. The annotations and the properties of the annotated graph elements, i.e., nodes or edges, bear the information based on which the algorithm shall learn to predict the differences humans would detect in previously not seen directed acyclic graph pairs.\\
The learning problem can be defined as follows: $D_{HumanLikeDetectedDifferences}$ shall be the dataset annotated with our DFS-algorithm. The learning algorithm shall now minimize the loss $L$ of predicting actually not different graph elements, i.e., graph elements annotated with a $0$, as being different. As our annotations consist of $0$ and $1$ -- two classes -- this is a binary classification problem.

\subsubsection{Graph-Based and Image-Based Change Detection Approaches} 
\label{ssub:graph_based_and_image_based_change_detection_approaches}
\begin{figure}[tb]
    \centering
    \scalebox{.5}{\includegraphics{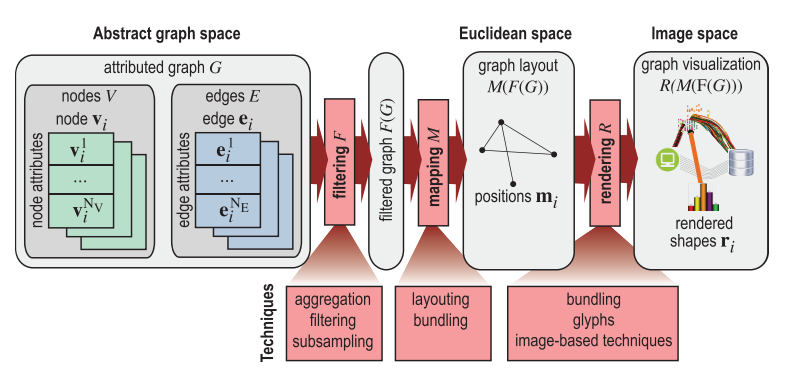}}
    \caption[Transformation of non-euclidean graph-structured data ($G$) into the euclidean space ($M(F(G))$) via laying out the graph data with a layout algorithm \cite{Telea2018}]{Transformation of non-euclidean graph-structured data ($G$) into the euclidean space ($M(F(G))$) via laying out the graph data with a layout algorithm \cite{Telea2018}. Then, the graph embedding is rendered ($R(M(F(G)))$). For our work, we ignore the aggregation techniques since we use small directed acyclic graphs. (Figure taken from \cite{Telea2018})}
    \label{fig:image-based}
\end{figure}

The question of the approach type becomes pivotal since this sets the focus of the learning method. While graph-based methods focus on the structural information of the graph data, image-based methods of a stronger focus on the representational image information. Graph-structured data and images are different in their geometric localization \cite{Grattarola2019,Telea2018,Wu2019}. Images  are directly representable in the euclidean space due to their 2D pixel locations. Graphs are per se non-euclidean \cite{Grattarola2019,Wu2019}. A commonly used circumvention of the non-euclidean nature of graph data is to use the graph embedding \cite{Wu2019}. Graph layout algorithms embed the nodes by assigning them positions, usually 2D or 3D positions, on the canvas. The result of this graph embedding is a transformation of the actually non-euclidean graph into the euclidean space. Figure~\ref{fig:image-based} visualizes this relation.
\FloatBarrier

\paragraph{Graph-Based Change Detection Approaches.} 
\label{par:graph_based_change_detection_approaches}
The goal of detecting differences with graph-based learning approaches is relatively new \cite{Grattarola2019}. Currently, the existing approaches rather focus on dynamic graphs. As explained in Section~\ref{sub:definitions_and_research_context} -- \texttt{Change Detection Task}, work dealing with dynamic graphs is transferable into the domain of our research question since one can always make one graph of the time series be the base graph and another one be the alternative. Difference learning on graph data is a classification problem \cite{Bhagat2011}. The machine learning algorithm learns labels for nodes and edges based on a labeled training dataset \cite{Bhagat2011}.

The current body of work suggest many learning algorithms for pure node classification \cite{Bhagat2011,10.1145/2939672.2939754,Hamilton2017,Kazienko2012,10.1145/2736277.2741093,Xie2019,Xu2018}. Algorithms employing random walk learn the class labels by computing probabilities using a transition matrix which contains probabilities for reaching neighboring nodes. More recent approaches rather employ graph neural networks \cite{Wu2019,Liu2019}. graph neural networks are usually categorized as \cite{Liu2019}:
\begin{enumerate}
	\item graph convolution networks
	\item graph attention networks
	\item graph auto-encoders
	\item graph generative networks
	\item graph spatial-temporal networks
\end{enumerate}

For the topic of change detection, there are solutions which use graph convolutional networks \cite{Liu2019a} and graph auto-encoders \cite{Liu2019}. graph convolutional networks base on the adaption of convolutional operations tp graph-structured data \cite{Wu2019,Xu2018,Kipf2016,KTENA2018431,Ma2018a}. Like neural networks, graph convolutional networks capture local and global features and also learn based on these features. Most graph convolutional network-based solutions lack the consideration of node positions which result from a graph layout. The majority of solutions perform their own embeddings by applying learning algorithms (cf. e.g., \cite{Wu2019}). The research of Danel et al. \cite{Danel} is an exception. Their solution considers node positions to augment the data by rotating or translating the graphs. Graph auto-encoders are an adaption of graph adversarial networks (GANs) \cite{Goodfellow2014,Zhang2018b}. GANs were created to generate graphs.

\paragraph{Image-Based Change Detection Approaches.} 
\label{par:image_based_change_detection_approaches}
Change detection based on images is also a binary classification problem ($0$: ``no difference'', $1$: ``change/addition''). Class $0$ is also called background. Class $1$ contains all connected pixels which represent a change. We call these image objects also difference objects since they contain all pixels which visualize a graph difference. In the current body of work, there are traditional, i.e., non-machine-learning-based, approaches (cf. e.g., \cite{Avanaki2009,Wang2004}) and learning based approaches (cf. e.g., \cite{Ji2019,Koch2015}). As already explained at the beginning of this Section, learning human detected differences requires a learning based approach for the differentiation of which of the differences are those which humans would detect. Consequently, we focus here on learning based approaches.

\textbf{Change Detection and Object Detection Approaches.} The objective of change detection can be completed by actual change detection algorithms and also object detection algorithms.

\emph{\textbf{Change Detection}} has the objective to detect differences in a set of given images \cite{1395984,Ji2019}. These differences are represented as a set of pixels which are significantly different when comparing the images of the given set. These significantly different pixels are called change masks. Change detection algorithms can be categorized into \cite{HUSSAIN201391}:
\begin{enumerate}
	\item pixel-based approaches
	\item object-based approaches
\end{enumerate}
Hussain et al. \cite{HUSSAIN201391} provide an encompassing overview of both pixel- and object-based approaches. There are learning-based algorithms for both categories. They employ i.a. support vector machines, neural networks, or decision trees to learn the change masks for the set of input images based on supervised feedback which is a target change mask.\\
The object-based approaches detect regions of connected pixel which are denoted as objects by comparing ``spectral information, geometric properties, or [...] semantic features'' \cite{Ji2019}. Object-based change detection has two sub-categories:
\begin{enumerate}
	\item object comparison
	\item object stacking
\end{enumerate}
Object comparison approach firstly extract connected pixel regions from all images via, for instance, segmentation. Before the extracted objects can be compared several preprocessing steps are necessary -- i.a. calculation of the segmentation masks and alignment of the segmented objects. Then, the segmented regions are compared and the decision is made whether they are different or not. The comparison step, however, leads to decision on all present differences which does not make it possible to differentiate between actually existing differences and those humans would spot. Consequently, object comparison approaches are not applicable for our objective.\\
Object stacking approaches are usually applied for dynamic data \cite{1395984,Ji2019}. As explained in Section~\ref{sub:definitions_and_research_context} -- \texttt{Change Detection Task}, work dealing with dynamic graphs is transferable into the domain of our research question since one can always make one graph of the time series be the base graph and another one be the alternative. Consequently, an object stacking approach might be a suitable learning algorithm for our objective. This type of learning approach predicts changes based on a dynamic set of images via ``stacking bi-temporal images along the RBG-channel \cite{AlenaMA}'' \cite{HUSSAIN201391,1395984,Ji2019}. This means: The algorithms stacks all images together to form a unit. Then, objects are extracted for the entire unit. By stacking the images are aligned and thus their geometric properties coincide. For learning change objects -- here: visualized graph changes -- this means that the algorithm learns change objects which spatially coincide based on the objects' consistent properties like location, shape, or size. Object stacking has the advantages, compared to object comparison, that it only needs one segmentation mask and that it provides the possibility of a binary change/no-change classification \cite{HUSSAIN201391}. The classifier learns  based on geometrical features \cite{HUSSAIN201391}. So, the learning algorithm learns based on features which are also used by humans for visual comparisons with respect to differences \cite{DiagramsDifferencePerception,10.1145/3335082.3335083}. So, object-based change detection approaches are suitable for learning human detected differences. The learning algorithm class which works with stacked images are instance segmentation and classification.\\

\emph{\textbf{Object Detection.}} The objective of object detection is to recognize and localize specific classes of object in images \cite{Agarwal2018,Zou2019}. Classification is responsible for the recognition part and bounding box prediction provides the localization. Object detection is a well-known research field in computer vision. Zou et al. \cite{Zou2019} give an encompassing overview over the developments in object-detection in the past $20$ years. Neural networks are usually the learning algorithms of choice \cite{Zou2019}. More specifically, neural networks are the learning algorithm of choice when it comes to image-based object-detection. Agarwal et al. \cite{Agarwal2018} provide a survey on this type of learning algorithm. Object detection and visual comparison with respect to differences have the same objective -- they aim to detect, recognize and locate, specific visual objects. Object detection is usually a supervised learning problem which allows to include data reflecting human behavior, notions, or perception \cite{Agarwal2018}.\\
 ``Spot the difference'' is commonly known as a children's game. Wu et al. \cite{SpotDifference} propose an object-detection-based learning approach for playing ``spot the difference''. They create their training and test dataset based on $6$-channel images which result from stacking  RBG-image pairs. To train Faster R-CNN in a supervised manner they pre-labeled their training dataset -- $0$: ``different'', $1$: ``same''. Then, they train a Faster R-CNN \cite{7485869} to solve this binary classification problem. Faster R-CNN \cite{7485869} is the first end-to-end trained object detector which runs near to real-time \cite{Zou2019}. Faster R-CNN employs region proposals to predict the bounding boxes of an object to be detected. The region proposals are provided by a so called region proposal network (RPN). They are candidate boxes which are likely to be the bounding box of the object to be detected. The detection network -- a second neural network -- takes the candidate boxes and evaluates them to find the best bounding box. It is possible to use the bounding boxes for the injection of the human notion on differences (cf. Section~\ref{sub:dataset_creation_algorithm_for_enriching_dag_data_with_human_like_detected_graph_differences}).  Wu et al. \cite{SpotDifference} use end-to-end training. It has the advantage that the RPN's parameters can be updated with detection information from the detection network throughout the backpropagation step \cite{Agarwal2018}. This leads to an increase in quality of the RPN \cite{Agarwal2018}. ``Spot the difference'' and visual comparison with respect to differences have the same objective -- they aim to detect, recognize and locate, specific visual objects. We interpret the differences of directed acyclic graphs as visual objects since our research is based on humans processing a node-link visualization of two directed acyclic graphs. Furthermore, both works are based on a pair of two images while their set of comparison items differ. ``Spot the difference'' generally aims to detect different classes of objects -- e.g., animals, humans, or cars -- and visual comparison with respect to differences wants to find out which nodes and/or edges are different between the base and the alternative graph. However, the precision of object detection is too coarse for the detectors of differences in visualized directed acyclic graphs. Object detection detects objects based on their bounding boxes. Bounding boxes estimate the pixel minimum and maximum of the object. Graph differences, however, are rather small (nodes) and narrow (edges) objects. The combination of these two facts leads to bounding boxes being to large for a unique identification of just one change object. It most likely can happen that other graph structure parts or, even worst, other difference are (partially) inside the bounding box. Consequently, purely bound-box-based methods are not suitable. Instance segmentation -- an extension of object detection which learns the pixels belonging to the change object \cite{Agarwal2018} -- mitigate this drawback and are thus suitable.

\emph{\textbf{Instance Segmentation.}} The extension of instance segmentation to pure object detection is the additional prediction of segmentation masks\footnote{Following the terminology He et al. \cite{He_2017_ICCV}: Object detection prediction results are bounding boxes not masks. Semantic segmentation results in pixel-wise differences and instance segmentation in object-wise differences.} Instance segmentation is a combination of semantic segmentation and object-based instance detection \cite{Garcia-Garcia2018}. Semantic segmentation is a pixel-wise class segmentation while instance segmentation does the classification for connected groups of pixels, also called object \cite{Garcia-Garcia2018}. Garcia-Garcia et al. \cite{Garcia-Garcia2018} provide an overview over instance segmentation approaches. Since the domain is dynamic, the Garcia-Garcia et al.'s survey does not contain all recent approaches to instance segmentation.\\
Early approaches employ segmentation proposals. Pinheiro \cite{Pinheiro2015} and Hariharan et al. \cite{10.1007/978-3-319-10584-0_20} learn segmentations which they then classify. While the former uses an end-to-end trained neural network, the latter uses a support vector machine for classification. This approach is comparable to the preprocessing step of change detection. Consequently, it share the limitations of change detection's preprocessing. It is slow and inaccurate \cite{He_2017_ICCV}. SharpMask \cite{Pinheiro2015} is an improvement of DeepMask. It increases speed and accuracy via combining low- and high-level features  stemming from shallow resp. deep layers.\\
Current approaches rather use neural networks -- i.a. Mask R-CNN \cite{He_2017_ICCV}, U-Net \cite{10.1007/978-3-319-24574-4_28}, and FC-Siam-diff \cite{CayeDaudt2018}. Caye Daudt et al. \cite{CayeDaudt2018}, with their FC-Siam-diff architecture, propose to use a fully convolutional network architecture. This architecture combines a fully convolutional encoder-decoder structure with siamese networks. Furthermore it uses skip connections as U-Net does it \cite{10.1007/978-3-319-24574-4_28}. The siamese network processes the image pairs in parallel and the skip connections improve the spacial pixel accuracy via the combination of low- and high-level features. All of this, however, is being done pixel- and not object-wise which is a drawback due to the afore explained reasons. Ji et al. \cite{Ji2019} detect changes on buildings by extracting the buildings with Mask R-CNN and detecting the changes with U-Net.\\
Mask R-CNN is the improvement of Faster R-CNN which was proposed by Wu et al. \cite{SpotDifference} to solve the task of object detection in the context of the popular children's game ``Spot the difference''. While Faster R-CNN only predicts bounding boxes and classifies the detected objects, Mask R-CNN predicts segmentations mask in addition to its predictions of bounding boxes and classification results. Currently, due to its outstanding training time and performance \cite{Ji2019,Agarwal2018,Zou2019}, Mask R-CNN is the state of the art technology for instance segmentation. Unlike other networks such as DeepMask \cite{Pinheiro2015}, it does separately segment and classify the instances. Mask R-CNN does it together.
The fine-grained nature of Mask R-CNN's segmentation masks solve the issue of object detection algorithms' bounding boxes being too coarse for the small (nodes) and narrow (edges) graph elements of directed acyclic graphs visualized as node-link diagrams. Furthermore, these segmentation masks allow us to inject the human notion of differences into the network. The work of Radke et al. \cite{1395984} confirms this. Given the afore discussed advantages of instance segmentation and that the work of Wöhler et al. \cite{wohler2019learning} further substantiates the suitability of instance segmentation for learning human notions, we proceed with instance segmentation with Mask R-CNN as our learning approach.

\emph{\textbf{Our Learning Approach Decision.}} The decision for Mask R-CNN means a decision for image-based training. While this is a decision for not the most efficient approach with respect to training time, it is a decision for the approach which is closest to how humans do their comparison -- i.e., humans work with the visualized directed acyclic graphs when they visually compare them. Consequently, it is necessary to learn on the images, since the visualization resp. the visualization design choices are known to influence human processing \cite{Zhang2017}. In case, one would like to have a certain independence of the visual design choices, one could follow the approach of Wöhler et al. \cite{wohler2019learning} who combined an image- and a structural-data-based network. Consequently, they were were able to train on both and achieved a certain independence of the visualization. This is also feasible for us in the future. At the moment, we refrained from doing this since we identified our difference factors based on the standard design -- circular nodes and arrow-headed edges -- for node-link diagrams and have no information how the factors change for varying node-link diagram designs.

\subsection{A Convolutional Neural Network for Learning Human Detected Structural Differences in Directed Acyclic Graphs} 
\label{sub:a_convolutional_neural_network_for_learning_human_detected_differences_in_dags}
In this Section, we first explain our training strategy and performance evaluation metric for our convolutional neural network based on Mask R-CNN as both set the framework for our network's performance. Then, we explain the general architecture of Mask R-CNN in more detail as this information is relevant for understanding our adaptions of Mask R-CNN to make it applicable to our problem domain -- learning human detected structural differences in directed acyclic graphs. As basis implementation, we used Mask R-CNN's torchvision implementation since it is an optimized version and developed for research \cite{Marcel2010}. Further, we used PyTorch Version 1.4.0 and CUDA 10.1. Our adaptions encompass minor architectural changes and major changes for the data and transfer learning, and also changes of the hyper-parameter optimization. Finally, we discuss our performance evaluation results.

 \subsubsection{Performance Evaluation Metrics} 
 \label{ssub:performance_evaluation_metrics}
 The evaluation metrics for Mask R-CNN are influenced by object detection since Mask R-CNN is an instance segmentation approach which builds upon object detection. Furthermore, we only consider metrics for object-based approaches since Mask R-CNN is an object-based one.

 Commonly, the mean average precision (mAP) is used to evaluate object detection. Mask R-CNN extends this metric to the mask mAP \cite{Agarwal2018,Zou2019,Henderson2016}. Further frequently used metrics are precision, recall, average precision (AP) and the F1 score \cite{Ji2019,Liu2019b,Zhu2018}. The mAP is calculated as the mean of the classes' average precision (AP). The AP of the individual classes is the area under the precision/recall (PR) curve. The PR-curve related correct predictions (true positives) and false predictions (false positives/negatives) by calculating the precision and the recall. Object detection and instance segmentation usually consider predictions as correct if the overlap with the target by an intersection over union (IoU) threshold of $>0.5$ (cf. i.a. \cite{Ji2019,Agarwal2018,He_2017_ICCV}). For details on the definition please refer to Figure~\ref{fig:IoU} and Section \texttt{Region Proposal Network (RPN)} -- \texttt{Intersection over Union (IoU)}.

\paragraph{Precision} is a measure of prediction accuracy as it calculates the ratio of correct predictions (true positives) to all predictions (true and false positives) (cf. Equation~\eqref{eq:Precision}). For learning to predict differences humans would detect, this is the ratio of the predicted human detected differences to all predicted differences. Predicted human detected differences are the true positives while the rest of the predicted differences count as false negatives. A high precision correlates with the prediction of only human detected differences and no other differences -- e.g., GT differences. We also set the IoU threshold to $>0.5$.
 \begin{eqfloat}[h]
 	\begin{equation}
 		\begin{aligned}
 			precision = \frac{TP}{TP+FP}\\
 			TP = \text{ true positives}\\
 			FP = \text{ false positives}
 		\end{aligned}
 	\end{equation}
 	\caption{Performance evaluation -- precision formula.}
	 	\label{eq:Precision}
 \end{eqfloat}
 \myequations{Performance evaluation -- precision formula}

\paragraph{Recall} is a measure of many target objects are detected \cite{Teufel}. It is the ratio of detected targets to the number of all targets (cf. Equation~\eqref{eq:recall}). For predicting human detected differences, it is the ratio of the predicted human detected differences to those human detected differences not predicted by the network. A high recall is an indication for the detection of all target objects.
 \begin{eqfloat}[h]
 	\begin{equation}
 		\begin{aligned}
 			recall = \frac{TP}{TP+FN}\\
 			TP = \text{ true positives}\\
 			FN = \text{ false negatives}
 		\end{aligned}
 	\end{equation}
 	\caption{Performance evaluation -- recall formula.}
	 	\label{eq:recall}
 \end{eqfloat}
 \myequations{Performance evaluation -- recall formula}

\paragraph{F1 Score.} The network's recall could be improved by increasing the number of returned predictions as the predictions are the sum of true and false positives. However, this bears the risk of a decrease in precision since it is defined as the ratio of true positives to all predictions. This phenomenon is called the precision-recall tradeoff. A metric for the balance of this tradeoff is the F1 score (cf. Equation~\eqref{eq:F1Score}) \cite{Teufel}.
 \begin{eqfloat}[h]
 	\begin{equation}
 		\begin{aligned}
 			F1=\frac{1}{2}(\frac{1}{precision}+\frac{1}{recall})
 		\end{aligned}
 	\end{equation}
 	\caption{Performance evaluation -- F1 score formula.}
	 	\label{eq:F1Score}
 \end{eqfloat}
 \myequations{Performance evaluation -- F1 score formula}

\paragraph{PR-Curve.} The PR-curve is the plot of the ranked precision-recall pairs for varying probability thresholds with precision being on the y-axis and and recall being on the x-axis. The AP for each class is represented by the area under the PR-curve. Since the precision and recall values are between zero and one, the AP is also between zero and one. For the AP, we still have an IoU threshold of $0.5$ (AP@50).

\textbf{Our Performance Evaluation Metrics Decision.} Since our work is located in the research field of change detection, we use metrics which are similar to those of other change detection experiments in the machine learning domain. Herewith, we want to foster comparability. An example is the work of Ji et al \cite{Ji2019}. They used also an IoU threshold of $>0.5$. Further, they used AP, precision, recall, and the F1 score. As we use AP with an IoU threshold of $>0.5$, precision, recall, and the F1 score, our results are comparable with other work from the research fields of object and change detection and instance segmentation. Comparability to these research fields is suitable since, as already mentioned before, our objective is related to the objective of these research fields. Since our results for segmentation masks and bounding boxes differ only marginally ($+/-0.01$), we only provide the metric results calculated for our segmentation masks. The masks were chosen because our goal is to predict the segmentation masks and not the boxes.

\subsubsection{Model Training for Instance Segmentation} 
\label{ssub:model_training_for_instance_segmentation}
Training Mask R-CNN means feeding an image into the network. After processing the image, Mask R-CNN's results are a bounding box, a label, a classification score, and a segmentation mask. For the evaluation of Mask R-CNN's training performance these results are compared with the target which is fed into the network as supervised feedback. The target also consists of a bounding box, a label, a classification scare, and a segmentation mask for the instance to be segmented. The comparison of target and training result works via the calculation of the network loss. Hyperparameters determine the loss' convergence \cite{Agarwal2018}. Furthermore, there is the possibility to improve Mask R-CNN's performance by applying transfer learning or data augmentation.

\paragraph{Loss.} 
\label{par:loss}

The loss computation compares the training result to the target. Since the objective of instance segmentation consists of the classification of the object to segment, the localization of this object, and its segmentation, the loss of this training objective is a so called multi-task loss \cite{He_2017_ICCV}. Multi-task losses are combined losses -- i.e., the (weighted) sum of the individual losses. In the case of instance segmentation, the loss consists of a loss for the identification of the object to be segmented (classification), for the predicted bounding box (regression), and for the predicted segmentation mask. Formally, this is the sum of the normalized classification loss $L_{cls}$, the normalized regression loss $L_{box}$, and the normalized segmentation mask loss $L_{mask}$ (cf. Equation~\eqref{eq:MultiTaskLoss}).
\begin{eqfloat}[H]
	\begin{equation}
		\begin{aligned}
			L=\frac{1}{N_{cls}}*L_{cls}+\frac{1}{N_{box}}*L_{box}+L_{mask}\\
			\\
			N_{cls}=\text{ Mini batch size -- here: }2\\
			N_{box}=\text{ Number of predicted boxes}\\
			\\
			L_{cls}=\text{ Binary cross entropy loss (cf. Equation~\eqref{eq:Cls_Loss})}\\
			L_{box}=\text{ $smooth_{L_1}$ loss as defined in \cite{Girshick2015} (cf. Equation~\eqref{eq:Box_Loss})}\\
			L_{mask}=\text{ Pixel-wise cross entropy loss}\\
		\end{aligned}
	\end{equation}
	\caption{Instance segmentation -- multi-task loss. Formally, this is the sum of the normalized classification loss $L_{cls}$, the normalized regression loss $L_{box}$, and the normalized segmentation mask loss $L_{mask}$.}
		\label{eq:MultiTaskLoss}
\end{eqfloat}
\myequations{Instance segmentation -- multi-task loss}
Like Fast R-CNN \cite{Girshick2015} and Faster R-CNN \cite{He_2017_ICCV}, the architectures Mask R-CNN is based on, Mask R-CNN uses a weighting factor of $1$ for the individual losses. We follow this approach. However, we also tested other weights for the individual losses to uncover the full picture. Our results showed that weights $\neq1$ are worse. The model tends to overfit with such weights.

The \textbf{classification loss} $L_{cls}$ quantifies how well the objects to be segmented are identified. Via the binary cross entropy loss (cf. Equation~\eqref{eq:Cls_Loss}), or log loss \cite{Girshick2015}, the network learns to reduce the uncertainty in the class prediction outcome by minimizing the distance to the target class label $y$.
\begin{eqfloat}[H]
	\begin{equation}
		\begin{aligned}
		    L_{cls}(p, y) = -\sum_{k=0}^{K}y_k*log(p_k)\\
			\\
			y_k = \text{$k^{th}$ target class label}\\
			p_k = \text{ probability of the $k^{th}$ class}\\
			K = \text{ Number of classes}
		\end{aligned}
	\end{equation}
	\caption{Instance segmentation -- classification loss $L_{cls}$. Via the binary cross entropy loss, or log loss \cite{Girshick2015}, the network learns to reduce the uncertainty in the class prediction outcome by minimizing the distance to the target class label $y$.}
		\label{eq:Cls_Loss}
\end{eqfloat}
\myequations{Instance segmentation -- classification loss $L_{cls}$}

The \textbf{regression box loss} calculates by how much the coordinates of the predicted bounding boxes diverge from the coordinates of the target bounding box coordinates (cf. Equation~\eqref{eq:Box_Loss}). Here, the training objective is to minimize this divergence. $L_{box}$ is the $smooth_{L_1}$ loss as defined in \cite{Girshick2015}. The $smooth_{L_1}$ loss is a type of $L_1$ loss and thus not so sensitive to outliers as the $L_2$ loss \cite{Girshick2015}. Each target bounding box $t^t$ is labeled with the respective target class label $y$. $L_{box}$ is only calculated for target bounding boxes labeled with $y=1\text{ = ``difference''}$.  The predicted bounding boxes are $t^b = [t_x^b, t_y^b, t_w^b, t_h^b]$ with $x,y$ being the coordinates of the box's center and $w,h$ being the box's width and height.
\begin{eqfloat}[H]
	\begin{equation}
		\begin{aligned}
			L_{box}(t^b, t^{t}, y) = \sum_{i\in\{x,y,w,h\}}^{ } y * smooth_{L_1}(t_i^b - t_i^t)\\
			\\
		    smooth_{L_1}(x) = \begin{cases}0.5x^2 & if |x| < 1\\
		    |x| - 0.5 & otherwise \\
		    \end{cases}\\
			\\
			t^b = \text{ predicted bounding box}\\
			t^t = \text{ target bounding box}\\
			y = \text{ target class label}\\
			x,y = \text{ coordinates of the box's center}\\
			w,h = \text{ being the box's width and height}
		\end{aligned}
	\end{equation}
	\caption{Instance segmentation -- classification loss $L_{box}$. The regression box loss calculates by how much the coordinates of the predicted bounding boxes diverge from the coordinates of the target bounding box coordinates. Here, the training objective is to minimize this divergence. $L_{box}$ is the $smooth_{L_1}$ loss as defined in \cite{Girshick2015}.}
		\label{eq:Box_Loss}
\end{eqfloat}
\myequations{Instance segmentation -- classification loss $L_{box}$}

The \textbf{mask loss} $L_{mask}$ determines how well the pixels of the predicted segmentation mask $m$ match the pixels of the target mask $m^{t}$. Consequently, $L_{mask}$ is a classification loss like $L_{cls}$ -- just on pixels. Since segmentation masks are pixel based, $L_{mask}$ is defined as the pixel-wise cross entropy loss of all predicted masks \cite{He_2017_ICCV} (cf. Equation~\eqref{eq:Mask_Loss}).
\begin{eqfloat}[H]
	\begin{equation}
		\begin{aligned}
			L_{mask}(m,m^{t}) = \sum_{k = 0}^{K} pixelLoss(m,m^{t})\\
			\\
			pixelLoss(m,m^{t}) = \sum_{i = 0}^{M} \sum_{j = 0}^{M} -m[i,j]*log(m^{t}[i,j])-(1-m[i,j])*log(1-m^{t}[i,j])\\
			\\
			m = \text{ predicted segmentation mask}\\
			m^{t} = \text{ target segmentation mask}\\
			K = \text{ number of classes -- here two: }\\
			1 = \text{ segmentation mask pixel}; 0 = \text{ pixel not belonging to segmentation mask}\\
			M = \text{ size of the segmentation masks}
		\end{aligned}
	\end{equation}
	\caption{Instance segmentation -- mask loss $L_{mask}$. The mask loss $L_{mask}$ determines how well the pixels of the predicted segmentation mask $m$ match the pixels of the target mask $m^{t}$.}
		\label{eq:Mask_Loss}
\end{eqfloat}
\myequations{Instance segmentation -- mask loss $L_{mask}$}

\textbf{Loss Specifics for End-to-End Training.} In case of end-to-end training, which is especially beneficial for region proposals of object detectors, the regional proposal network's losses become part of the multi-task loss (cf. Equation~\eqref{eq:MultiTaskLoss} and~\eqref{eq:MultiTaskLoss_EndToEnd}).
\begin{eqfloat}[H]
	\begin{equation}
		\begin{aligned}
			L=\frac{1}{N_{cls}}*L_{cls}+\frac{1}{N_{box}}*L_{box}+L_{mask}+(L_{rpn\_cls}+L_{rpn\_box})\\
			\\
			N_{cls}=\text{ Mini batch size -- here: }2\\
			N_{box}=\text{ Number of predicted boxes}\\
			\\
			L_{cls}=\text{ Binary cross entropy loss (cf. Equation~\eqref{eq:Cls_Loss})}\\
			L_{box}=\text{ $smooth_{L_1}$ loss as defined in \cite{Girshick2015} (cf. Equation~\eqref{eq:Box_Loss})}\\
			L_{mask}=\text{ Pixel-wise cross entropy loss}\\
			L_{rpn\_cls}=\text{  Binary cross entropy loss of the regional proposal network -- }\\
			\text{calculated as shown in Equation~\eqref{eq:Cls_Loss}}\\
			L_{rpn\_box}=\text{ $smooth_{L_1}$ loss of the regional proposal network -- }\\
			\text{calculated as shown in Equation~\eqref{eq:Box_Loss}}
		\end{aligned}
	\end{equation}
	\caption{Instance segmentation -- multi-task loss for end-to-end training.}
		\label{eq:MultiTaskLoss_EndToEnd}
\end{eqfloat}
\myequations{Instance segmentation -- multi-task loss for end-to-end training}
{\scshape Our Model Training Decision.} Due to the beneficial effects of end-to-end training on object detectors, we train our Mask R-CNN architecture end-to-end. Consequently, our multi-task loss function is the same as Equation~\eqref{eq:MultiTaskLoss_EndToEnd}.

\paragraph{Hyperparameters.} 
\label{par:hyperparameter}
Object detection and, consequently, instance segmentation are highly-convex optimization problems which include thousands of parameters \cite{Agarwal2018}. In the current body of work, there are numerous options of optimizers for neural networks, for instance, the Adam optimizer or RMSProp. The Mask R-CNN architecture employs a stochastic gradient descent (SGD) optimizer \cite{Agarwal2018}. This optimizer type uses information of previous gradients to achieve an oscillation reduction when being close to the minimum \cite{Agarwal2018}. Furthermore, it is one of the standard optimizers in the machine learning domain \cite{bishop2006pattern}. Certainly there are more advanced optimizers proposed by the current body of work. But still, object detectors commonly employ plain stochastic gradient descent optimizers ``without putting much thought into it '' \cite{Agarwal2018}.\\
\textbf{Our Model Training Decision.} We follow this common direction of the object detection resp. instance segmentation research field (cf. i.a. \cite{7485869,He_2017_ICCV,Girshick2014}) and thus also employ SGD. Another advantage of employing SGD as an optimizer is that our results become comparable to existing work.

Of the thousands of parameters to be optimized for loss convergence, learning rate and batch size are considered the most important for object detection.

\textbf{Learning Rate.} The learning rate is the factor which determines the amount by how much the gradient is updated during optimization \cite{bishop2006pattern}. According to Agarwal et al. \cite{Agarwal2018}, in their survey, there is no concrete procedure which will lead to the optimal or at least good learning rate. The quality pf the learning rate depends on different factors -- i.a. the employed optimizer, he model's architecture, and the batch size. So, for now, there is no other option than a trial-and-error procedure. Machine learners try different learning rates, observe the respective loss curves, and take the learning rate with the lowest loss. In case, they would like to use a learning rate from literature, the training procedures need to be comparable.

A learning rate too small is likely to result in a long training that could get stuck, whereas too large one is likely to result in learning sub-optimal weights too fast or in an unstable training. A common counter measure for these issues is learning rate annealing \cite{Koch2015,Zeiler2012}. This strategy describes the adaption of the learning rate after a fixed number of epochs or after the validation loss has not changed after a fixed number of epochs (plateau annealing). After $120.000$ epochs, Mask R-CNN reduces the learning rate by $10$ starting off with an initial rate of $0.02$.

The learning rate also depends on the transfer learning approach (cf. Section \texttt{Transfer Learning}). If the network learns from scratch the learning rate is higher due to the random initialization of the model parameters. If the model is fine-tuning a pre-trained model, a smaller learning rate is advisable since the parameters are already pre-trained with respect to the optimization objective -- at least to some degree \cite{Garcia-Garcia2018}. Consequently machine learner commonly use a learning rate of $0.001$ for fine-tuning (cf. i.a. \cite{Girshick2014}).\\
\emph{\textbf{Our Model Training Decision.}} As we will also use fine-tuning\footnote{ for details on the rationals please refer to the Sections \texttt{Transfer Learning} and \texttt{MS COCO Dataset}} -- i.e., we will have a comparable training procedure -- we use a learning rate of $0.001$.
	%

\textbf{Batch Size.} The batch size determines how many tensors are batched during training. Batching tensors means training multiple data items together by concatenating them. Larger batch sizes reduce the training time while extra large batch sizes like $8192$ also have a beneficial effect on the performance \cite{Agarwal2018}. Such large batch sizes, however, are not feasible for instance segmentation since instance segmentation works on the entire image and so on large tensors \cite{Agarwal2018}. Moreover, Mask R-CNN's additional calculations such as the anchors raise the amount of needed memory. So, it is likely to run in out-of-memory errors for large batch sizes. This leads to the object detection and instance segmentation community training with small batch sizes \cite{Agarwal2018}. Small means here a value range of $1$ to $16$ $3$-channel images. Mask R-CNN trains two images per GPU and it uses eight GPUs in total which results in a mini match size of $16$ \cite{He_2017_ICCV}.\\
\emph{\textbf{Our Model Training Decision.}} To identify our mini batch size, we stepwise increased it and set it to the maximum not leading to out-of-memory errors. This procedure resulted in a batch size of $4$ images per GPU. As GPU we used a  NVIDIA GTX 1660. This results in a tensor size of $[4,800,800,6]$.

\paragraph{Transfer Learning.} 
\label{par:transfer_learning}
All state of the art instance segmentation architectures commonly use transfer learning, also called pre-training \cite{Ji2019,Agarwal2018,7485869,Garcia-Garcia2018,He_2017_ICCV,Girshick2015}. Transfer learning became popular since it reduces training time and increases loss convergence. It builds upon the idea to use an existing model which is trained based on a dataset similar to the own for the initialization of the parameters \cite{Agarwal2018}. Instance segmentation models are typically pre-trained on the MS COCO dataset (cf. i.a. \cite{He_2017_ICCV,Lin2014}). The COCO dataset was created specifically for object segmentation. For details on the COCO dataset please refer to the following Section -- Section \texttt{MS COCO Dataset}.

Transfer learning can be categorized into:
\begin{itemize}
	\item no transfer -- i.e., learning from scratch
	\item transfer learning -- i.e., fine-tuning the parameters of an already trained network
	\item no learning -- i.e., feature extraction
\end{itemize}

\textbf{Learning from scratch} requires random initialization of the networks parameters \cite{He_2017_ICCV}. This is not circumventable in case the dataset to train the network on and the pre-trained dataset differ too much \cite{Agarwal2018}. Learning from scratch might even be preferable when the dataset to be trained is large enough \cite{Agarwal2018}. However, usually a pre-trained model should be preferred in contrast to random initialization \cite{Agarwal2018}.

\textbf{Transfer learning} or \textbf{fine-tuning} uses the parameter of a pre-trained network as initialization -- i.e., the start of the training. Typically, during transfer learning, the network parameters are updated with a smaller learning rate since the pre-training coarsely updates the parameter and transfer learning updates them on a detailed level. It is possible to update all or only a subset of the network's weights. This depends on the learning objective. For instance, R. Girshick \cite{Girshick2015} found for the Fast R-CNN architecture that the first layer of the neural network is data and task independent, however, the remainder of Fast R-CNN's convolutional backbone requires fine-tuning for the specific task's objective. Similarity, Appalaraju et al. \cite{Appalaraju2017} only had to fine-tune the last two convolutional layers for learning the similarity of images. Faster R-CNN \cite{7485869} and Mask R-CNN \cite{He_2017_ICCV} start their fine-tuning at the fourth convolutional layer. The necessity to only fine-tune the last, or deeper, layers is explainable by: The deeper layers hold high-level information. They learn this high-level information by aggregating information from shallow layers. Fine-tuning only deeper layers uses the same information as the pre-trained network from shallow layers, however, creates different high-level features from this low-level information. These different high-level features lead to an adaption of the model to different objects and datasets.\\
\emph{\textbf{Domain-specific fine-tuning}} deals with the fine-tuning of a network on domain data \cite{Girshick2014}. Ji et al. \cite{Ji2019} fine-tuned Mask R-CNN on the SC-2016 dataset which contains buildings in order to learn  the prediction of changes in buildings.

\textbf{Feature extraction} is also based on a pre-trained model. But, in contrast to transfer learning, it keeps most parameters fixed. Throughout the training, the gradients of these fixed parameters are not calculated and consequently the parameters are not learned. Instead, the features of the pre-training are used. This is beneficial for training scenarios with small datasets which are comparable to the dataset used for pre-training. Additionally, feature extraction can save training time.

\textbf{Our Model Training Decision.} We will use transfer learning since our directed acyclic graphs and our task of difference detection has domain specific properties -- e.g., the influence of the influence factors -- which cannot be learned from other datasets. However, due to the shapes of our graph elements the COCO dataset is suitable for pre-training -- for details cf. Section \texttt{MS COCO Dataset}.

\paragraph{MS COCO Dataset -- Pre-Training Dataset.} 
\label{par:ms_coco_dataset_pre_training_dataset}
The COCO dataset is one of the largest datasets for instance segmentation \cite{Lin2014}. Further, it is publicly available and machine learners commonly employ it for pre-training. It contains $328.000$ $256x256$ images with $>2.500.000$ labeled instances of $80$ categories. Each category contains $>10.000$ labeled instances. The images are daily life images which depict, for instance, cats, dogs, or a soccer match.

Since the images of the COCO dataset are daily life images, their features differ from our directed acyclic graph images. The factor of \texttt{shape} which is the most dominant influence factor sets this into perspective. Our dataset contains nodes and edges which we visualizes as circles (nodes) and lines with arrowheads at the end (edges). These shapes also appear in the COCO dataset. Circles, for instance, appear in images containing a ball and clear edges in images with knifes. If we consider the shape of the entire directed acyclic graph -- nodes and edges combined -- there are manifold shape options which, e.g., resemble a house.\\
\textbf{Our Model Training Decision.} So, a dataset like the MS COCO dataset is a suitable pre-training dataset for our objective of learning human detected structural differences in directed acyclic graphs.

\paragraph{Training Dataset Size.} 
\label{par:training_dataset_size}
\begin{table}[tb]
    \centering
    \begin{tabular}{|c|c|c|c|c|}
		\hline
        Training Size &  Precision & Recall & F1 Score  & AP\\
        \hline
        $200$ & $0.639$ & $0.418$ & $0.921$ & $0.575$ \\
        $1000$ & $0.765$ & $0.620$ & $0.894$ & $0.732$ \\
        $2000$ & $0.873$ & $0.677$ & $0.93$1 & $0.784$ \\
        $3000$ & $0.899$ & $0.738$ & $0.937$ & $0.826$ \\
		\hline
    \end{tabular}
    \caption[Experiment and analysis -- varying training dataset sizes]{Experiment and analysis -- varying training dataset sizes: small training datasets already show relatively good performance. Increasing dataset size improves the performance metrics. (Table taken from \cite{AlenaMA})}
    \label{tab:trainingSizeEval}
\end{table}

\begin{figure}[tb]
    \centering
    \includegraphics[scale=0.35]{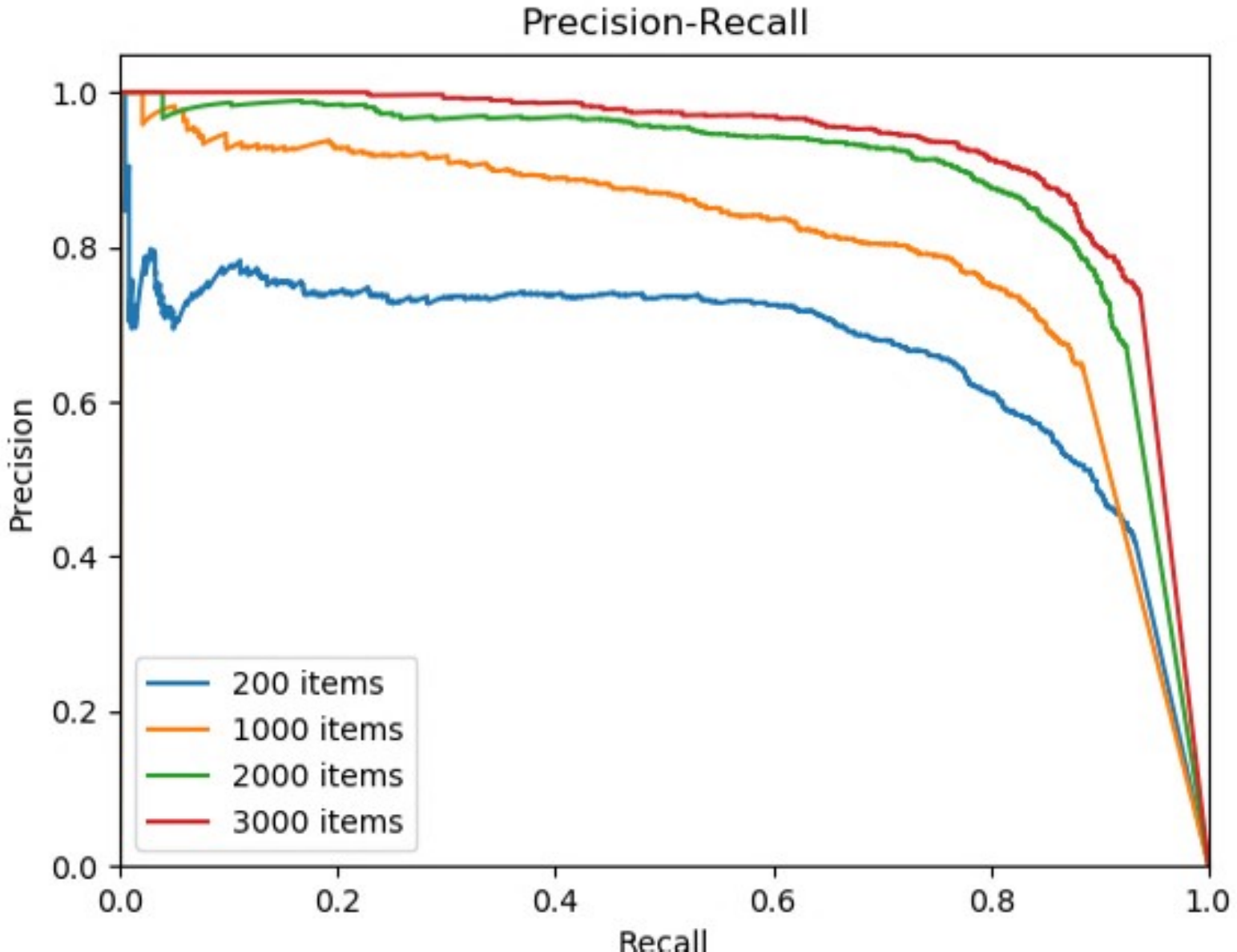}
    \caption[PR-curves for varying training dataset sizes]{PR-curves for varying training dataset sizes -- a larger training dataset size improves the PR-curve and thus the AP (average precision) by increasing both precision and recall. (Figure taken from \cite{AlenaMA})}
    \label{fig:PRTraining}
\end{figure}

The decision of how many data items of the entire dataset -- in our case $10.000$ annotated directed acyclic graph pairs (cf. Section~\ref{sub:a_dataset_for_learning_human_detected_differences_in_dags_based_on_graph_data_enriched_with_human_inspired_detected_graph_difference}) -- to use for training the machine learning architecture is challenging like the decision for hyperparameters like the learning rate. As for the learning rate, there is no concrete procedure for determining the training dataset size. While the current body of work from the early 90s found that the training objective influences the sample size \cite{Patuwo1993,Raudys1991}, there is the rule-of-thumb amongst image classification practitioners that $1.000$ data items are sufficient for image classification objectives \cite{Warden2017}. In case of the application of data augmentation the overall training dataset size is reduced since the affine transformation of data augmentation result in additional data items and thus additional data diversity. The benchmark dataset ImageNet \cite{10.1145/3065386} is the source of this rule-of-thumb. ImageNet contains daily life pictures and $1.000$ instances per class. These images depict multiple shapes, colors, sizes, or occlusions and are thus more complex than our images. Our images only contain white background pixels, black pixels for the depiction of the graph elements -- nodes and edges -- and blue pixels, in case of the difference images, to depict the differences. As our images are less complex, we expect that we will achieve already relatively good results with less than $2.000$ images.

\textbf{Experiment \& Analysis.} Table~\ref{tab:trainingSizeEval} shows our experiment results. With a training dataset size of $200$ we already achieve an F1 score of $0.921$ and an AP of $0.575$. An increased number of data items positively affects especially the precision, recall, and AP (average precision) of our network. The AP for a training dataset size of $2.000$ is higher due to both an increased precision and an increased. The PR-curve in Figure~\ref{fig:PRTraining} shows this since the \PRGreencolor{green} PR-curve is higher than the \PRBluecolor{blue} and the \PRYellowcolor{yellow} one.\\
\emph{\textbf{Our Model Training Decision.}} As an increase to a training dataset size of $3.000$ rather marginally improves our performance metrics (cf. Figure~\ref{fig:PRTraining} -- \texttt{\PRRedcolor{red} PR-curve}), we choose a training dataset size of $2.000$.
\FloatBarrier

\paragraph{Data Augmentation.} 
\label{par:data_augmentation}
\begin{table}[tb]
    \centering
    \begin{tabular}{|c|c|c|c|c|c|}
		\hline
        directed acyclic graph Type & Augmentation -- yes/no? &  Precision & Recall & F1 Score & AP\\
        \hline
        tree & data augmentation & $0.868$ & $0.657$ & $0.930 $& $0.770$ \\
        tree & \textbf{no} data augmentation & $0.864$ & $0.760$ & $0.913$ & $0.830$ \\
        sparse & data augmentation & $0.846$ & $0.595$ & $0.959$ & $0.736$ \\
        sparse & \textbf{no} data augmentation & $0.895$ & $0.727$ & $0.955$ & $0.825$ \\
		\hline
    \end{tabular}
    \caption[Experiment and analysis -- data augmentation]{Experiment and analysis -- data augmentation: for our learning objective data augmentation is not beneficial. It especially negatively affects the network's recall for both directed acyclic graph types -- tree-like and sparse. (Table taken from \cite{AlenaMA})}
    \label{tab:dataAugmentationEval}
\end{table}

\begin{figure}[tb]
        \centering
		\subfloat[Including data augmentation for tree-like directed acyclic graphs does \newline decrease the network's recall and thus its AP.]{\includegraphics[width=.5\linewidth]{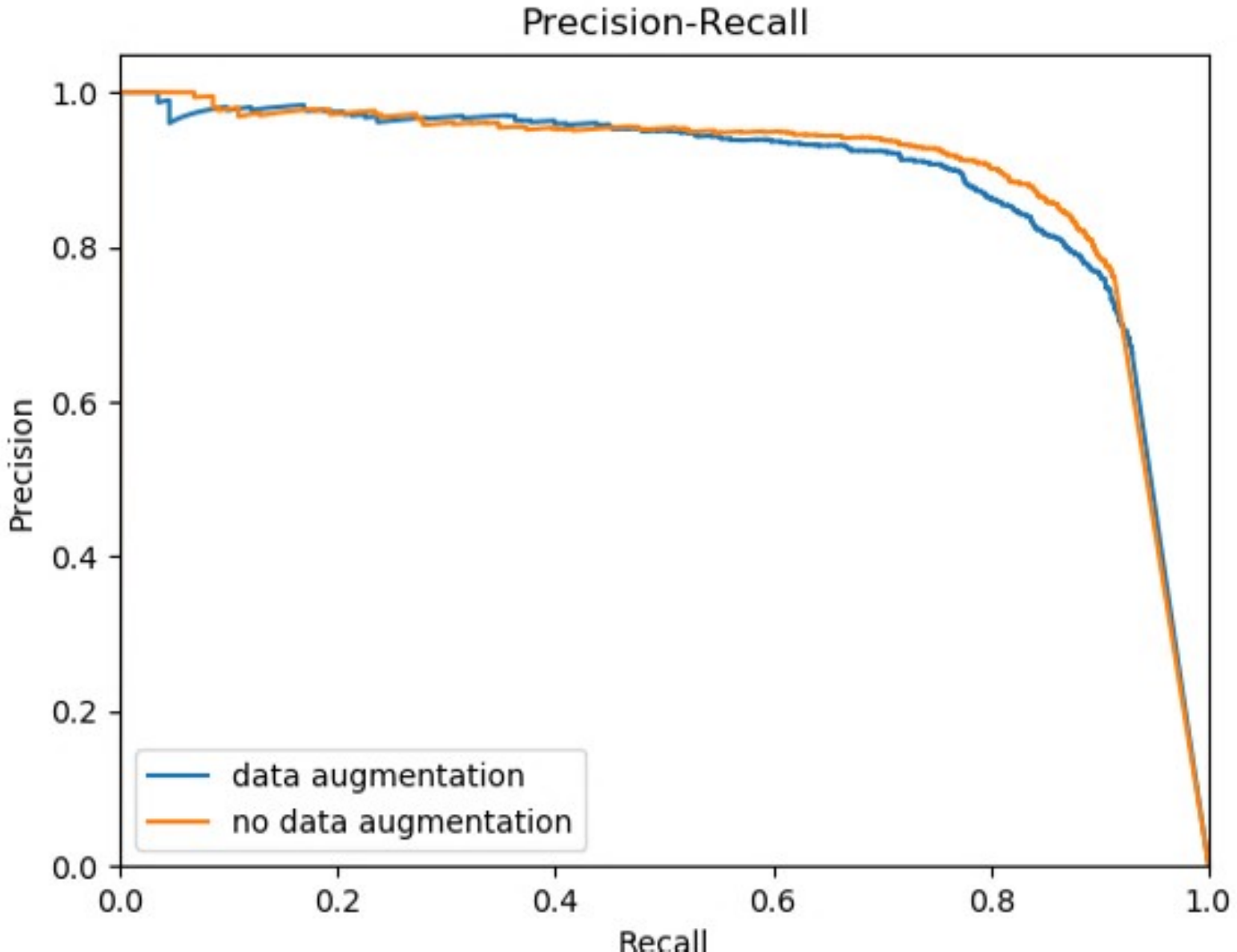}}
        \centering
		\subfloat[Including data augmentation for sparse directed acyclic graphs has even a stronger decreasing effect than the effect for tree-like directed acyclic graphs. Refraining from data augmentation for sparse directed acyclic graphs improves AP by $0.05$ and precision by $0.13$.]{\includegraphics[width=.5\linewidth]{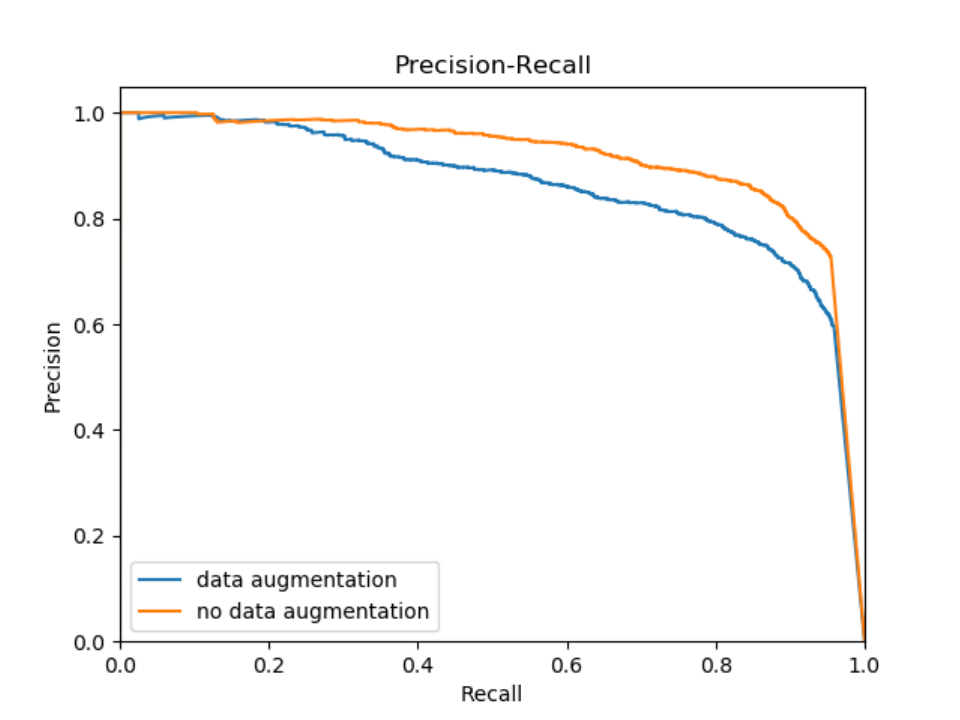}}
    \caption[PR-curves for the evaluation of (no) data augmentation for tree-like and sparse data]{PR-curves for the evaluation of (no) data augmentation for tree-like and sparse data. (Figure taken from \cite{AlenaMA})}
    \label{fig:PRAugmentation}
\end{figure}

Data augmentation creates data diversity and enlarges the database which helps to avoid overfitting and improves the network's generalization abilities \cite{Agarwal2018}. Data augmentation applies affine transformations to the images of the dataset. Examples of such transformations are rotation, translation, or scaling. There are many possibilities for affine transformations, however, not all are always useful for every dataset \cite{Agarwal2018}. Vertical flip of a traffic sign, for instance, may change the semantic meaning to the one of a different, existing traffic sign \cite{5565311}. Consequently, domain knowledge is required to decide which transformation are useful. For our directed acyclic graphs, only affine transformations are valid which could also be produced by a hierarchical, Sugiyama-like, top-rooted layout algorithm because this is the layout algorithm type of our choice (cf. Section \label{sub:definitions_and_research_context} -- \texttt{Graph Drawings}). Transformations which such a layout algorithm could not produce would lead to a tampering of the learned features. So, rotations and and vertical flips are excluded. A flip in vertical direction, for instance, would result in a bottom-rooted directed acyclic graph. Furthermore, we avoid augmentations like image cropping that would affect the visualization of the data. This is in line with the approach of Wöhler et al. \cite{wohler2019learning} who deal with scatterplot visualizations. Varying the brightness, hue and saturation setting had no impact. So, the augmentation methods we applied are: horizontal flip, scaling, and translation. We apply these transformations in a predefined range with random parameters to avoid inducing an unintended pattern which the network may learn and to increase the data variability (cf. \cite{Appendix_Diss_KB} for details).\\
However, as, for instance, according to Kobayashi et al. \cite{8616375}, ``in many cases, data augmentation is effective; however, in some cases, it not be'', we evaluate the benefits of data augmentation for our learning objective first before we finally the decide on whether to use the above discussed strategy or not.

\textbf{Experiment \& Analysis.} We evaluated data augmentation for both of our directed acyclic graph types -- tree-like an sparse directed acyclic graphs. As Table~\ref{tab:dataAugmentationEval} shows, data augmentation especially negatively affects the network's recall which consequently results in a lower AP. For tree-like directed acyclic graphs, data augmentation decreases the AP by $0.06$ (cf. Figure~\ref{fig:PRAugmentation} (a) -- \texttt{\PRBluecolor{blue} curve}). For sparse directed acyclic graphs, it decreases the AP by $0.089$ (cf. Figure~\ref{fig:PRAugmentation} (b) -- \texttt{\PRBluecolor{blue} curve}). This negative impact of data augmentation for our learning objective is explainable as follows: The transformations, supposably, especially the scaling and translations produce data items which are  too different from the not augmented data items in spite of us having chosen the transformation types and ranges based on domain knowledge. The difference of augmented and not augmented data items is likely to result from the strict node placement and layer assignment rules of hierarchical, Sugiyama-like layouts. And translating the directed acyclic graph influences the node placement while scaling the directed acyclic graph influences node placement and layer distance.\\
\emph{\textbf{Our Model Training Decision.}} As a consequence, we refrain from employing data augmentation.
\FloatBarrier

\paragraph{Validation Dataset Size.} 
\label{par:validation_dataset_size}
As for the training dataset size or for the setting of hyperparameters like the learning rate, there is no concrete procedure for determining the validation dataset size. As, for instance, the research of Nguyen et al. \cite{Nguyen:2021aa}, various researchers work on procedure guidelines or size ratio recommendations. However, the influence of the dataset characteristics which is used for the formulating the guidelines resp. recommendations  is not to be neglected. For our data and its characteristics we decided for a validation dataset size of $1.000$.


\subsubsection{Mask R-CNN -- Architectural Details} 
\label{ssub:mask_r_cnn_architectural_details}

\begin{figure}[tb]
    \centering
    \scalebox{.35}{\includegraphics{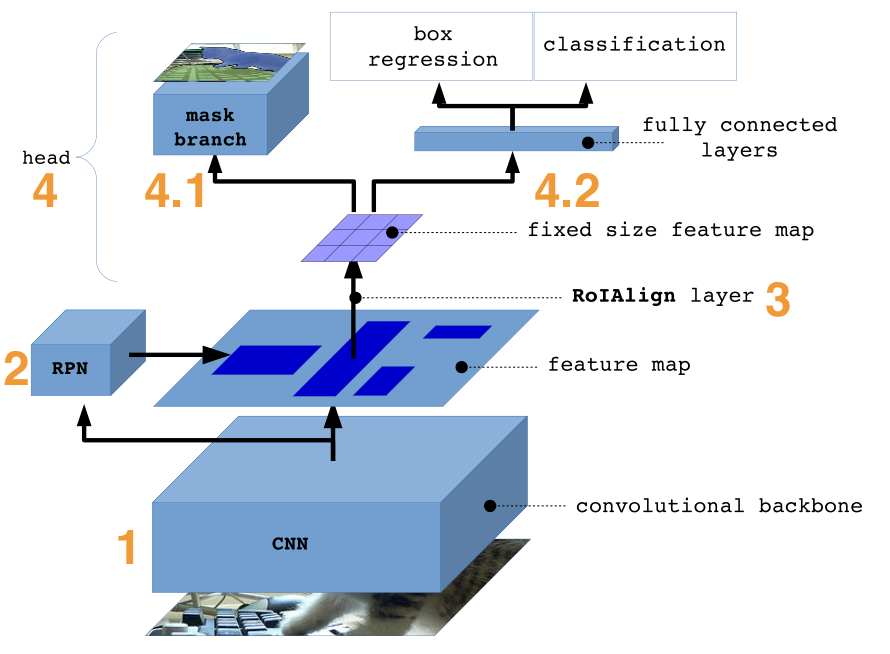}}
    \caption[Schema of Mask R-CNN's architectural components]{Schema of Mask R-CNN's architectural components. (Figure taken from \cite{Galesso})}
    \label{fig:MaskR-CNN}
\end{figure}

  \vspace{-0.6cm}
\begin{figure}[tb]
	  \vspace{-0.6cm}
  \centering
  \vspace{-0.6cm}
    \includegraphics[height=23.2cm]{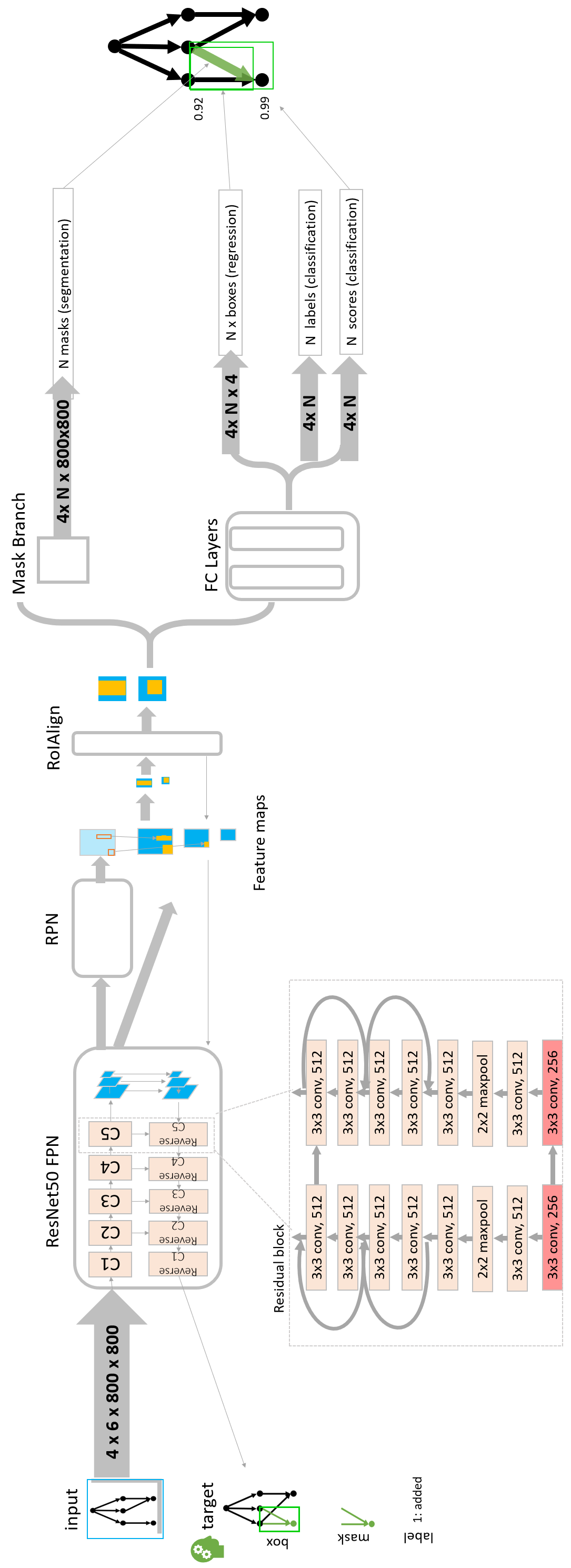}
    \caption[Detailed schematic representation of \textbf{our} Mask R-CNN architecture]{Detailed schematic representation of \textbf{our} Mask R-CNN architecture for the prediction of human detected structural differences in directed acyclic graphs. (Figure taken from \cite{AlenaMA})}
    \label{fig:detailedMaskR-CNN}
\end{figure}

%

Mask R-CNN is an extension of Faster R-CNN \cite{7485869}. So, it not only predicts bounding boxes and classification but also segmentation masks (cf. Figure~\ref{fig:MaskR-CNN} -- \texttt{mask branch}). Due to Mask R-CNN being an extension of Faster R-CNN knowledge from object detection is also applicable to Mask R-CNN.

As Figure~\ref{fig:MaskR-CNN} shows, Mask R-CNN consists of four parts:
\begin{enumerate}
	\item[\orange{1}] a neural network as the backbone
	\item[\orange{2}] a regional proposal network (RPN)
	\item[\orange{3}] a layer aligning the regions of interest (RoIAlign)
	\item[\orange{4}] the head which subdivides into
	\begin{enumerate}
		\item[\orange{4.1}] the mask branch which calculates the segmentation masks
		\item[\orange{4.2}] the fully connected layers which
		\begin{itemize}
			\item predict the best fitting bounding boxes (box regression) and
			\item do the classification
		\end{itemize}
	\end{enumerate}
\end{enumerate}

Figure~\ref{fig:detailedMaskR-CNN} shows our architecture of Mask R-CNN which we used to reach the objective of learning human detected differences for directed acyclic graphs.
\FloatBarrier

\paragraph{Input Data} 
\label{par:input_data}
The input data size of our basis Mask R-CNN architecture is $[800,800,3]$.

\textbf{Our Architectural Decision.} We change the input data size to $[800,800,6]$. This is our implementation of image stacking along the RBG-channel to enable Mask R-CNN to process image pairs. Herewith, we follow the approach of Wu et al. \cite{SpotDifference} who proposed an object-based change detection approach for the popular children's game ``Spot the difference''. Our task is not that different from this game. While this game aims to detect different object classes -- cats, dogs, every day objects etc. -- we aim to detect differences with respect to just two object classes -- nodes and edges.\\
During training, since we use a batch size of $4$ per GPU, the input data size is $[4,800,800,6]$. The model inference phase (testing) only uses the image tensor with a size of $[800,800,6]$.

\emph{\textbf{Target Dictionary -- Supervised Feedback.}} The target dictionary is a look-up table in form of a json file holding the supervised feedback for the training phase. It consists of:
\begin{enumerate}
	\item the directed acyclic graphs (path to the GraphML files of the directed acyclic graphs of the directed acyclic graph pair -- base graph ($G_1$) and alternative ($G_2$))
	\item the images $I_1$ and $I_2$ of the respective directed acyclic graphs belonging to the respective directed acyclic graph pair -- base graph ($G_1$) and alternative ($G_2$) (path to the PNG images of the respective directed acyclic graph pair)
	\item the difference image $I_{diff}$ which is difference graph $G_{diff}$ (path to the PNG images $I_{diff}$)
	\item the segmentation masks (path to the binary mask images which we save as NPY files)
	\item the bounding boxes (list of bounding boxes per directed acyclic graph pair)
	\item the labels (list of labels per bounding box -- $1$: ``contains difference''; $0$: ``no difference contained'')
\end{enumerate}
For details on the calculation of the segmentation masks, bounding boxes and bounding box labels of the supervised learning feedback, please refer to Section \texttt{Calculation of the Segmentation Masks, Bounding Boxes, and Bounding Box Labels of the Supervised Learning Feedback} of the supplementary material of this Section in \cite{Appendix_Diss_KB}.

\paragraph{Convolutional Backbone.} 
\label{par:convolutional_backbone}
The neural network, as the backbone, extracts image features and returns them in feature maps (cf. Figure~\ref{fig:MaskR-CNN} -- \texttt{\orange{1}}). Feature maps are vectors -- also called tensors -- containing information about the images learned. These tensors are of varying size since they contain either low- or high-level features. High-level feature tensors are smaller because they are learned on deeper layers and so more convolutions are already executed on the tensors. Due the neural networks task of feature learning, the neural network choice is crucial for the entire instance segmentations architecture \cite{Agarwal2018}.

Mask R-CNNs discriminative power is a measure of its ability to classify, localize, and segment instance \cite{Zou2019}. The objective of classification aims to learn information on ``difference'' and ``no difference'' -- i.e, high-level semantic information \cite{Zou2019}. While the task of localization is the discrimination of position and scale changes segmentation has to differentiate change regions from other image regions which are not of interest for difference detection based in instance segmentation \cite{Zou2019}. Due to the varying tasks, classification. localization, and segmentation require features with varying characteristics. Since an object should be still detectable when it is scaled or rotated, classification requires invariant\footnote{invariant = features do not change when geometric transformations such as scaling or rotation are applied} features \cite{10.1007/s11263-018-1098-y}. Since the difference object's position and their segmentation depend on its rotation and scaling, localization and segmentation require equivariant\footnote{equivariant = $f: S \rightarrow T$ is equivariant with respect to g: for every $s$ in $S$ when $g_Tf(s) = f(g_Ss)$ with $g_T$ = transformation in feature space of the CNN, $g_S$ = transformation in image space, $S$ = image space features, $T$ = feature space of the CNN \cite{10.1007/s11263-018-1098-y}} features. Consequently, instance segmentation requires both invariant and equivariant features. This feature combination, also called feature fusion \cite{Zou2019} is achievable with deep residual neural networks (ResNet) \cite{7780459}. Such ResNet architectures introduce skip connections and skipped layers. These skip connections pass features from shallow layers to deeper layers without changing them. Mask R-CNN combines ResNet with a feature pyramid network (FPN) which leads to a significant increase in speed and performance \cite{He_2017_ICCV}.

\textbf{Feature Pyramid Network (FPN).} Feature pyramid networks are the neural network based version of traditional feature pyramids. They exploit the implicitly defined feature hierarchy of neural networks. The hierarchy or pyramid is defined by the fact that shallow layers learn low-level features while deeper layers learn higher-level features \cite{Wu2019}. FPNs build the image pyramid via returning the different scales \cite{Lin2017}. FPNs have a top-down and a bottom-up path \cite{Lin2017}. In the feed-forward pass, the bottom-up path builds the implicit image pyramid via the execution of several convolutions at different scales ob a single-scaled input. This leads to multiple feature maps with varying scale. The top-down path combines multiple feature maps with the introduction of lateral connections. These merge feature maps by scaling up the smaller high-level feature maps to the same size of the bigger low(er)-level feature maps. FPNs became state of the art since they lead to a significant localization improvement due to the presence of high-level semantic at all scales \cite{Zou2019}. These multi-scale feature maps are beneficial for our learning objective since our differences and their size range from nodes to large(r) connected components.\\
\emph{\textbf{Our Architectural Decision.}} Since our images have the same size as those of Lin et al. \cite{Lin2017}, we also train end-to-end on a Mask R-CNN architecture pre-trained on the MS COCO dataset, and we also have a detection objective, it is fair to assume that Lin et al.'s settings for the FPN are also valid for our scenario. The parameter settings are:
\begin{itemize}
	\item Upscaling of the feature maps by a factor of $2$
	\begin{itemize}
		\item resulting feature map scales: ${F*1, F*2, F*4, F*8}$
	\end{itemize}
	\item Number of output channels: $256$
\end{itemize}
We then evaluated how well node and edge features are detectable with these settings. As the features from the FPN are used by the regional proposal network to predict the RoIs, we evaluate the quality of the feature extraction as a result of the RoI predictions' quality. So, please refer to the \texttt{Experiment \& Analysis} Paragraph of the \texttt{Regional Proposal Network (RPN)} Section.

\textbf{Experiment \& Analysis -- ResNet50 vs. ResNet101.} For the ResNet part of the convolutional backbone, we compared ResNet50 and ResNet101. Both are common choice for the feature extraction task \cite{Agarwal2018}. RestNet101 is twice as deep than ResNet50. So, ResNet101 remarkably increases the number of parameters, training time and memory consumption while it only increases the AP by $0.0.2$.\\
\emph{\textbf{Our Architectural Decision.}} Consequently, we chose ResNet50. Another advantage, we gained with this decision is that ResNet50 is available pre-trained on the COCO dataset -- both only the convolutional backbone and the complete Mask R-CNN. The end-to-end pre-trained Mask R-CNN includes a pre-trained regional proposal network and a pre-trained head.

\begin{table}[tb]
    \centering
    \begin{tabular}{|c|c|c|c|c|}
		\hline
        Pre-Training & Precision & Recall & F1 Score & AP\\
        \hline
        no & $0.003$ & $0.021$ & $0.005$ & $0.000$\\
        yes & $0.160$ & $0.981$ & $0.276$ & $0.397$ \\
		\hline
    \end{tabular}
    \caption[Experiment and analysis -- pre-training]{Experiment and analysis -- pre-training: pre-training Mask R-CNN on the MS COCO dataset increases recall significantly ($+ 0.961$) and precision by $+0.157$. (Table taken from \cite{AlenaMA})}
    \label{tab:pretrainedMaskRCNN}
\end{table}

\begin{figure}[tb]
    \centering
    \scalebox{.65}{\includegraphics{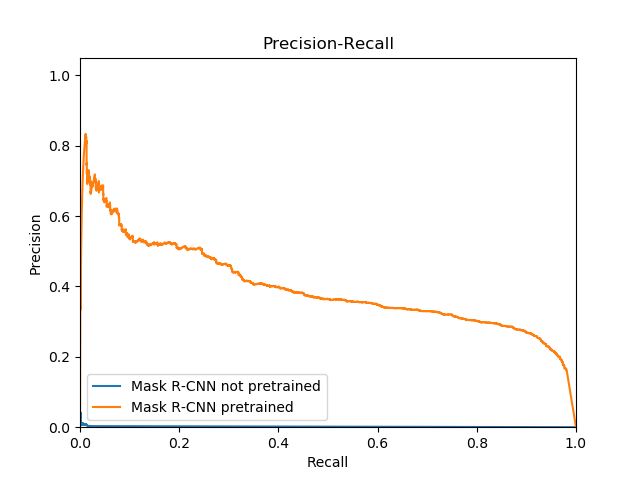}}
    \caption[PR-curve -- pre-training]{PR-curve -- pre-training: A pre-trained Mask R-CNN clearly outperforms a not pre-trained one. (Figure taken from \cite{AlenaMA})}
    \label{fig:PRpretrainedMaskRCNN}
\end{figure}

\textbf{Experiment \& Analysis -- (No) Pre-Training.} Here, we evaluated our assumption that the MS COCO dataset is suitable as a pre-training dataset. Figure~\ref{fig:PRpretrainedMaskRCNN} and Table~\ref{tab:pretrainedMaskRCNN} clearly show that this assumption is confirmed. The PR-curve of a \PRYellowcolor{pre-trained} network clearly outperforms the one of a \PRBluecolor{not pre-trained} network. The pre-trained Mask R-CNN achieves an AP of $0.397$ while the not pre-trained one has an AP of $0.000$. Furthermore, pre-training improves Mask R-CNN's precision and recall -- especially the recall with $+0.960$. Pre-training has also a significant effect on the loss convergence. The pre-trained Mask R-CNN's loss converged much faster. This is an indication that the pre-trained parameters are already good for our directed acyclic graph image pairs. To ensure that we have no issue with overfitting, we tested the AP. The AP increases significantly which can be seen in the values of Table~\ref{tab:pretrainedMaskRCNN} -- i.e., we do not overfit by using pre-training.\\
\emph{\textbf{Our Architectural Decision.}} Since pre-training is clearly beneficial for the network's performance, we use the on the COCO dataset pre-trained ResNet50 backbone.

The implementation of our backbone can be found in \texttt{mask\_rcnn.py}.
\FloatBarrier

\paragraph{Regional Proposal Network (RPN).} 
\label{par:regional_proposal_network}
\begin{figure}[tb]
    \centering
    \scalebox{.5}{\includegraphics{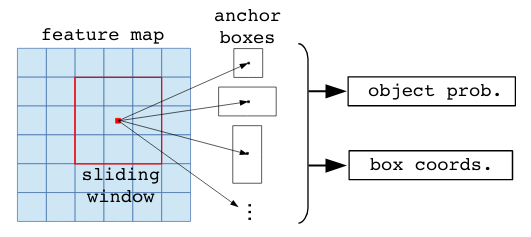}}
    \caption[Schematic representation of the regional proposal network]{Schematic representation of how the regional proposal network generates the anchors -- i.e., the bounding box proposals. (Figure taken from \cite{Galesso})}
    \label{fig:RPN}
\end{figure}

\begin{figure}[tb]
    \centering
    \scalebox{.5}{\includegraphics{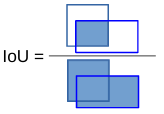}}
    \caption[Schematic representation of the intersection over union metric]{Schematic representation of the intersection over union metric. (Figure taken from \cite{AlenaMA})}
    \label{fig:IoU}
\end{figure}

 \begin{eqfloat}[h]
 	\begin{equation}
 		IoU = \frac{a_{t} \cap a_a}{a_{t} \cup a_a}
 	\end{equation}
 	\caption{Intersection over union metric formula.}
	 	\label{eq:iou}
 \end{eqfloat}
 \myequations{Intersection over union metric formula}

After the extraction of the features, instance segmentation consists of two parts \cite{Agarwal2018}:
\begin{enumerate}
	\item \textbf{the proposal part}\\
	This network part learns to predict bounding box proposal for the objects to be detected.
	\item \textbf{the classification, regression, and segmentation part}\\
	The second part then learns to predict the objects' classes, to choose the best fitting bounding box compared to the target bounding box, and to segment pixels.
\end{enumerate}

Here, we elaborate on the proposal part -- the regional proposal network (cf. Figure~\ref{fig:MaskR-CNN} -- \texttt{\orange{2}}). The second part -- also called head -- is discussed in one of the following paragraphs.

The RPN is a small network which takes the extracted features as input and its output are bounding box proposals -- localization of the rectangular boxes and classification. These depend on the actual image features. This means, if there are, for instance, only small objects depicted in the images a correctly functioning RPN would only predict small bounding boxes surrounding the images objects. The RPN feeds the feature maps through a convolutional layer \cite{7485869}. This result then is piped into two sibling fully connected layers. The first one predicts the bounding box coordinates. The second one -- the classification layer -- predicts the probability, also called the networks confidence, of an object to be detected being inside the predicted bounding box proposal. The bounding box predictions are given as $d^b = [d_x^b, d_y^b, d_w^b, d_h^b]$ with $d_x,d_y$ being the deviation of the box center and $d_w,d_h$ being the box width and height. The probabilities $p=(p_0, p_1...p_K)$ are predicted with a Softmax function over $K$ classes.

The RPN works based on the sliding window approach. Furthermore, since the objects to be detected can be differently sized and have different aspect ratios, the RPN employ anchors. Anchors are potentially matching bounding boxes which the RPN generates for varying scales and aspect ratios. The RPN generates at each sliding window center position anchors for an in advance defined range of aspect ratios and scales (cf. Figure~\ref{fig:RPN}). To give an example -- an anchor scale of $[32^2,64^2]$ pixels and anchor aspect ratios of $1:1$ and $2:1$ produce $4$ anchors per sliding window position -- i.e, $32x32$px, $64x64$px, $32x16$px, $64x32$px. In case an image has a size of $800x800$px this leads to the generation of $2.560.000$ anchors per image.

\textbf{Intersection over Union (IoU).} In the training phase, the RPN evaluates for all anchors how well they match the target bounding boxes $t^b$. To evaluate this, the RPN calculates the overlap of all anchors and the target bounding box for all image objects. The overlap is defined as the ratio of the intersection of both the anchor and the target bounding box and the union area of both (cf. Equation~\eqref{eq:iou} and Figure~\ref{fig:IoU}). This metric is referred to as the ``Intersection over Union Metric''. It is also part of the performance evaluation metrics (cf. Section \texttt{Performance Evaluation Metrics}).\\
Anchors with an overlap greater than an in advance defined IoU threshold are classified as containing an object (class $1$). Mask R-CNN classifies anchors with an overlap $\geq 0.7$ as containing an object \cite{7485869}. Anchors with $<0.3$ overlap are classified as not containing an object (class $0$) \cite{7485869}. Further, Mask R-CNN ignores anchors which are neither classified as containing an object nor classified as \textbf{not} containing an object \cite{7485869}.  Due to the generation of multiple anchors per sliding window center position, the overall amount of anchor classes are biased to negative -- class $0$ -- anchors. To solve this bias, $256$ anchors are randomly sampled from the class $1$ and class $0$ anchors \cite{7485869}. The sampling aims for a $1:1$ ratio of both anchor classes \cite{7485869}. The sampled $256$ anchors, or Regions of Interest (RoIs), are then the input of Mask R-CNN's head.

Mask R-CNN's RPN generates anchors with 5 scales -- $32^2$, $64^2$, $128^2$, $256^2$, $512^2$ -- and three aspect ratios -- $1:1$, $1:2$, $2:1$. Herewith, it follows the architectural decision of Faster R-CNN \cite{He_2017_ICCV}. Also for the RPN, it is possible to train it separately or end-to-end \cite{He_2017_ICCV}. End-to-end training results in high quality region proposals for object detection learning objectives \cite{7485869}.

If the convolutional backbone contains an FPN, as in the case of Mask R-CNN, the RPN  gets a set of feature maps at multiple scales as input. The RPN adapts its region proposals to those feature maps with the best match to the anchors' scale and aspect ratio. Figure~\ref{fig:detailedMaskR-CNN} visualizes this with the red boxes -- anchors -- and the outgoing gray arrows referring to the feature map with the best match. Since Mask R-CNN compares anchors to feature maps, the RPN and the head can share their weights \cite{Agarwal2018}. Sharing weights denotes the reuse of learned weight parameters of one model architecture part for another one \cite{He_2017_ICCV,Schmidhuber2015}. In Figure~\ref{fig:detailedMaskR-CNN}, the weight sharing is illustrated by two gray arrows going out of ResNet50-FPN.

\textbf{Non Maximum Suppression (NMS).} Since RPN generates anchors of multiple sizes at each sliding window position, the region proposals overlap significantly. To reduce this redundancy, we apply non-maximum suppression (NMS) during training and inference. NMS is a post-processing step in which duplicate proposals are rejected by thresholds \cite{Agarwal2018}. Only the proposals with the highest confidence and above an certain IoU threshold are retained ($IoU > 0.5$ = value for Mask R-CNN). The use of NMS does not sacrifice detection accuracy \cite{7485869}, but it significantly reduces the number of proposals and thus the training resp. inference time. Fast inference time is important when models are applied in tools, such as visual comparison tools. For example, a fast prediction time of a model applied in a visual comparison tool is important to provide immediate feedback to the user. Since we also generate many redundant anchors, we also apply NMS.

\textbf{Our Architectural Decision.} Due to our decision to train end-to-end like Lin et al. \cite{Lin2017}, we can also employ the settings of Lin et al. \cite{Lin2017} for our RPN. The settings are those:
\begin{itemize}
	\item anchor area size: ${32^2, 64^2, 128^2, 256^2, 512^2}$ px
	\item anchor aspect ratios: ${1:2, 1:1, 2:1}$
	\item IoU threshold for RoIs containing an image object (class $1$): $IoU>0.7$ -- in our case: a human detected structural difference
	\item IoU threshold for RoIs \textbf{not} containing an image object (class $0$): $IoU<0.3$ -- in our case: background
\end{itemize}
Due to the generation of multiple anchors per sliding window center position, we also encounter the bias of the anchors towards anchors of class $0$. As a consequence, we employ the already above explained sampling solution. We also sample $256$ anchors with a $1:1$ ratio of class $1$ and class $0$ anchors \cite{7485869,Lin2017}.

\emph{\textbf{Non Maxima Suppression (NMS).}} Faster R-CNN \cite{7485869} uses an $IoU > 0.7$ while Mask-R-CNN \cite{He_2017_ICCV} uses an $IoU > 0.5$. Since a higher IoU threshold leads to fewer boxes and faster training resp. inference time, we use an IoU threshold of $IoU > 0.7$. Furthermore, since the number of RoIs is too large, we set the number of RoIs to randomly selected $2000$ RoIs before applying NMS and $1000$ RoIs after NMS. This is in line with the approach of \cite{Lin2017}.

\begin{figure}[tb]
        \centering
        \subfloat{\includegraphics[width=.15\linewidth]{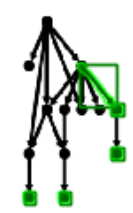}}
        \centering
        \subfloat{\includegraphics[width=.15\linewidth]{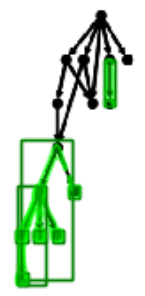}}
    \caption[Two examples of final predictions of our Mask R-CNN architecture]{Two examples of final predictions of our Mask R-CNN architecture -- segmentation masks and bounding boxes are detected precisely around the node and edge shapes of the human detected directed acyclic graph changes. (Figure taken from \cite{AlenaMA})}
    \label{fig:RPNfeatures}
\end{figure}

\textbf{Experiment \& Analysis -- RPN.} We conducted our evaluation with the above mentioned settings. The focus of our evaluation was whether the network learns precise bounding boxes and masks as the RPN provides the region proposals which are the candidate boxes in which the nodes and/or edges resp. the connected components are likely to be located in -- i.e., the quality of the final predictions allows to draw conclusions about the quality of the RPN's proposals. The final predictions, Figure~\ref{fig:RPNfeatures} depicts two examples, show that the segmentation masks are closely around the objects to be detected -- nodes and/or edges resp. connected components and also the final bounding box predictions surround the entire objects to be detected. In conclusion, our network learns precise node, edge, and connected component features.\\
\emph{\textbf{Confirmation of Our Architectural Decision.}} Since our network is able to precisely predict human detected structural difference, we keep the above explained settings.
\FloatBarrier



\paragraph{Head} 
\label{par:head}
 In alignment with Figure~\ref{fig:MaskR-CNN}, the head covers the RoIAlign layer, the fully connected layers and the mask branch.

\textbf{RoI Alignment Layer (RoIAlign, cf. Figure~\ref{fig:MaskR-CNN} -- \texttt{\orange{3}}).} The RoI alignment layer brings the differently sized feature maps to fixed spatial extent \cite{He_2017_ICCV}. Mask-R CNN does this by the RoIAlign operation. The RoIAlign operation is a further development of the RoIPool operation \cite{Girshick2015}. RoIPool is based on aggregated bins which provoke issues for the localization and for pixel-accurate segmentation masks \cite{He_2017_ICCV}. So, RoIAlign replaces the aggregation inside the bins with bilinear interpolation. The RoIAlign layer of Mask R-CNN brings the feature maps to a spatial extent of $[7x7]$ for the fully connected layers and to a spacial extent of $[14x14]$ for the mask branch.\\
\emph{\textbf{Our Architectural Decision.}} For the RoI alignment layer, we keep the settings of He et al. \cite{He_2017_ICCV} as it achieves state of the art performance for instance segmentation. We set the sampling size -- the number of the grid interpolation points -- to $2$. Herewith, we follow the standard practice \cite{Chen2019a}.

\textbf{Mask Branch (cf. Figure~\ref{fig:MaskR-CNN} -- \texttt{\orange{4.1}}).} For the calculation of the segmentation masks, Mask R-CNN pixel-wise applies a Sigmoid function. The Sigmoid function assigns the value $1$ to all pixels which a pixels of the image object to be detected and $0$ to all other pixels. According to Ch. M. Bishop \cite{bishop2006pattern}, Sigmoid functions are standard for binary classifications. Mask R-CNN has a separate mask branch for predicting masks independent of the FC layer probabilities for avoiding a mask competition amongst classes \cite{He_2017_ICCV} preferring masks with high probabilities over others. Masks either can be predicted as one mask per class (class-specific) or one mask per RoI (class-agnostic) \cite{He_2017_ICCV}. The segmentation masks are predicted in the mask branch by a classification layer. Mask R-CNN predicts class-specific segmentation masks\footnote{class-specific masks are a list of $k$ binary images with $k$ being the number of classes in $K$ -- in our case $k=2$}. But performance-wise, class-agnostic\footnote{class-agnostic masks are a list of $n$ binary images with $n$ being the  number of RoIs} masks barely differ from the class-specific ones. Since the mask branch does not convolute the feature maps to short output vectors, it keeps the spacial layout of the feature maps by exploiting the pixel correspondence of the feature maps and the masks which i.a. leads to a higher mask accuracy \cite{He_2017_ICCV}. The mask branch assigns a value of $1$ to every pixel of the feature map with a Sigmoid result of $>0.5$ \cite{He_2017_ICCV}.\\
\emph{\textbf{Our Architectural Decision.}} Here, we also follow the settings of He et al. \cite{He_2017_ICCV} as it achieves state of the art performance for instance segmentation.

\textbf{Fully Connected Layers (FC Layer, cf. Figure~\ref{fig:MaskR-CNN} -- \texttt{\orange{4.2}}).} The FC layers predict, based on information coming from the RoIAlign layer, the bounding box coordinates, a class label, and a probability -- also called classification score. The prediction of the bounding box coordinates is a regression problem. Hereby, the box's deviation of the box center -- $x$, $y$ -- is predicted and also the box' width $w$ and height $h$ \cite{He_2017_ICCV}. The FC layer responsible for classification predicts for each RoI whether it contains an image object to be detected (class $1$) or not (class $0$). The classification score is a discrete probability for each of the $k$ classes -- $p=(p_0,p_1,...,p_k)$ with $k$ being the number of classes in $K$ -- in our case: $p=(p_0,p_1)$. The classification score is the result of the application of a Softmax function. The Softmax function is commonly used for the prediction of classification scores (cf. i.a. \cite{Garcia-Garcia2018,bishop2006pattern,7780459,Keskar2016}). The classification score is a quantification of the network's confidence in the respective RoI containing an image object to be detected. A score of $90\%$ expresses a high confidence.\\
\emph{\textbf{Our Architectural Decision.}}
After features extraction and region proposal, the FC layers of the head combine information and predict for each proposed RoI four outputs: label, probability (prediction confidence), bounding box and segmentation mask. The label expresses the class -- $0$: ``background/no change'', $1$: ``change''. The discrete probability distribution assigns a probability to each class prediction $p=(p_0, p_1)$. This prediction expresses the network's confidence that the RoI either contains a change (class $1$) or not (class $0$). The regression layer outputs the parameterized form of the predicted bounding boxes for the regions which contain detected differences. The mask layer predicts the segmentation masks -- i.e., all pixels belonging to one difference. We follow our predecessor implementations Fast R-CNN \cite{Girshick2015}, Faster R-CNN \cite{7485869}, and Mask R-CNN \cite{He_2017_ICCV} and use two non-linear (ReLU-acitvated) fully connected layers -- fc6 and fc7 layer -- which combine the information of the RoIAlign layer. The output of the FC layers is fed into the predictor\footnote{The term predictor denotes the mask branch, the box regression and the classification part of the network.} which is a network part that flattens the FC layers and predicts graph differences.\\
In contrast to the rest of the network, the predictor is randomly initialized. While parameters for region proposals and aligned RoIs benefit from pre-training, the predictor should be specific to our dataset and hence, not be pre-trained.\\
Similar to the region proposal network (RPN), the head evaluates RoI proposals based on the IoU overlap with the target differences. We assign equal probabilities to differences and background by setting the $IoU > 0.5$. We follow parameter settings of Girshick et al. \cite{Girshick2014} since all preceding architectures \cite{Girshick2015, He_2017_ICCV, 7485869} that base on the initial work of Girshick et al. \cite{Girshick2014} kept the head parameters the same.\\
As positive RoIs -- RoIs containing directed acyclic graph changes -- are extremely rare compared to negative RoIs -- RoIs with no differences -- we ensure a minimum fraction of $0.25$ positive RoIs. Following Lin et al. \cite{Lin2017}, we set the batch size of detections to $512$. The number of box detections depends on the number of changes in the data. Since the number of changes in our dataset is limited to $[1-8]$ additions per directed acyclic graph pair, the maximum number of detections is limited to $8$ changes per image. However, a human could highlight more differences than actually exist. Consequently, the network could predict additional differences. Therefore we set the upper limit of the number of detection, by including a generous buffer of $\approx 90$ items, to $100$ allowing the network to predict additional boxes if required.\\
During inference, we control the number of predictions with a lower probability bound threshold. We set this threshold to $p > 0.92$ as we only want to have the differences in which the network has a high confidence. To set the threshold, we compared the AP which was maximized by $p > 0.92$. Overlapping boxes are reduced with NMS IoU threshold of $IoU > 0.5$.
\FloatBarrier

\subsubsection{Further Evaluation Insights} 
\label{ssub:further_evaluation_insights}

In the following, we elaborate on further insights which we gained throughout the evaluation.

\paragraph{Class Imbalance.} 
\label{par:class_imbalance}
\begin{table}[tb]
    \centering
    \begin{tabular}{|c|c|c|c|c|}
		\hline
        Class Weights & AP & Precision & Recall & F1 Score \\
        \hline
        $[0.5,0.5]$ & $0.744$ & $0.894$ & $0.781$ & $0.834$ \\
        $[0.2,0.8]$ & $0.805$ & $0.758$ & $0.860$ & $0.806$ \\
        $[0.1,0.9]$ & $0.857$ & $0.676$ & $0.918$ & $0.779$ \\
		\hline
    \end{tabular}
    \caption[Class imbalance -- evaluation of three class weights as an imbalance countermeasure]{Class imbalance -- evaluation of three class weights as an imbalance countermeasure: class weights of $[0.1,0.9]$ achieved highest AP even though precision and recall are least balanced which is indicated by a lower F1 score. (Table taken from \cite{AlenaMA})}
    \label{tab:classWeights}
\end{table}

\begin{figure}[tb]
    \centering
    \scalebox{.3}{\includegraphics{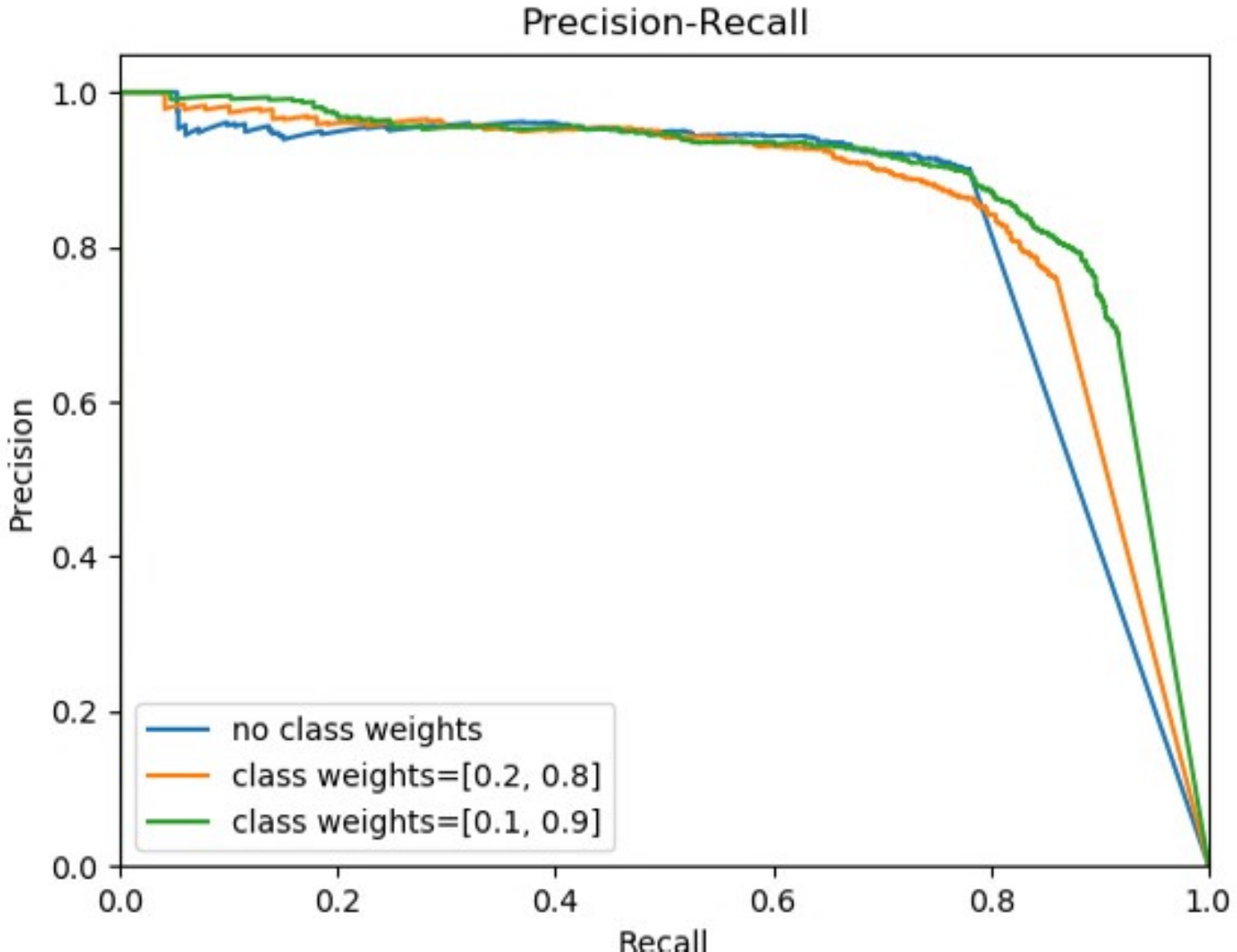}}
    \caption[PR-curve -- class imbalance]{PR-curve -- class imbalance:  evaluation of three class weights as an imbalance countermeasure. The use of class weights increase the recall but decreases the precision. Class weights of $[0.5,0.5]$ are equal to having no class weights (cf. \PRBluecolor{blue} PR-curve). Class weights of $[0.1,0.9]$ lead to the best performance (cf. \PRGreencolor{green} PR-curve). (Figure taken from \cite{AlenaMA})}
    \label{fig:classWeights}
\end{figure}

Our dataset is imbalanced towards background areas. This is a problem which commonly occurs for object detection \cite{Agarwal2018}. More precisely, we consider all image areas that do not cover a directed acyclic graph difference as background -- i.e, this applies to all areas which which cover common graph elements. As our data items only have $1-8$ changes and changes consist of small and narrow objects, only a small part of the image actually cover differences. Consequently, RoIs covering background (negative RoIs) dominate RoIs covering directed acyclic graph differences (positive RoIs). This issue is called class imbalance \cite{Agarwal2018}. In order to set a focus on differences, i.e., the positive RoIs, we balanced the class entropy loss $L_cls$, $L_{rpn\_cls}$. Herewith, we follow the proposal of Agarwal et al. \cite{Agarwal2018}. We do not balance the binary entropy loss $L_mask$, as masks are only considered for positive RoIs. We tested class weights for $[0.1,0.9]$, $[0.2,0.8]$, $[0.5,0.5]$ for the classes $0$ and $1$. Class weights of $[0.5,0.5]$ expresses having no class differences as classes are equally weighted. Class weights of $[0.1,0.9]$ foster class $1$ to a maximal extent and still consider negative RoIs -- background -- with $10\%$. We weight background RoIs with a factor $0.1$ to not ignore them. Class weights of $[0.2,0.8]$ show the trend between having no class weights to class weights of the maximal extent. The values are determined by maximizing evaluation metrics. Class weights of $[0.1,0.9]$ improved the network recall from $0.781$ to $0.918$ and thus the AP -- i.e., those weights lead to the best performance (cf. Figure~\ref{fig:classWeights} -- \texttt{\PRGreencolor{green} PR-curve}). As a consequence, we chose those weights. Since a higher class weight for class $1$ -- differences -- corresponds to favoring differences over background boxes, the number of positive boxes compared to the number of negative boxes is increased. Differences count as detections. Consequently, a higher class weight for differences increases the number of detections. As discussed before, an increased number of detections increases the network's recall but decreases its precision. This results in a lower F1 score since precision and recall are less balanced. Table~\ref{tab:classWeights} also reflects that.
\FloatBarrier

\paragraph{Hyperparameter Optimization.} 
\label{par:hyperparameter_optimization}

For the hyperparameter optimization we observed the training and validation loss. Herewith, we evaluated the model's learning speed and whether the hyperparameters support learning.

\textbf{Optimizer.} We employ a stochastic gradient descent optimizer \cite{LeCun1989c}. Herewith, we are in line the current body of work of the object detection domain \cite{Agarwal2018}. We set the optimizer's momentum to $0.9$ and its weight-decay to $0.0005$. Herewith we follow the work of He et al. \cite{He_2017_ICCV} who initially proposed the Mask R-CNN architecture.

\textbf{Learning Rate.} As we employ transfer learning, we start with a learning rate of $0.001$. This learning rate is typical for training pre-trained models \cite{7299016,Dosovitskiy2016a,Grattarola2019,SpotDifference}. With this small learning rate we ensure that the knowledge of the already trained parameters is conserved \cite{7485869}. We reduce the learning rate by a factor of $10$ if the validation loss does not improve anymore after a certain amount of epochs. We decreased the learning rate by factor $10$ after $10.000$ iterations as the validation loss stagnated. This setting was best with respect to convergence and non-overfitting.\\
To be complete, we tested also higher learning rates. They lead to $NaN$ values which are an indication that the gradient is too large. As the gradient gets multiplied with the learning rate during backpropagation, a reduction of the learning rate also reduces the gradient. We
Higher learning rates of 0.1 have been tested. They resulted in nan values which indicates a too large gradient. As the learning rate multiplies the gradient during backpropagation, reducing the learning rate reduces the gradient as well.

\textbf{Loss Convergence, Epochs, Early Stopping.} Our training loss converged after $6.000$ iterations. This is yet another indication that transfer learning accelerates loss convergence and that a small learning rate of $0.001$ is appropriate in our context. Ji et al. \cite{Ji2019} reported similar number. Their loss also converged after $6.000$ for the Mask R-CNN architecture domain-specifically fine-tuned on the SC-2016 dataset. Also De Luca et al. \cite{10.1007/978-3-030-35802-0_38}, who trained on graph drawing images, reported similar loss convergence numbers. Consequently, we see our convergence values as comparable with respect to pre-training and architecture and as reasonable in the context of training on graph drawing images using a pre-trained Mask R-CNN architecture. As additional epochs increased precision, we did not stop after $6.000$ iterations -- i.e. $12$ epochs -- but trained for $40$ epochs with a mini-batch size of $4$ image pairs which took about $15$ hours. Initially, we set our epoch to $50$. As change detection publications which employed Mask R-CNN reported convergence values of $5-40$ epochs and literature training neural networks on graph drawing imaged reported convergence values of up to $20$ epochs, we wanted to have a buffer and therefore decided for $50$ epochs. However, our validation loss started increasing again after $40$ epochs which is the reason why we employed early stopping and stopped after $40$ epochs.


\subsubsection{Domain-Specific Fine-Tuning and Final Architecture} 
\label{ssub:domain_specific_fine_tuning_and_final_architecture}

\begin{figure}[tb]
  \centering
    \includegraphics[width=.9\textwidth]{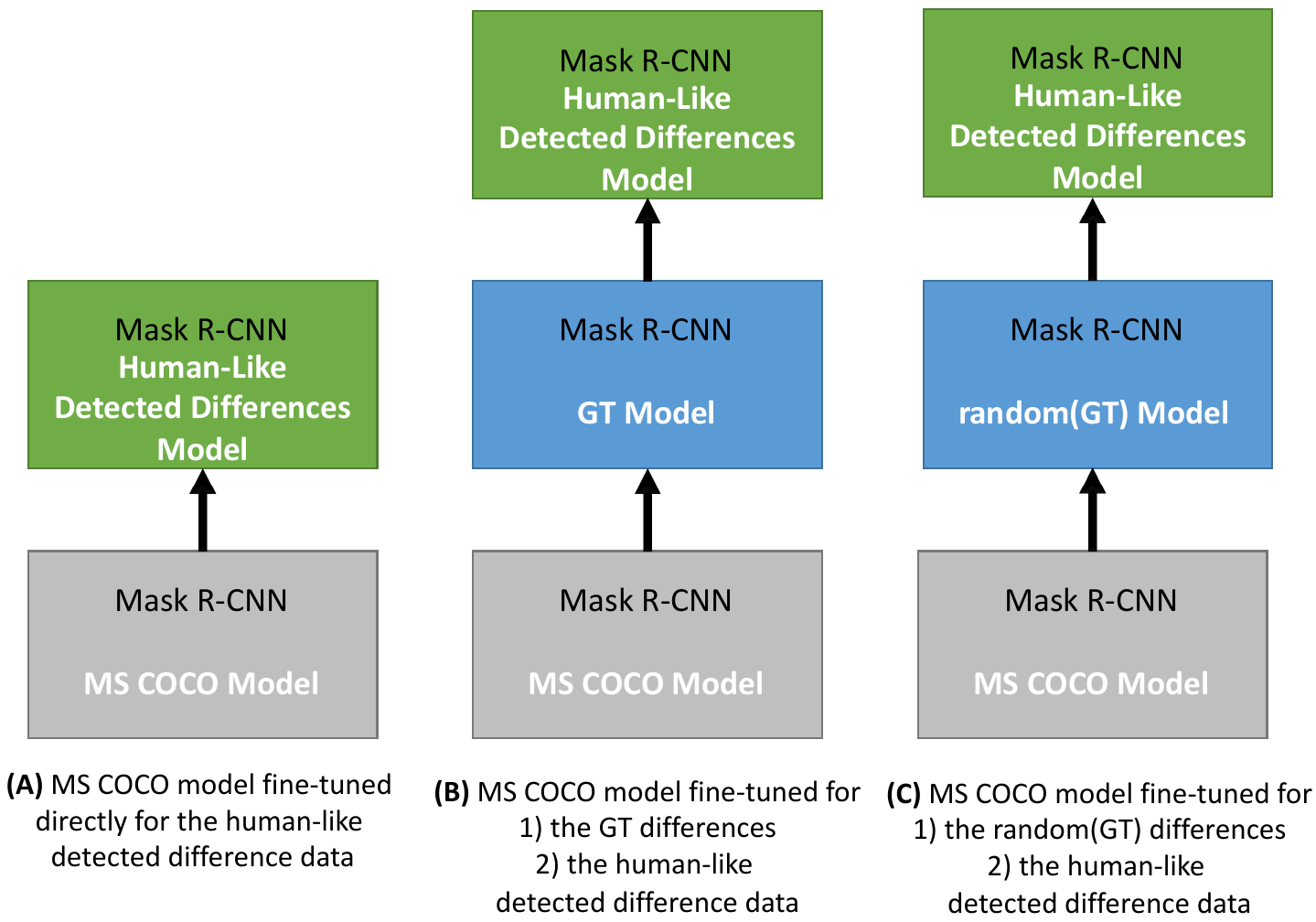}
  \caption[Schematic representation of the domain-specific fine-tuning approach we evaluated]{Schematic representation of the domain-specific fine-tuning approach we evaluated. (Figure based on original Figure from \cite{AlenaMA})}
  \label{fig:imgs_TrainingChains}
\end{figure}

To adapt our Mask R-CNN architecture pre-trained on the COCO dataset to the characteristics of our domain data -- directed acyclic graphs visualized as node-link diagrams with circular nodes and arrow-headed edges -- we employ domain-specific fine-tuning. To figure out the appropriate fine tuning, we tested in total three different approaches -- see Figure~\ref{fig:imgs_TrainingChains}:
\begin{itemize}
	\item[\textbf{(A)}] MS COCO model fine-tuned directly for the human-like detected difference data
	\item[\textbf{(B)}] MS COCO model fine-tuned for
	\begin{itemize}
		\item[1)] the GT differences
		\item[2)] the human-like detected difference data
	\end{itemize}
	\item[\textbf{(C)}] MS COCO model fine-tuned for
	\begin{itemize}
		\item[1)] the random(GT) differences
		\item[2)] the human-like detected difference data
	\end{itemize}
\end{itemize}

Our data to be fine-tuned initially consisted of two datasets:
\begin{enumerate}
	\item the GT dataset
	\item the human-like detected differences dataset
\end{enumerate}
We introduced both in Section~\ref{sub:a_dataset_for_learning_human_detected_differences_in_dags_based_on_graph_data_enriched_with_human_inspired_detected_graph_difference} -- \texttt{Dataset Creation}.
\FloatBarrier

\begin{table}[tb]
    \centering
    \begin{tabular}{|c|c|c|c|c|c|}
		\hline
        directed acyclic graph Density & Architecture & AP & Precision & Recall & F1 Score \\
        \hline
        tree-like & (A) & $0.862$ & $0.689$ & $0.920$ & $0.788$ \\
        tree-like & (B) & $0.310$ & $0.595$ & $0.468$ & $0.524$ \\
        tree-like & (C) & $0.857$ & $0.670$ & $0.922$ & $0.776$ \\
        sparse & (A) & $0.844$ & $0.605$ & $0.959$ & $0.742$ \\
		 sparse & (C) & $0.864$ & $0.663$ & $0.944$ & $0.779$ \\
		\hline
    \end{tabular}
    \caption[Experiment and analysis -- domain-specific fine-tuning]{Experiment and analysis -- domain-specific fine-tuning: Domain-specific fine-tuning the human-like detected differences with an advance fine-tuning on the GT difference dataset (architecture (B)) clearly underperforms compared to the other approaches. For tree-like directed acyclic graphs there is only a marginal difference for directly fine-tuning on the human-like detected differences (architecture (A)) or an in advance fine-tuning on the random(GT) dataset (architecture (C)). However, for sparse directed acyclic graphs, architecture (C) is recognizably better. (Table taken from \cite{AlenaMA})}
    \label{tab:architectures}
\end{table}

\begin{figure}[tb]
        \centering
        \subfloat[Architecture (B) recognizable underperforms compared \newline to architecture (A) and (C). There is only a marginal \newline difference in performance for training on tree-like directed acyclic graphs \newline when it comes to architecture (a) and (C).]{\includegraphics[width=.5\linewidth]{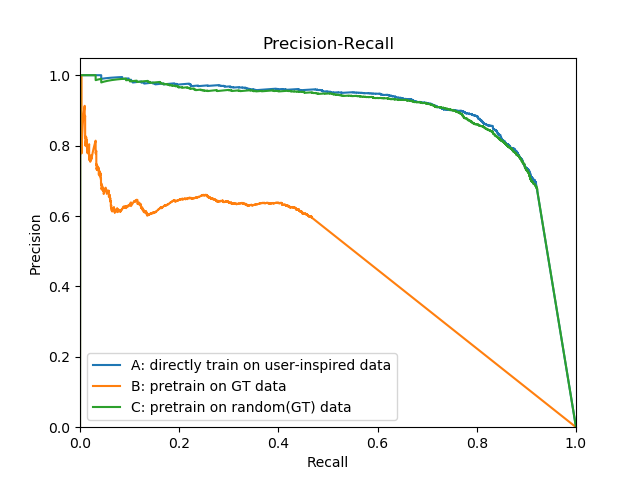}}
        \centering
        \subfloat[For sparse directed acyclic graphs, architecture (C) is recognizably better than architecture (A).]{\includegraphics[width=.5\linewidth]{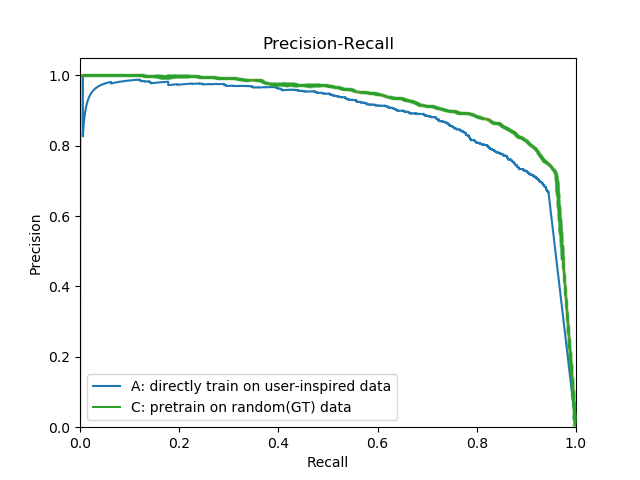}}
    \caption[PR-curve -- domain-specific fine-tuning]{PR-curve -- domain-specific fine-tuning: Comparison of the different domain-specific fine-tuning approach shown in Figure~\ref{fig:imgs_TrainingChains}. (Figure taken from \cite{AlenaMA})}
    \label{fig:PRarchitectures}
\end{figure}

\textbf{Experiment \& Analysis -- Domain-Specific Fine-Tuning.} Architecture (A) directly fine-tunes on the human-like detected differences dataset. Herewith, architecture (A) requires only one training. As especially real human detected differences data may be limited, We came up with training approach (B). Here, we operate under the idea that a two-stage fine-tuning -- 1) fine-tuning on GT dataset, 2) fine-tuning on human-like detected differences dataset -- is beneficial as features characterizing nodes and edges can be learned from the larger GT differences dataset and then be used throughout the adaption to the differences humans would detect.
During training, we realized that architecture (B) learns to detect all graph theoretic difference, however, is not able to adapt to the differences humans would detect when the GT model is fine-tuned based on the human-like detected differences. This is supported by the PR-curve of architecture (B). In Figure~\ref{fig:PRarchitectures} a), we can clearly see that the \PROrangecolor{orange} PR-curve is way worse than those of architecture (A) and (C). As a consequence, we introduced the random(GT) dataset and architecture (C). Architecture (C) follows the same idea as architecture (B) but uses the random(GT) dataset which is a random sampling of the GT differences which finally makes the random(GT) more comparable to the human-like detected differences dataset than the GT dataset (cf. Section~\ref{sub:a_dataset_for_learning_human_detected_differences_in_dags_based_on_graph_data_enriched_with_human_inspired_detected_graph_difference} -- \texttt{Dataset Creation} for details).\\
For tree-like data, we can see in Figure~\ref{fig:PRarchitectures} a) that there is only a marginal difference in the performance of architecture (A), the \PRBluecolor{blue} PR-curve, and (C) -- the \PRGreencolor{green} PR-curve. Table~\ref{tab:architectures} supports this insight. However, when it comes to sparse directed acyclic graphs, architecture (C), the \PRGreencolor{green} PR-curve, is recognizably better than architecture (A) -- the \PRBluecolor{blue} PR-curve (cf. Figure~\ref{fig:PRarchitectures} b)). This may result from the fact that the network's ability to generalize for more complex directed acyclic graphs -- here: sparse directed acyclic graphs -- benefits from the increased diversity brought by the random(GT) dataset. Agarwal et al. \cite{Agarwal2018} also discuss this effect of diversity on the ability to generalize in their survey. As we cover with our work both densities, cf. Section~\ref{sub:a_dataset_for_learning_human_detected_differences_in_dags_based_on_graph_data_enriched_with_human_inspired_detected_graph_difference} -- \texttt{Density} for details, we choose architecture (C).
\FloatBarrier


\begin{table}[tb]
    \centering
    \begin{tabular}{|c|c|c|c|c|}
		\hline
        Dataset & AP & Precision & Recall & F1 Score \\
        \hline
        tree-like & $0.857$ & $0.670$ & $0.922$ & $0.776$ \\
        sparse & $0.844$ & $0.605$ & $0.959$ & $0.742$ \\
        \makecell{tree-like \textbf{\&} sparse} & $0.862$ & $0.647$ & $0.940$ & $0.767$\\
		\hline
    \end{tabular}
    \caption[Experiment and analysis -- combination of tree-like and sparse directed acyclic graphs]{Experiment and analysis -- combination of tree-like and sparse directed acyclic graphs: trained separately and especially combined, our network achieves a performance which is comparable to related change detection work with Mask R-CNN. An example is the work of Ji et al. \cite{Ji2019} who achieve an AP of $0.858$ for the detection of changes of buildings. (Table taken from \cite{AlenaMA})}
    \label{tab:appComparison}
\end{table}

\begin{figure}[tb]
    \centering
    \subfloat[Architecture (C) achieves an \newline average precision of $0.857$ on tree-\newline like dataset with a precision \newline of $0.670$.]{\includegraphics[width=0.333\linewidth]{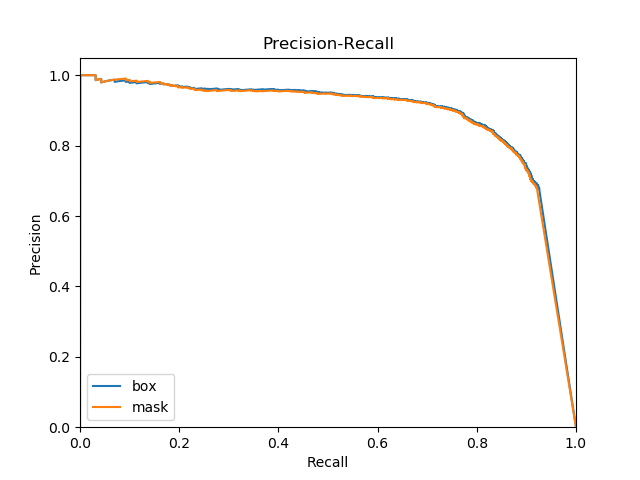}}
    \centering
    \subfloat[Architecture (C) achieves an \newline average precision of $0.844$ on \newline sparse dataset with a precision \newline of $0.605$.]{\includegraphics[width=0.333\linewidth]{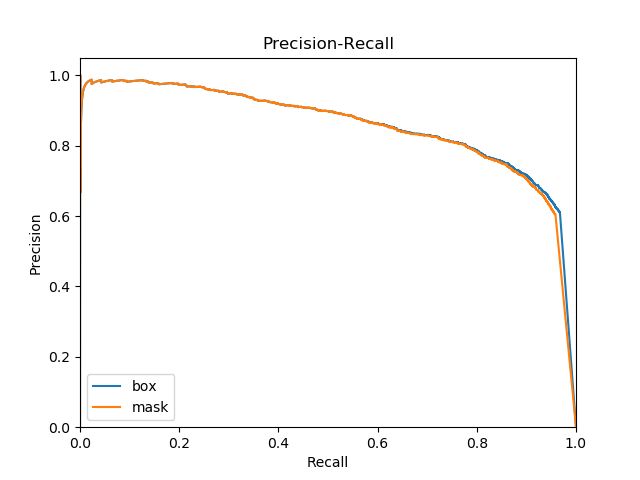}}
    \centering
    \subfloat[Architecture (C) achieves an \newline average precision of $0.862$ trained \newline on tree and sparse data with a \newline precision of $0.647$.]{\includegraphics[width=0.333\linewidth]{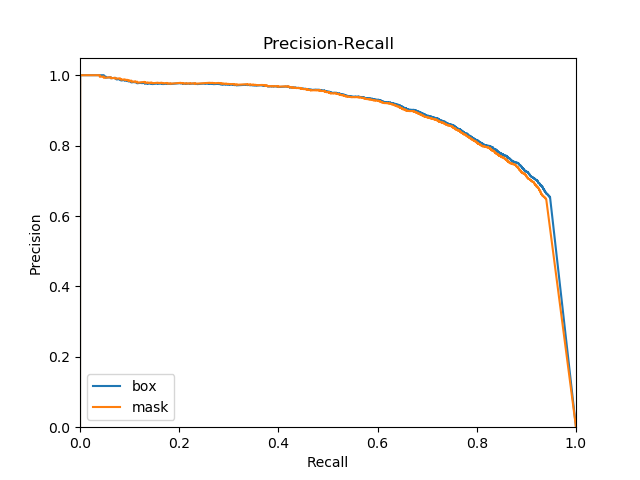}}
\caption[PR-curves for the experiment and analysis of the combination of tree-like and sparse directed acyclic graphs]{PR-curves for the final training experiment and analysis -- training of the final architecture (C) for tree-like, sparse, and combined -- tree-like \& sparse -- directed acyclic graphs. (Figure taken from \cite{AlenaMA})}
\label{fig:PRCurves}
\end{figure}
\paragraph{Final Training Experiment \& Analysis -- Combination of Tree-Like and Sparse directed acyclic graphs.} 
\label{par:final_training_combination_of_tree_like_and_sparse_dags}
As tree-like and sparse directed acyclic graphs may appear together in a real world application, we conduct a final experiment on our final architecture and compare the networks performance on tree-like and sparse directed acyclic graphs separately and combined. The results for the model trained on the combined dataset -- tree-like and sparse directed acyclic graphs -- shows that our chosen hyperparameters and architectural decisions work for the data characteristics we aimed to cover. As we can see in Table~\ref{tab:appComparison}, our model achieves for all datasets and $AP\approx0.8$ -- tree-like directed acyclic graphs: $AP=0.857$; sparse directed acyclic graphs: $AP=0.844$; tree-like \textbf{\&} sparse directed acyclic graphs combined: $AP=0.862$. The PR-curves in Figure~\ref{fig:PRCurves} support this. For the combined dataset it achieves the best results over all evaluation metrics:  $AP=0.857$, $precision=0.647$, $recall=0.940$, $F1 \text{ }Score=0.767$ (cf. Table~\ref{tab:appComparison}).

Since our model has a high average recall of $0.940$ and recall is a quantification of whether all differences humans would detect are detected by the network, we can state that our model is able to predict human detected differences of directed acyclic graph pairs. However, our average precision indicates that there are still a certain amount of false positives (cf. Table~\ref{tab:appComparison}). An in depth discussion follows in Section~\ref{sub:discussion_conclusion_and_future_work}. The slight imbalance of our network's  precision and recall can also be seen in the F1 score of $0.767$. But still, our high recall indicates that the human detected differences are well predicted by our network. With our average precisions -- tree-like directed acyclic graphs: $AP=0.857$; sparse directed acyclic graphs: $AP=0.844$; tree-like \textbf{\&} sparse directed acyclic graphs combined: $AP=0.862$ -- we achieve a performance which is comparable to  the performance of related change detection work with Mask R-CNN. An example is the work of Ji et al. \cite{Ji2019} who achieve an AP of $0.858$ for the detection of changes of buildings.
\FloatBarrier


\subsection{Discussion, Conclusion, and Future Work} 
\label{sub:discussion_conclusion_and_future_work}

Our discussion, conclusion and elaboration on future work are three-partite. It covers the learning of human detected structural differences, image-based learning, and capturing the human notion of structural differences.

\subsubsection{Learning Human Detected Structural Differences in Directed Acyclic Graphs} 
\label{ssub:learning_human_detected_structural_differences_in_directed_acyclic_graphs}

\paragraph{Result Summary.} 
\label{par:result_summary}
As the task of change detection is similar to our learning objective and as we decided to learn on picture due to humans also processing the visualizations of the directed acyclic graphs, our basis architecture for learning the human detected structural differences in directed acyclic graph pairs was a Mask R-CNN pre-trained on the COCO dataset. One of our core architectural adaption was to make Mask R-CNN work on image pairs. We achieved this by implementing image stacking. Our final Mask R-CNN architecture is a domain-specific fine-tuned network. Special to the fine-tuning is that before the network is fine-tuned to the human-like detected differences dataset it is fine-tuned to the random(GT) difference dataset. We introduced both datasets in Section~\ref{sub:a_dataset_for_learning_human_detected_differences_in_dags_based_on_graph_data_enriched_with_human_inspired_detected_graph_difference}. The idea was that the random(GT) dataset increases the training dataset size as especially human detected differences data may be limited. This idea was beneficial -- especially for the denser sparse directed acyclic graphs. The final architecture has a constant and high prediction performance across datasets -- tree-like directed acyclic graphs: $AP=0.857$; sparse directed acyclic graphs: $AP=0.844$; tree-like \textbf{\&} sparse directed acyclic graphs combined: $AP=0.862$.

\paragraph{Discussion.} 
\label{par:discussion}
With the AP of $0.862$ of our final architecture we can compete with related work in the change detection domain which also employs Mask R-CNN and fine-tunes it for domain-specific data. The work of Ji et al. \cite{Ji2019} is an example. The authors' learning objective was the detection of building changes. They achieve an AP of $0.858$. Ji et al. also trained with simulated data as we did with our human-like detected difference segmentation masks. Our precision of $0.647$, opposed to Ji et al.'s \cite{Ji2019} ($0.833$) is worse by $0.186$. Our recall is comparable  -- ours: $0.940$; Ji et al.'s: $0.922$. Our high recall is an indication that nearly all human detected differences are predicted. The decrease in precision may be due to the difference of our change detection task and the one of Ji et al. \cite{Ji2019}. While Ji et al. aimed to detect all existing building changes we wanted to predict all human detected structural differences which not necessarily have to be all existing structural differences. This results from the influence factors having a hindering effect. However, it might also be that the human detected differences encompass all existing changes -- i.e., the GT changes. This mostly happens when there are no dense regions in the directed acyclic graphs. Because of that, the network learns features and selection patterns which are more dominantly representative for the human detected differences but also for the GT differences and this may lead to the prediction of more changes than the human detected ones leading to a decrease of the network's precision. In conclusion, this suggests that the final differentiating feature is still to be detected.

\paragraph{Future Work.} 
\label{par:future_work}
As a high precision of is a quality measure for only predicting  the human detected differences, the decrease in our precision indicates that our network predicts also additional differences. These additional differences are other combinations of nodes and edges. Currently, there is a fixed IoU threshold determining whether the prediction is a true positive or not. It might be a worthwhile to investigate whether it is an option to use an adaptable IoU threshold which reacts to where in the directed acyclic graph the prediction is. This would mean that also the IoU threshold respects the human notion of differences thus, for instance, in dense regions it is more likely that smaller differences are spotted by humans as dense regions are hindering and tend to mask edges while in sparser regions connected changes are more likely to be seen by humans. However, in case the IoU is not adaptable and is set to a relatively large number, e.g., to $>0.5$, small predictions are considered as false positives in spite of them being reasonable due to the density of the region. This, in turn, then negatively impacts the network's precision. There is no one right answer when it comes to the interpretation of commonalities, differences, or similarities -- neither from a graph theoretical nor a human standpoint. And an adaptable IoU threshold, as we believe, could be a means to consider that. Furthermore, it might be also worthwhile to further investigate what differentiates human selection patterns and GT differences and under which circumstances human selection patterns resemble the GT differences. Knowledge on that would shed light on differentiating features which may have the potential to avoid additional predictions decreasing precision.

\subsubsection{Image-Based Learning} 
\label{ssub:image_based_learning}
\paragraph{Result Summary} 
\label{par:result_summary2}
With Mask R-CNN we chose an image-based learning approach. While this is a decision for not the most efficient approach with respect to training time, it is a decision for the approach which is closest to how humans do their comparison -- i.e., humans work with the visualized directed acyclic graphs when they visually compare them. The visualization resp. the visualization design choices are known to influence human processing \cite{Zhang2017,holten:hal-00696823}. Consequently, it is necessary to learn on the images.

\paragraph{Discussion} 
\label{par:discussion2}
In case, one would like to have a certain independence of the visual design choices, one could follow the approach of Wöhler et al. \cite{wohler2019learning}. The authors combined an image- and a structural-data-based network to learn the correlation perception of humans. Consequently, they were were able to train on both images and the structural data and achieved a certain independence of the visualization. For pure image-based learning approaches, the learning approach depends on the visual design choices. Consequently, it is not fair to assume that the model would work for a varying visual design. This, however, is currently no limitation of our work. In fact, at the moment, we refrained from doing this since we identified our difference factors based on the standard design -- circular nodes and arrow-headed edges -- for node-link diagrams and have no information how the factors change for varying node-link diagram designs.

\paragraph{Future Work.} 
\label{par:future_work2}
However, it is an interesting future work question whether the integration of a graph-based learning approach -- e.g., a graph convolutional neural network -- would result in a more robust architecture. Maybe the integration of additional structural graph information is even able to solve the precision issue we discussed in Section \texttt{Learning Human Detected Structural Differences in Directed Acyclic Graphs}. Pos-GCN \cite{Danel} might be promising since it also processes euclidean node positions. Since our graph data is laid out in advance, we have the required euclidean embedding. This is a significant difference to other graph-based learning approaches which mostly assume the graphs not to be embedded in the euclidean space.


\subsubsection{Capturing the Human Notion of Structural Differences} 
\label{ssub:capturing_perception}

\paragraph{Result Summary and Discussion.} 
\label{par:result_summary3}
Our results for capturing the human notion of structural directed acyclic graph differences rest on two legs:
\begin{enumerate}
	\item the data augmentation algorithm for the creation of a large enough training dataset to train a robust and generalizable network
	\item the selection, adaption, and evaluation of a suitable learning approach -- Mask R-CNN
\end{enumerate}

\textbf{Data Augmentation.} Our DFS-algorithm for data augmentation is a knowledge-based algorithm which generates the supervised feedback -- segmentation masks, bounding boxes, and labels -- for our learning approach. The augmented data encodes human-like detected differences. By human-like we mean difference annotations which are similar to those which an actual human would do. The similarity is achieved by using a knowledge-based approach. Our DFS-algorithm gains its knowledge from our studies on the visual comparison with respect to differences (cf. \cite{10.1007/978-3-319-73915-1_20,JGAA-467}). Our difference-coined comparison studies identified six dominant factors which influence humans -- some of them are hindering while others foster the detection of differences. Our DFS-algorithm uses these factors to decide which of the GT differences, or actually existing differences, a human would detect. While the calculation of some of the factors like edge crossing or visual symmetry is entirely clear and also substantiated by the current body of work, the calculation of of other factors like density shows room for improvement.

Our implementation of \textbf{Mask R-CNN} has shown to learn the patterns based on which humans select resp. spot structural differences. Our recall indicates that nearly all human detected differences are detected by our network, while precision indicates that there are additional predictions which are not human detected differences. For this we already discussed reasons and future work in Section \texttt{Learning Human Detected Structural Differences in Directed Acyclic Graphs}.
%
%

\begin{figure}[tb]
        \centering
		\subfloat[The hindering factor \newline of density is the reason \newline that in the orange highlighted area only the nodes \newline are annotated as human- \newline like detected differences.]{\includegraphics[width=.222\linewidth]{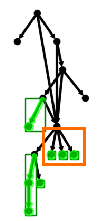}}
        \centering
		\subfloat[Example 1 of an \newline area which would \newline rather obfuscate \newline edges.]{\includegraphics[width=.222\linewidth]{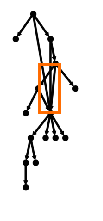}}
        \centering
		\subfloat[Example 2 of an area which would rather obfuscate edges.]{\includegraphics[width=.222\linewidth]{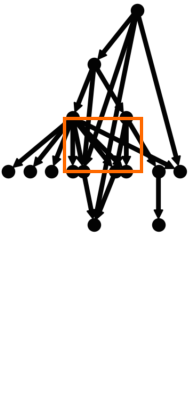}}
    \caption[Example which substantiates my future work idea to make the data augmentation algorithm's the decisions more sophisticated]{Example which substantiates my future work idea to make the data augmentation algorithm's the decisions more sophisticated -- especially those based on the hindering factors. The behavior shown in Figure~\ref{fig:DataAugMoreSophisticated} a) is exactly the correct behavior of my algorithm since this area is an area with a high proportion of colored pixels. However, given the visual impression, this area would rather not obfuscate any edges for a human. Figure~\ref{fig:DataAugMoreSophisticated} b) and c) show examples of areas which would obfuscate edges for humans. (Figure based on original Figures from \cite{AlenaMA})}
    \label{fig:DataAugMoreSophisticated}
\end{figure}

\paragraph{Future Work.} 
\label{par:future_work3}
For our \textbf{data augmentation}, an interesting future work step would be to make especially the decision based on the hindering factors more sophisticated. To explain this, we will use the example of density. In Figure~\ref{fig:DataAugMoreSophisticated}, the hindering factor of density is the reason that in the orange highlighted area only the nodes are annotated as human-like detected differences. This is exactly the correct behavior of our algorithm since this area is an area with a high proportion of colored pixels. However, given the visual impression, this area would rather not obfuscate any edges for a human. An area which would obfuscate edges for humans would rather look like the orange highlighted area in Figure~\ref{fig:DataAugMoreSophisticated}. Consequently, it would be interesting and beneficial to incorporate this details into data augmentation. To achieve this, there is still a knowledge gap to close since, again exemplarily explained using density, even the pixel density thresholds used for the decision on density are an approximation inspired by research on scatterplots \cite{Bertini2005}. We had to use this approximation, since, to the best of our knowledge, it is currently not known in the current body of work what the precise human notion of pixel density with respect to graph density is. When this knowledge gap is closed, the parameters of our DFS-algorithm can be adapted to the research insights. It is fair to draw inspiration from scatterplot research thus the nodes of graphs and scatterplots are quite similar to each other.

Currently, \textbf{Mask R-CNN} learned the average selection patterns of humans for structural differences in directed acyclic graph pairs. However, assuming that each person shows certain differences in the differences they recognize, it would be interesting to adapt our Mask R-CNN implementation to the individual user in the future, in addition to using our learned model in an interactive system. For this, the training phase of the network would have to be reactivated and the supervised feedback would have to come from the user. The user's first response could be whether she agrees with the prediction. If yes, the prediction is also taken as supervised feedback. If no, the user has to mark the differences she detects and then these are taken as supervised feedback for the network. Here, the challenges reside mostly in the calculations of the supervised feedback based on the user's feedback.
\FloatBarrier


%
%
%
%
%
%
%
%

\subsubsection{Conclusion} 
\label{ssub:conclusion}
We propose an image- and segmentation-based change detection approach that is able to predict and visually highlight structural differences -- comparable as to related visual comparison systems such as TreeJuxtaposer \cite{10.1145/1201775.882291}, DifferenceMaps \cite{archambault2009structural} or egoComp \cite{Liu2017a}. In contrast to those systems, we propose an approach to incorporate the human notion by learning the human detected differences. As we use a supervised neural network, the learning phase requires the graph differences as labeled segmentation masks. After the network is trained and then used only in the same context, e.g., for the same user or for the same class of graph, the difference computation can be skipped and the trained model can be used directly for prediction.

	\section*{Acknowledgements}

	We greatly benefited from the feedback of Prof. Ch. Kersting and Prof. G\"{u}nther Wallner. We would like to thank both of you very much for this.

	\bibliographystyle{abbrvurl}
	\bibliography{meine_Diss_references}

\end{document}